Accepted, *Neural Computation*.

# Derandomizing Dense Binary Hypervector Codebooks for Quantized Scalars

Dmitri Rachkovskij[1,2], Evgeny Osipov[1], Olexander Volkov[2], Denis Kleyko[3,4], Vaclav Snasel[5]

[1]Department of Computer Science, Electrical and Space Engineering, Luleå University of Technology, 971 87 Luleå, Sweden
[2]Institute of Information Technologies and Systems, 03187 Kyiv, Ukraine
[3]AI, Robotics and Cybersecurity Center (ARC) and Department of Computer Science, Örebro University, 70182 Örebro, Sweden
[4]Intelligent Systems Lab, Research Institutes of Sweden, 16440 Kista, Sweden
[5]Department of Computer Science, VSB - Technical University of Ostrava, 708 00 Ostrava-Poruba, Czech Republic

## Abstract

Hyperdimensional computing and vector symbolic architectures often represent quantized scalar levels by dense binary codebooks whose level-to-level similarity is intended to follow a prescribed function of scalar separation. At finite dimensionality, randomized scalar codebook constructions deviate from this target because of sampling noise, random-start imbalance, update-count fluctuations, component dependence, and finite-capacity effects.

We develop a transition-based derandomization framework for dense binary scalar codebooks across two target-similarity families, with similarity decaying exponentially or linearly with level separation. The framework separates the target similarity law, the derandomization variant, and the concrete generator construction, making explicit how initialization, selection, update, and capacity-handling mechanisms shape the induced similarity profile. We formalize derandomization variants that separately constrain initial Hamming weight, update-count variability, and update balance, thereby controlling distinct sources of finite-dimensional error. For each family and variant, we derive the induced mean similarity, identify realization-wise and mean target-matching regimes, and derive exact finite-dimensional expressions for bias, variance, and root-mean-square error.

Simulations across dimensions, quantization ranges, reference scalar levels, and generator constructions validate the theory and show how each constraint removes or reduces a specific source of similarity mismatch. The results provide practical guidance for choosing scalar codebook generators that more closely match a desired similarity law under finite-dimensional and hardware-relevant constraints.

**Keywords:** Hyperdimensional computing; Dense binary hypervectors; Scalar codebook generation; Derandomization; Similarity-preserving encodings

# 1 Introduction

**Hyperdimensional Computing / Vector Symbolic Architectures** (HDC/VSA) (Kanerva, 2009), (Gayler, 2004), (Neubert et al., 2019), (Thomas et al., 2021), (Aygun et al., 2023), (Clarkson et al., 2023), (Thomas et al., 2023), (Stock et al., 2024), (Reimann, 2025) constitute a multi-faceted framework at the intersection of symbolic and neural-network based Artificial Intelligence. Beyond conventional machine learning (ML) such as classification (Kleyko et al., 2018), (Ge & Parhi, 2020), (Dewulf et al., 2025), HDC/VSA has been explored in resource-constrained scenarios, including embedded and edge computing. This interest is reinforced by the fact that HDC/VSA provides a computational paradigm for emerging hardware (Kleyko, Davies, et al., 2022).

A central task in HDC/VSA is to transform data (scalars, vectors, sequences, graphs) into hypervectors (HVs) so that HV similarity reflects task-relevant relationships in the input space. These transformations are typically randomized and nonlinear. The required dimensionality and HV format depend on the application: complex structured representations, such as graphs, may require thousands of dimensions, whereas edge computing benefits from compact representations as well as energy- and compute-efficient formats such as binary HVs. This makes the behavior of HDC/VSA representations at practical finite dimensionalities an important design consideration.

For numeric data, HDC/VSA commonly uses fixed randomized nonlinear transformations designed so that nearby inputs produce similar HVs while distant inputs produce dissimilar HVs. Prior work distinguishes three broad approaches for mapping numeric data to HVs: compositional, explicit receptive-field-based, and random-projection-based (Rachkovskij et al., 2013), (Kleyko, Rachkovskij, et al., 2022).

In this paper, we focus on the scalar representations under the **compositional approach** (Rachkovskij & Fedoseyeva, 1990), (Kussul et al., 1991a, 1991b) as summarized in (Rachkovskij, Slipchenko, Kussul, et al., 2005; Rachkovskij, Slipchenko, Misuno, et al., 2005). In this approach, each component of a numeric input vector (a scalar) is first transformed into a similarity-preserving scalar HV. These scalar HVs can subsequently be bound to their respective feature identifiers and superimposed to encode numeric feature vectors or used as constituents of more general compositional representations (e.g., (Plate, 2000), (Rachkovskij, 2001), (Kahawala et al., 2026)), where preserving controlled similarity relations at the scalar level remains important. Here we isolate the scalar-codebook stage and study the accuracy and implementation properties of its generator constructions.

The original scalar codebook constructions were developed for highly randomized distributed representations and specialized wide-word hardware, where noise tolerance and compositional capacity were central design considerations (Kussul et al., 1991a), (Kussul et al., 1991b), (Rachkovskij, Slipchenko, Kussul, et al., 2005). Accordingly, they did not enforce constant Hamming weight or fixed numbers of component updates between scalar levels.

Current HDC deployments increasingly use dense binary HVs (but see (Haputhanthri et al., 2026)) directly in similarity search and classification and emphasize predictable behavior and implementation efficiency. In this setting, uncontrolled initialization and update randomness can produce finite-dimensional (finite-D) deviations from the intended similarity law. Existing work has largely focused on proposing or applying scalar-codebook constructions rather than systematically characterizing these deviations and the mechanisms that produce them. This motivates a systematic treatment of dense binary scalar generators that progressively reduce randomness while targeting a prescribed similarity law, together with finite-D analysis of the resulting bias and variability.

**To address this gap**, we study dense binary scalar codebooks with target one-density $p = 1/2$ under two prescribed similarity-law families, with exponential or linear dependence on level separation. We introduce a common set of derandomization variants that constrain initial Hamming weight (i.e., the number of 1-components), the number of component updates per transition, and the balance between $1 \rightarrow 0$ and $0 \rightarrow 1$ updates. For each family, we specify concrete generator constructions, derive their induced mean similarity and finite-D bias, variance, and root-mean-square error (RMSE), and validate the predictions by simulation. The present study is restricted to scalar codebooks; full vector encodings and task-level behavior are left to companion work.

Our main **contributions** are as follows:

1. We develop a unified derandomization framework for dense binary scalar hypervector codebooks, covering exponential and linear target-similarity families and a common set of constraints on initialization, update count, and update balance.
2. We specify family- and variant-specific generator constructions and analyze how component selection, reuse, update, and capacity handling mechanisms shape the induced similarity behavior, including conditions for realization-wise or mean target matching.
3. We develop exact finite-D average-case theory for the induced similarities, deriving their means and variances and the resulting bias and root-mean-square error relative to the target similarity law.
4. We validate the theory empirically across similarity families, generator constructions, dimensionalities, quantization ranges, and reference scalar levels, and identify how the derandomization mechanisms affect similarity accuracy, providing mechanism-based guidance for generator and parameter selection.

# 2 Background and Related Work

## 2.1 HDC/VSA: Hypervector Formats and Similarity

In HDC/VSA, data items are represented by HVs whose dimensionality (number of components) is typically in the hundreds to tens of thousands, depending on the application. The underlying VSA model determines the HV format: common choices include real-valued representations such as Holographic Reduced Representations (Plate, 1995), complex-valued variants (Plate, 2003), bipolar $\{-1,+1\}$ (Multiply-Add-Permute) (Gayler, 1998), dense binary (Binary Spatter Code) (Kanerva, 1994), (Kanerva, 1996), sparse binary (Sparse Binary Distributed Representations) (Rachkovskij, 2001), (Laiho et al., 2015).

A standard design principle is that unrelated / dissimilar data items (such as symbols) are assigned independently generated HVs. Similarity is then controlled by construction for related items. Similarity measures depend on the format: dot-product or cosine similarity is common for real-valued HVs, while for binary HVs similarity is often evaluated via Hamming distance (the number of different components) or overlap-based dot-product similarity. Throughout this paper, we focus on dense binary HVs with target one-density $p = 1/2$.

## 2.2 Scalar Hypervectors in the Compositional Approach

Under the compositional approach, each quantized scalar input is assigned an HV from a codebook constructed to enforce a desired similarity profile over scalar levels, such as exponential or linear decay with level separation. To represent a full numeric feature vector, scalar value HVs are bound to their corresponding feature identifiers, and the resulting HVs are aggregated by superposition. The scalar codebook contributes to the similarity structure inherited by the final vector-level HV, and thus influences performance and implementation requirements. This paper isolates the scalar stage and studies scalar codebook generators with progressively reduced randomness. We analyze how closely the resulting similarity profiles follow their intended targets at practical dimensionalities and quantization levels.

## 2.3 Previous and Related Work

### 2.3.1 Historical Context and Motivation

The original scalar codebook constructions of (Rachkovskij, Slipchenko, Kussul, et al., 2005) were developed under assumptions that differ from many current HDC deployments. They targeted specialized “associative–projective neurocomputers” (Kussul et al., 1991a), (Kussul et al., 1991b) supporting efficient bit-wise operations on very wide machine words of 512 and 2048 bits, potentially extending to the full HV dimensionality, and tolerated unreliable bits in memory and even in registers. They were also motivated by brain-inspired, fully-distributed high-dimensional representations, where variability and noise are intrinsic, and strict constraints on Hamming weight (i.e., the number of 1-components) or update size are not naturally justified. Accordingly, the original constructions were highly randomized and did not enforce constant Hamming weight in either the HVs or their updates. These randomized high-dimensional hypervectors were intended primarily as constituents for constructing larger compositional structures, including nodes, edges, subgraphs, and graphs (Kahawala et al., 2026), without changing dimensionality. This motivated large initial dimensionalities both to support composition and to stabilize the statistical properties of randomized representations. The original work was developed in the setting of Sparse Binary Distributed Representations (Rachkovskij & Kussul, 2001), where sparsity is a central design feature, although the underlying ideas extend to other HDC formats.

Subsequent work has adapted this compositional approach to different HV formats and basic operations used to construct compositional representations, with increasing emphasis on direct use of HVs in classification or similarity search rather than primarily as constituents of larger structured representations. This shifted the design priorities from neurobiological plausibility and highly randomized, noise-tolerant representations, toward more deterministic, hardware-friendly constructions with predictable behavior in similarity search and classification and efficient implementations in resource-constrained systems.

These changes make several open issues practically important. Dense binary HVs can support reduced dimensionalities and simpler hardware datapaths, while uncontrolled randomness and Hamming-weight drift can translate into variability of the similarity structure and consequently of task performance. Existing work has largely focused on proposing and deploying scalar constructions, rather than systematically analyzing their finite-D behavior. This motivates scalar generators that reduce or remove selected sources of randomness while targeting a prescribed similarity law, together with finite-D accuracy analysis of realized similarities under practical dimensionalities and run budgets.

Early schemes for similarity-preserving scalar HVs were proposed independently by (Rachkovskij & Fedoseyeva, 1990) and (Smith & Stanford, 1990). These and related constructions (Rachkovskij, Slipchenko, Kussul, et al., 2005), (Purdy, 2016) can be interpreted as implementing a "component flow" from the HV of one quantization level to the next. From this perspective, whether components are selected for update with or without replacement determines the resulting similarity-law family.

#### 2.3.2 Exponential Target Similarity Law

Scatter codes (Smith & Stanford, 1990) and the Subtract–Add scheme in (Rachkovskij, Slipchenko, Kussul, et al., 2005), also used in (Rachkovskij & Fedoseyeva, 1990), generate correlated scalar HVs by iteratively updating a random reference HV. At each level, selected components are flipped and remain eligible for selection at subsequent levels. This with-replacement mechanism yields an expected similarity that decreases geometrically with level separation, producing an exponential target similarity law. Applications to classification were discussed, for example, in (Rachkovskij & Fedoseyeva, 1990) and (Kleyko et al., 2018).

#### 2.3.3 Linear Target Similarity Law

Here, linear refers to dependence on the nonnegative level separation; for a fixed reference level, the similarity profile over comparison levels is triangular (piecewise linear). The concatenation construction of (Rachkovskij, Slipchenko, Kussul, et al., 2005) assigns independent random reference HVs to the minimum and maximum levels. At level $q$, the first $(Q-q)D/Q$ components are taken from the first reference HV and the last $qD/Q$ components from the second, producing a linear similarity profile between the two references. Related constructions obtain linear similarity by progressively flipping components of a single reference HV without replacement (Widdows & Cohen, 2015), (Rahimi et al., 2016), (Kleyko et al., 2018). This can be viewed as a two-reference construction where the second reference is the complement of the first, yielding a linear decay to zero similarity at the maximum level.

Many implementations flip a fixed number of previously unused components per level: $D/(2Q)$ as in (Rahimi et al., 2016), (Smets et al., 2023) yields a linear slope toward the similarity between the initial and final references. While flipping $D/Q$ components, as in (Hsiao et al., 2021), (Huang et al., 2021), (Smets, Rachkovskij, Osipov, Volkov, et al., 2025), (Smets, Rachkovskij, Osipov, Van Leekwijck, et al., 2025) reaches zero similarity at $q = Q$. Related interpolation schemes were proposed in (Cohen et al., 2013), (Widdows & Cohen, 2015), (Kleyko et al., 2018). Piecewise-linear concave similarity profiles can be formed by combining multiple reference HVs with different slopes.

#### 2.3.4 Deterministic Localist Encodings with Linear Similarity

Deterministic or partially distributed binary encodings provide low-randomness counterparts to distributed constructions. In “float” (closeness, bar, pointer, sliding) codes (Penz, 1987), level $q$ is represented by $W$ consecutive active units starting at position $q$, requiring $Q+W$ units in total. The dot-product similarity decreases linearly from $W$ to 0 between the codes of $q$ and $q+i$ for $i \in [0,W]$, and is zero for larger separations. For $W = Q$, each code has half of the units active, and each level transition swaps one active and one inactive component, preserving constant Hamming weight. Multifloat codes (Kussul et al., 1993) combine floats of different lengths to obtain piecewise-linear concave similarity profiles.

In thermometer (unary) codes (Penz, 1987), (Jackel et al., 1987), $q$ of $Q$ units are active, so each level transition activates one additional component. The stack code of (Healy & Caudell, 1997) appends the complement ($0 \leftrightarrow 1$) of the thermometer code, doubling dimensionality and yielding a float code with $W = Q$.

Recent work on compositional pipelines includes float encoding followed by multiplicative binding with feature-ID HVs: (Vergés et al., 2023) termed it “flocet” following a typo in (Rachkovskij, Slipchenko, Kussul, et al., 2005). Unary or thermometric encodings in compositional pipelines have also been considered in (Kleyko et al., 2021), (Roodsari et al., 2024), (Moghadam et al., 2024), (Masum et al., 2025), (Moghadam et al., 2025).

## 3 Methodology and Derandomization Framework

We propose and study a unified derandomization framework for constructing dense binary HV codebooks for quantized scalar levels. The framework distinguishes the prescribed target similarity law, the derandomization variant that specifies which sources of finite-D variability are constrained, and the generator construction that implements those constraints. Section 3.1 develops the common derandomization hierarchy used for both exponential and linear similarity families; section 3.2 defines the notation, similarity measures, and target laws; and section 3.3 defines the error metrics and evaluation protocol used in our analysis.

### 3.1 Framework and Derandomization Hierarchy

For scalar representations, we begin with highly randomized generator constructions similar in spirit to the Subtract–Add scheme of (Rachkovskij, Slipchenko, Kussul, et al., 2005), in which both the initial hypervector and the components updated at each level transition are sampled probabilistically. Consequently, in the Baseline constructions, both $|\mathbf{x}_q|$ and the number of components updated in a level transition vary across realizations and, generally, across levels within a realization. The derandomization hierarchy introduced below progressively constrains selected sources of this variability. $q$,

Let $\{\mathbf{x}_q\}_{q=0}^{Q}$, with $\mathbf{x}_q \in \{0,1\}^D$, denote the hypervector codebook, and consider the level transition from $\mathbf{x}_q$ to $\mathbf{x}_{q+1}$. For the constructions studied here, the principal distinction between the two target-similarity families is whether components may be reused across level transitions. In the exponential family, components remain eligible for selection in later transitions; selection is therefore with replacement across levels. In the linear family, a component selected in one transition is not selected again; so selection is without replacement across levels. In the constructions analyzed later, these two across-level selection rules lead, respectively, to exponential and linear dependence of expected similarity on level separation.

Throughout this paper, we consider dense binary hypervectors with target one-density $p = 1/2$, so the target $|\mathbf{x}_q| = pD = D/2$ at every level. Depending on the derandomization variant, this target is satisfied either in expectation or in every realization. Within either target family, derandomization progressively controls three sources of variability: (1) the initial value $|\mathbf{x}_0|$; (2) the total number of components updated in each level transition; (3) the balance between $1\to 0$ and $0\to 1$ updates within a transition.

Fixing $|\mathbf{x}_0|$ removes random-start imbalance in the starting hypervector. Fixing the total update count removes update-count variability. Enforcing equal numbers of $1\to 0$ and $0\to 1$ updates prevents Hamming-weight drift and preserves $|\mathbf{x}_q|$ at every level. We call an update balanced in expectation when the expected numbers of $1\to 0$ and $0\to 1$ changes are equal. Such updates preserve $\mathbf{E}[|\mathbf{x}_q|]$, although $|\mathbf{x}_q|$ may vary across levels and realizations. Equal-swap updates enforce this balance in every transition. In the linear family, the Fixed-kpD constructions preserve exactly $D/2$ ones at every level and match the linear target law realization-wise. The deterministic linear constructions introduced in section 5 are instances of this variant.

Figure 1 summarizes the derandomization hierarchy considered within both target-similarity families, and Table 1 states the corresponding constraints. The choice between exponential and linear target laws is separate from this hierarchy and determines the across-level component-reuse rule. Each target family realizes these variants through its own generator constructions.

A derandomization variant does not uniquely determine a generator construction. Different constructions may satisfy the same variant constraints while differing in their concrete component-selection and update mechanisms. Such differences may preserve the same induced mean-similarity law while affecting finite-D bias and variance.

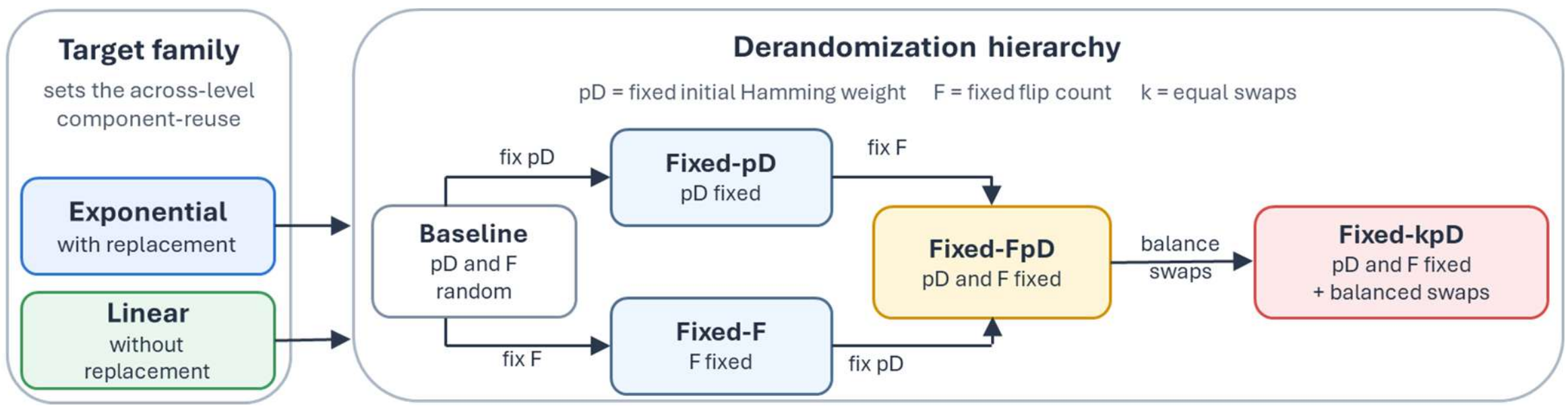

Figure 1: Scalar codebook derandomization hierarchy. The target family determines the across-level component-reuse rule: exponential-family constructions use with-replacement reuse, whereas linear-family constructions use without-replacement reuse. Both target families are organized by the same derandomization hierarchy, but each family realizes the variants through its own generator constructions. Baseline (Bernoulli) branches into the single-constraint variants Fixed-pD and Fixed-F; Fixed-FpD combines these constraints; Fixed-kpD additionally enforces equal-swap balance. Each variant may admit one or more family-specific generator constructions.

Table 1: Constraints defining the derandomization variants at $p = 1/2$. "Balanced in expectation" means $\mathrm{E}[k_{10}] = \mathrm{E}[k_{01}]$ for the constructions studied in sections 4 and 5. The quantities $D/2$, $F$, and $k$ are assumed to be integers whenever required.

| Variant | $\lvert\mathbf{x}_0\rvert$ | Updates per transition | Number of $1 \to 0/0 \to 1$ flips | $\lvert\mathbf{x}_q\rvert, q > 0$ |
|---|---|---|---|---|
| Baseline | Random; $\mathrm{E}[\lvert\mathbf{x}_0\rvert] = D/2$ | Random | Random; balanced in expectation | Not fixed; $\mathrm{E}[\lvert\mathbf{x}_q\rvert] = D/2$ |
| Fixed-pD | $D/2$ exactly | Random | Random; balanced in expectation | Not fixed; $\mathrm{E}[\lvert\mathbf{x}_q\rvert] = D/2$ |
| Fixed-F | Random; $\mathrm{E}[\lvert\mathbf{x}_0\rvert] = D/2$ | $F$ exactly | Random; balanced in expectation | Not fixed; $\mathrm{E}[\lvert\mathbf{x}_q\rvert] = D/2$ |
| Fixed-FpD | $D/2$ exactly | $F$ exactly | Random; balanced in expectation | Not fixed; $\mathrm{E}[\lvert\mathbf{x}_q\rvert] = D/2$ |
| Fixed-kpD | $D/2$ exactly | $2k$ exactly | $k$ in each direction | $D/2$ exactly |

### 3.2 Notation, Similarity Measures, and Target Laws

Table 2 summarizes the notation used throughout the theoretical and empirical analysis.

Table 2: Main notation.

| Symbol | Meaning |
|---|---|
| $D$ | hypervector dimensionality (number of components) |
| $p$ | target one-density (throughout this paper $p = 1/2$) |
| $q$ | quantization-level index |
| $Q$ | maximum quantization-level index |
| $a$, $b$ | generic level indices |
| $\mathbf{x}_0$ | initial hypervector |
| $\mathbf{x}_q$ | hypervector assigned to level $q$ |
| $\Delta := \lvert b - a \rvert$ | level separation |
| w.o.r. | without replacement |
| w.r. | with replacement |
| $r$ | nominal per-component flip rate in one level transition |
| ρ | one-step correlation coefficient (decay parameter) for the exponential family |
| $F$ | fixed total number of component flips in one level transition |
| $k_{10}$ | number of 1→0 flips in one level transition |
| $k_{01}$ | number of 0→1 flips in one level transition |
| $O(a,b)$ | Overlap between $\mathbf{x}_a$ and $\mathbf{x}_b$ |
| $S(a,b)$ | Normalized overlap similarity |
| $\mathrm{sim}_{\mathrm{ham}}(a,b)$ | Normalized Hamming similarity |

$|\mathbf{x}_q|$ changes according to $|\mathbf{x}_{q+1}| = |\mathbf{x}_q| - k_{10} + k_{01}$. Consequently, $\mathrm{E}[k_{10}] = \mathrm{E}[k_{01}]$ preserves $\mathrm{E}[|\mathbf{x}_q|]$, whereas $k_{10} = k_{01}$ preserves the realized $|\mathbf{x}_q|$ in every transition. If $\mathrm{E}[|\mathbf{x}_0|] = D/2$, balance in expectation gives $\mathrm{E}[|\mathbf{x}_q|] = D/2$ for all $q$. If $|\mathbf{x}_0| = D/2$ and every transition is exactly balanced, then $|\mathbf{x}_q| = D/2$ for every level and every realization. Constant Hamming weight is therefore a property of the complete generator construction, including both its initialization and its update rule, and is stronger than balance in expectation.

For binary HVs $\mathbf{u}$ and $\mathbf{v}$, define their overlap and Hamming distance by $\mathrm{sim}_{\mathrm{dot}}(\mathbf{u},\mathbf{v}) := \sum_i (u_i \wedge v_i)$ and $\mathrm{dist}_{\mathrm{Ham}}(\mathbf{u},\mathbf{v}) := \sum_i (u_i \oplus v_i)$, where $\wedge$ and $\oplus$ denote component-wise AND and XOR, respectively. For codebook levels $a$ and $b$, let $O(a,b) = \mathrm{sim}_{\mathrm{dot}}(\mathbf{x}_a,\mathbf{x}_b)$. The normalized overlap similarity is $S(a,b) = O(a,b)/(pD)$. At $p = 1/2$, this reduces to $S(a,b) = 2\,O(a,b)/D$. We also use the normalized Hamming similarity $\mathrm{sim}_{\mathrm{ham}}(a,b) = 1 - \mathrm{dist}_{\mathrm{Ham}}(\mathbf{x}_a,\mathbf{x}_b)/D$. The Hamming weight is $|\mathbf{u}| = \sum_i u_i$, the number of components equal to 1, and its one-density is $|\mathbf{u}|/D$.

For example, let $D = 8$, $p = 1/2$, and $\mathbf{x}_a = (1,1,1,1,0,0,0,0)$, $\mathbf{x}_b = (1,1,0,1,1,0,0,0)$. Both vectors have Hamming weight four, and one-density $1/2$. They overlap in three 1-components and differ in two components, so $O(a,b) = 3$, $\mathrm{dist}_{\mathrm{Ham}}(\mathbf{x}_a,\mathbf{x}_b) = 2$. Therefore, $S(a,b)$=3/4, $\mathrm{sim}_{\mathrm{ham}}(a,b) = 1 - 2/8 = 3/4$.

For a given generator construction, $T(a,b)$ denotes the intended normalized similarity between levels $a$, $b$. We call $T$ the target similarity law. Both target laws considered in this paper are shift-invariant: they depend on the level indices only through their separation, $T(a,b) = T(\Delta)$, $\Delta := |b - a|$. The exponential target law is

$$T_{\text{exp}}(a,b) = T_{\text{exp}}(\Delta) = 1/2 + (1/2)\,\rho^{\Delta}, \quad (3.1)$$

where ρ is the prescribed one-step decay parameter of the target law. Section 4 specifies how the update parameter $r$ of each exponential-family generator is chosen to reproduce this decay in the induced mean similarity. The linear target law is $T_{\text{lin}}(a,b) = T_{\text{lin}}(\Delta) = \max(0, 1 - \Delta \cdot F/D) = \max(0, 1 - \Delta/Q)$. Here, linear refers to the dependence on the nonnegative level separation $\Delta = |b - a|$; as a function of the signed difference $a - b$, this law is triangular and piecewise linear. Section 5 specifies how the linear-family generators realize this law using $F = D/Q$, equivalently $r = F/D = 1/Q$, including the finite-capacity corrections of the random-count constructions.

For a realized codebook, $S(a,b)$ is generally a random quantity. Its generator-induced mean similarity is $\mathrm{E}[S(a,b)]$, which we compare with the intended target law $T(a,b)$. For a fixed anchor level $a$, varying $b$ over $\{0, \dots, Q\}$ gives a similarity profile for $T(a,b)$, $S(a,b)$, or $\mathrm{E}[S(a,b)]$. Although the target laws are anchor-independent, realized similarities and some construction-specific induced mean profiles may exhibit anchor dependence.

### 3.3 Error Metrics and Evaluation Protocol

This section defines the error quantities and their finite-$N$ empirical estimates used to compare the generator constructions with the target similarity law. Unless stated otherwise, $\mathrm{E}[.]$ and $\mathrm{Var}(.)$ denote expectation and variance over independent generator realizations. Empirical quantities carry the subscript emp and are computed from $N$ independent realizations.

#### 3.3.1 Point-wise Quantities and Finite-$N$ Empirical Estimates

For fixed levels $a,b$, the realized similarity $S(a,b)$ is a random variable. Its mean and variance over generator realizations are $\mu(a,b) := \mathrm{E}[S(a,b)]$, $\mathrm{v}(a,b) := \mathrm{Var}(S(a,b))$. The corresponding point-wise bias and mean-squared error relative to the target similarity law are $\mathrm{bias}(a,b) := \mu(a,b) - T(a,b)$, $\mathrm{MSE}(a,b) := \mathrm{E}[(S(a,b) - T(a,b))^2]$.

For $N$ independent realizations, let $S_n(a,b)$ denote the similarity in realization $n$. Define $\mu_{\text{emp}}(a,b) := (1/N) \cdot \sum_{n=1}^{N} S_n(a,b)$, $\mathrm{v}_{\text{emp}}(a,b) := (1/N) \cdot \sum_{n=1}^{N} (S_n(a,b) - \mu_{\text{emp}}(a,b))^2$, $\mathrm{bias}_{\text{emp}}(a,b) := \mu_{\text{emp}}(a,b) - T(a,b)$, $\mathrm{MSE}_{\text{emp}}(a,b) := (1/N) \cdot \sum_{n=1}^{N} (S_n(a,b) - T(a,b))^2$. With the $1/N$ normalization used for $\mathrm{v}_{\text{emp}}(a,b)$, they satisfy the exact finite-sample identity $\mathrm{MSE}_{\text{emp}}(a,b) = \mathrm{v}_{\text{emp}}(a,b) + \mathrm{bias}^2_{\text{emp}}(a,b)$.

#### 3.3.2 Reference-Level Averaging and RMSE, Bias, and Variance Summaries

For a fixed anchor level $a$, define averaging over all comparison levels $b$ by $\mathrm{mean}_b[g(a,b)] := (1/(Q+1)) \sum_{b=0}^{Q} g(a,b)$. We report the corresponding root-mean-square (RMS) summaries over $b$; $\mathrm{BiasRMS}(a)$, $\mathrm{VarRMS}(a)$, and $\mathrm{RMSE}(a)$ denote the nonnegative square roots of the corresponding squared definitions: $\mathrm{BiasRMS}^2(a) := \mathrm{mean}_b[\mathrm{bias}^2(a,b)]$, $\mathrm{VarRMS}^2(a) := \mathrm{mean}_b[\mathrm{v}(a,b)]$, $\mathrm{RMSE}^2(a) := \mathrm{mean}_b[\mathrm{E}[(S(a,b) - T(a,b))^2]]$. Their empirical counterparts are $\mathrm{BiasRMS}^2_{\mathrm{emp}}(a) := \mathrm{mean}_b[\mathrm{bias}^2_{\mathrm{emp}}(a,b)]$, $\mathrm{VarRMS}^2_{\mathrm{emp}}(a) := \mathrm{mean}_b[\mathrm{v}_{\mathrm{emp}}(a,b)]$, $\mathrm{RMSE}^2_{\mathrm{emp}}(a) := \mathrm{mean}_b[\mathrm{MSE}_{\mathrm{emp}}(a,b)]$.

For each fixed $(a,b)$, the bias–variance decomposition gives $MSE(a,b) = \mathrm{v}(a,b) + \mathrm{bias}^2(a,b)$. Averaging over $b$ yields $\mathrm{RMSE}^2(a) = \mathrm{VarRMS}^2(a) + \mathrm{BiasRMS}^2(a)$. The empirical definitions satisfy the corresponding exact identity $\mathrm{RMSE}^2_{\mathrm{emp}}(a) = \mathrm{VarRMS}^2_{\mathrm{emp}}(a) + \mathrm{BiasRMS}^2_{\mathrm{emp}}(a)$.

#### 3.3.3 Finite-$N$ Interpretation and Comparison with Theory

The quantities $\mu(a,b)$, $\mathrm{v}(a,b)$, $\mathrm{bias}(a,b)$, $MSE(a,b)$ are defined over generator realizations and therefore do not depend on the number $N$ of Monte Carlo realizations. Their empirical counterparts are computed from $N$ independent realizations. The empirical mean and MSE are unbiased for their population counterparts, whereas the squared empirical bias and the empirical variance have finite-$N$ corrections because $\mathrm{v}_{\mathrm{emp}}$ is normalized by $1/N$; the corresponding expectation identities are given in supplementary note 3.1.

Section 6 provides the corresponding theoretical values $\mathrm{bias}_{\mathrm{th}}(a,b)$ and $\mathrm{v}_{\mathrm{th}}(a,b)$ for the generator constructions. Their anchor-wise summaries are computed using the same $b$-averaging definitions as in section 3.3.2: $\mathrm{BiasRMS}^2_{\mathrm{th}}(a) := \mathrm{mean}_b[\mathrm{bias}^2_{\mathrm{th}}(a,b)]$, $\mathrm{VarRMS}^2_{\mathrm{th}}(a) := \mathrm{mean}_b[\mathrm{v}_{\mathrm{th}}(a,b)]$, and $\mathrm{RMSE}^2_{\mathrm{th}}(a) := \mathrm{mean}_b[\mathrm{v}_{\mathrm{th}}(a,b) + \mathrm{bias}^2_{\mathrm{th}}(a,b)]$. Section 7 reports the corresponding empirical summaries, with the primary comparison between $\mathrm{RMSE}_{\mathrm{emp}}(a)$ and $\mathrm{RMSE}_{\mathrm{th}}(a)$ at the same anchor level.

# 4 Exponential-Similarity Scalar Codebook Generation

This section develops scalar codebook generators for the exponential family. Their defining structural principle is with-replacement component reuse across level transitions: components updated at earlier transitions remain eligible for selection at later transitions. We consider two update schemes, Subtract–Add (SA) and Single–Commit XOR (SC). For each scheme, we state the update rule, the density-preserving parameterization, the resulting one-step decay factor, and the vectorized implementation cost. Supplementary note 1 provides full derivations and properties of the SA and SC mean-similarity laws (1.1 and 1.2); comparison of the two schemes (1.3); and the complementary Markov-chain and eigenvalue interpretation (1.4). SA is included to connect the framework with classical scalar-codebook generators and to compare its exponential decay and implementation cost with SC. The derandomization variants and their finite-D analysis are developed for the SC scheme in sections 4.3 and 6.1.

## 4.1 Common Exponential Template

The exponential family uses with-replacement reuse across levels: every component remains eligible for selection at each transition $q \to q+1$, irrespective of earlier selections. The update rule is level-homogeneous and permutation-invariant over component indices. Conditional on $\mathbf{x}_q$, every component therefore has the same marginal transition rule. Fixed-count variants may introduce dependence among component selections within a transition; the mean-similarity analysis requires exchangeability rather than within-step independence.

**Initialization and Hamming weight control:** We initialize $\mathbf{x}_0$ either with i.i.d. $\text{Bernoulli}(1/2)$ components (so $\text{E}(|\mathbf{x}_0|) = D/2$) or with a constant-weight vector $\mathbf{x}_0$ with $|\mathbf{x}_0| = D/2$ exactly, chosen uniformly among all subsets of that size. Balanced-in-expectation updates preserve $\text{E}[|\mathbf{x}_q|] = D/2$, whereas equal-swap constructions enforce $|\mathbf{x}_q| = D/2$ at every level $q$ in every realization by design. Under these conditions, the induced mean similarity depends on $(a,b)$ only through $\Delta := |a - b|$, and follows the exponential law in Eq. (3.1), governed by a one-step decay factor ρ (also known as retention, persistence, correlation coefficient, lag-1 autocorrelation coefficient) that controls how rapidly similarity decays with level separation. Sections 4.2 and 4.3 specify ρ and the vectorized implementations for SA and SC. The Markov-chain interpretation is given in supplementary note 1.4.

## 4.2 Subtract–Add and Single–Commit Update Schemes

The SA scheme applies two substeps at each transition $q \rightarrow q+1$. Degrade changes each component currently equal to 1 to 0 with probability $r \in [0,1]$, while leaving components equal to 0 unchanged. Refill then changes each post-Degrade 0 to 1 with probability $\alpha := r/(1+r)$, while leaving components equal to 1 unchanged. At marginal one-density $1/2$, $(1 - r)/2 + \alpha\,(1+r)\,/2 = 1/2$, so the complete transition preserves the marginal one-density. The induced mean similarity is $\text{E}[S(a,b)] = T_{SA}(\Delta) = 1/2 + (1/2)\rho_{SA}^{\Delta}$, with $\rho_{SA} = (1 - r)/(1+r)$. The derivation and additional properties are given in supplementary note 1.1.

In vectorized form, $\mathbf{y} = \mathbf{x}_q \wedge \mathbf{M}_{\text{deg}}$, $\mathbf{x}_{q+1} = \mathbf{y} \vee \mathbf{M}_{\text{ref}}$, where $\mathbf{M}_{\text{deg}}$ has $\text{Bernoulli}(1 - r)$ and $\mathbf{M}_{\text{ref}}$ has $\text{Bernoulli}(\alpha)$ i.i.d. components. Both masks are sampled over all component indices; a sampled mask value changes a component only when its current state is eligible for the corresponding substep. The operation count per machine word is two Bernoulli-mask generations, one AND, and one OR.

The SC scheme applies one update at each transition $q \rightarrow q+1$. A mask $\mathbf{M}_q$ with i.i.d. $\text{Bernoulli}(r)$ components is generated, and $\mathbf{x}_{q+1} = \mathbf{x}_q \oplus \mathbf{M}_q$. Thus, each component flips independently with probability $r$ and at most once during a transition. At marginal one-density $1/2$, the $1 \rightarrow 0$ and $0 \rightarrow 1$ probabilities are equal, so the marginal one-density is preserved. The induced mean similarity is $\text{E}[S(a,b)] = T_{SC}(\Delta) = 1/2 + (1/2)\rho_{SC}^{\Delta}$, with $\rho_{SC} = 1 - 2r$. The derivation and additional properties are given in supplementary note 1.2. Because each component can flip only once per transition, executed and net flips coincide. The operation count per machine word is one Bernoulli-mask generation and one XOR.

## 4.3 Single–Commit Derandomization Variants and Generator Constructions

Section 3.1 defines the common derandomization variants. Here, we instantiate them for SC updates within the exponential family. At every level transition, the selected update set is committed once by XOR. The variants differ in their initialization, component-selection mechanism, and constraints on the update set. All SC XOR variants have the exponential mean-similarity form $\mathrm{E}[S(a,b)] = T_{SC}(\Delta) = 1/2{+}(1/2)\rho_{SC}^{\Delta}$, with $\rho_{SC} = 1 - 2r$, where $r$ is the marginal per-component flip probability ($r = F/D$ for fixed-count updates and $r = 2k/D$ for equal-swap). For Baseline SC, this mean similarity profile follows from section 4.2; the corresponding finite-D results for all variants are developed in section 6.1.

For Baseline SC, $\mathbf{x}_0$ has i.i.d. Bernoulli$(1/2)$ components and each transition uses a fresh i.i.d. Bernoulli$(r)$ mask. Fixed-pD changes only the initialization, sampling $\mathbf{x}_0$ uniformly subject to $|\mathbf{x}_0| = D/2$. Fixed-F instead retains Bernoulli initialization but selects exactly $F = rD$ components uniformly at each transition. Fixed-FpD combines $|\mathbf{x}_0| = D/2$ with this fixed-count update. In all four variants, the $1 \to 0/0 \to 1$ split is not fixed to be equal. A representative generator construction is illustrated in Figure 2. Fixed-kpD additionally enforces exact balance: starting from $|\mathbf{x}_0| = D/2$, each transition selects $k$ current 1-components and $k$ current 0-components uniformly, with $F = 2k = rD$ and flips their union in one XOR commit. Hence $|\mathbf{x}_q| = D/2$ for every $q$, with zero variance in the total flip count and the $1 \to 0/0 \to 1$ split.

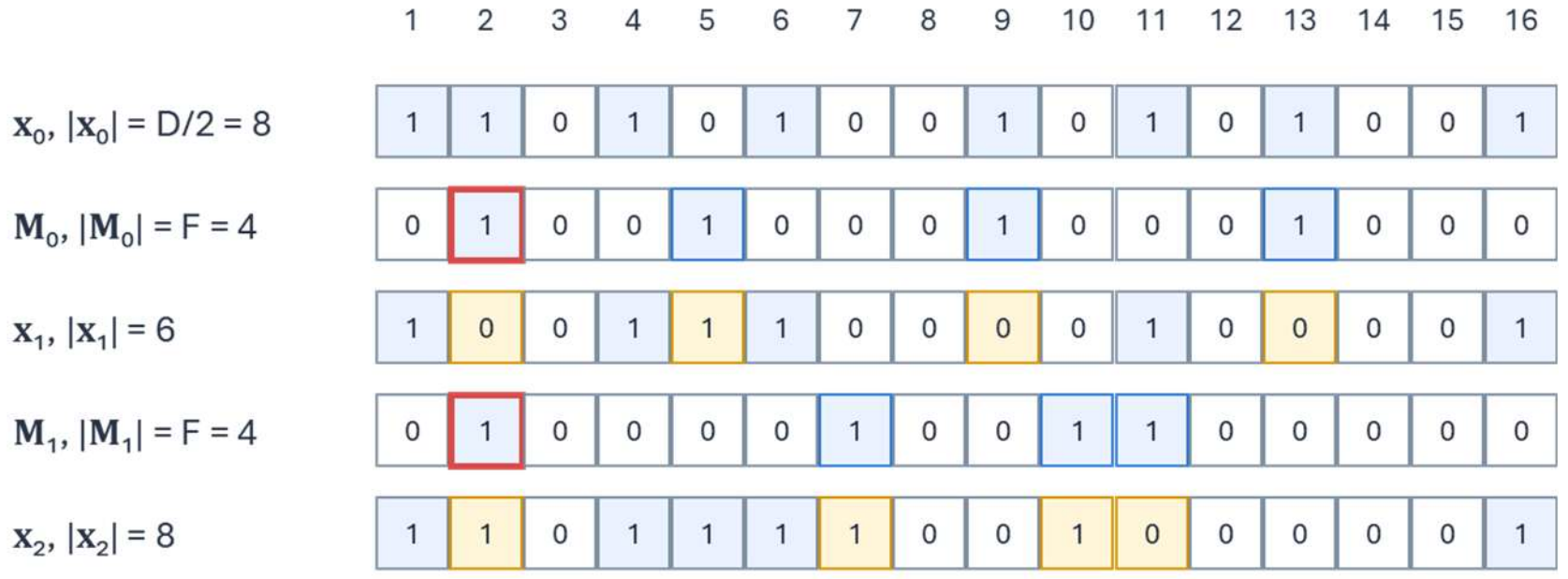

Figure 2. Exponential-family Single–Commit Fixed-FpD generator construction. The example uses $D = 16$, $|\mathbf{x}_0| = D/2 = 8$, and two successive updates with $|\mathbf{M}_0| = |\mathbf{M}_1| = F = 4$. Blue mask components are selected for flip; orange HV components change by XOR. The red outline marks the component selected in both masks, illustrating with-replacement reuse.

# 5 Linear-Similarity Scalar Codebook Generation

This section develops scalar-codebook generators for the linear family, whose defining structural principle is without-replacement component use across level transitions. Section 5.1 introduces this principle, the remaining-pool and deterministic equal-swap realizations, and the parameterization that yields linear decay over the full quantization range. Section 5.2 gives the corresponding mean-similarity analysis, including a general remaining-pool identity, random-count finite-capacity effects, and exactness conditions for fixed-count and equal-swap updates. Section 5.3 instantiates the five derandomization variants through their linear-family generator constructions. Proofs and detailed finite-capacity derivations are provided in supplementary note 2, while finite-D bias, variance, and RMSE are analyzed in section 6.2 and supplementary note 3.3.

### 5.1 Basic Idea

We construct scalar codebook generators that produce $\{\mathbf{x}_q\}_{q=0}^{Q} \subset \{0,1\}^D$ with linear target law $T_{\text{lin}}(\Delta) := \max(0,1-\Delta/Q)$. The key structural change in the linear family is without-replacement reuse across level transitions: each component can be flipped in at most one transition $q \to q+1$, $q = 0, \dots, Q-1$. The linear constructions preserve the SC update type used by their exponential SC counterparts, but change component reuse across levels: w.o.r. in the linear family, in contrast to w.r. reuse in the exponential family. This converts exponential decay into linear dependence on $\Delta$. How the linear target is attained depends on the construction: random-count constructions can have finite-capacity deviations near exhaustion, fixed-count constructions match it in mean overlap similarity and realization-wise normalized Hamming similarity, and Fixed-kpD matches it realization-wise for both similarities.

We instantiate this in two ways. The remaining-pool w.o.r. construction maintains a shrinking set $U_q$ of components not yet flipped by level $q$ (relative to the initial level $\mathbf{x}_0$). Each step flips a subset of $U_q$, and flipped components leave the pool (w.o.r. across levels). The resulting mean similarity is linear except for terminal deviations from truncation of a stochastic selection rule near pool exhaustion. Lemma 5.1 formalizes this. In deterministic equal-swap constructions, the updated components are fixed by design rather than sampled independently for each realization; this yields the linear similarity profile. We consider the five derandomization variants from section 3, specialized to the linear family via w.o.r., and the corresponding generator constructions with their update / selection / truncation mechanisms.

Throughout the linear-family constructions, $F$ denotes the nominal number of newly flipped components per level transition, with corresponding flip-rate parameter $r = F/D$. We use $F = D/Q$, equivalently $r = 1/Q$. For fixed-count without-replacement constructions, this choice uses all $D$ components exactly once over the $Q$ level transitions and exhausts the remaining pool at $q = Q$, yielding the linear similarity law $T_{\text{lin}}(\Delta) = 1 - \Delta/Q$. For random-count constructions, $F = D/Q$ is the nominal transition count; truncation near exhaustion can produce the finite-capacity corrections.

Other values of $F$ are not considered in this paper. If $F > D/Q$, fixed-count without-replacement updates exhaust the available components before $q = Q$, thereby shortening the usable quantization range. If $F < D/Q$, some components remain unflipped and therefore common even at the maximal separation $\Delta = Q$, underusing the available representation capacity (Smets, Rachkovskij, Osipov, Volkov, et al., 2025). The corresponding relation between transition size, capacity, and terminal similarity differs for $p < 1/2$, (Rachkovskij, Slipchenko, Kussul, et al., 2005), but that regime is outside the scope of this paper.

The linear constructions are related to existing work on Hyperdimensional Computing representations (Thomas et al., 2023), (Kleyko, Rachkovskij, et al., 2022), (Kleyko et al., 2023), on sequence-similarity encoders (Rachkovskij, 2024), (Rachkovskij & Kleyko, 2022), and on structured codes for HDC (Raviv, 2024). We do not rely on their specific constructions, but they provide broader context for the role of similarity profiles and structured HVs in HDC.

### 5.2 Analysis

In this section we state mean-similarity results for w.o.r. linear generators under assumptions on (i) how previously unused components are selected across levels and (ii) how selection is truncated when the remaining pool is near exhaustion. The statements are phrased entirely in terms of the remaining-pool process. Section 5.3 maps the generator constructions to these assumptions.

We first state Lemma 5.1, which expresses expected similarity $\mathrm{E}[S(a,b)]$ in terms of expected remaining-pool depletion. This lemma applies to any w.o.r. remaining-pool generator construction that flips only previously unused components, independently of the concrete selection mechanism, including fixed-order constructions that deterministically traverse a fixed ordering induced by an initial random permutation. The equal-swap constructions are treated separately because they yield the deterministic linear law. In that variant, each level transition additionally enforces $k_{10} = k_{01} = k = D/(2Q)$, and therefore $|\mathbf{x}_q| = D/2$ exactly for all $q$.

5.2.1 Lemma 5.1 (W.o.r. Identity for Expected Similarity)

Consider producing $\{\mathbf{x}_q\}_{q=0}^{Q} \subset \{0,1\}^D$ by per-level XOR updates driven by a per-level flip mask (a binary pattern indicating which components are flipped in the transition $q \rightarrow q+1$) under w.o.r. dynamics: each component is flipped in at most one transition $q \rightarrow q+1$, $q = 0, \dots, Q-1$. Let $U_q \subseteq \{1, \dots, D\}$ denote the remaining pool after forming level $q$ , i.e. the set of $\mathbf{x}_0$ component indices not yet flipped in transitions up to level $q$. Thus $U_0 = \{1, \dots, D\}$ and $U_{q+1} \subseteq U_q$. Under w.o.r., each $U_q$ is determined by the supports of the flip masks used up to level $q$ (i.e., by which components have been flipped up to that level).

Assume the following two conditions:

(A) Complement-invariant initialization: $\mathbf{x}_0$ and its component-wise complement $1 - \mathbf{x}_0$ have the same distribution. (Examples: $\mathbf{x}_0$ i.i.d. Bernoulli$(1/2)$; if $D$ is even, $\mathbf{x}_0$ uniform over all vectors in $\{0,1\}^D$ with exactly $D/2$ ones).

(B) Value-independent selection: For each transition $q \rightarrow q+1$, conditional on the current remaining pool $U_q$, the distribution of the flip mask used to form $\mathbf{x}_{q+1}$ from $\mathbf{x}_q$ does not depend on the component values of $\mathbf{x}_q$.

Let $J \sim \mathrm{Unif}(\{1, \dots, D\})$, independently of the generator randomness. All probabilities and expectations below are over the joint randomness of $J$, $\mathbf{x}_0$, and the generator. Without loss of generality assume $b \geq a$ (if $b < a$, swap $a$ and $b$). Then:

(i) $\Pr(x_a(J) = x_b(J)) = 1 - (\mathrm{E}[|U_a|] - \mathrm{E}[|U_b|])/D$, (ii) $\mathrm{E}[S(a,b)] = \Pr(x_a(J) = x_b(J))$.

#### 5.2.2 Random-Count Finite-Capacity Effects

Two random-count mechanisms are used in section 5.3. In Pool-Bernoulli, the expected executed flip count equals $F = D/Q$ while $|U_q| \geq F$, whereas all remaining components are selected once $|U_q| < F$. Its mean similarity therefore follows $T_{\text{lin}}$ closely away from pool exhaustion, with finite-capacity corrections becoming appreciable near the terminal levels; see Proposition 1 in supplementary note 2.2.1.

Random-count Perm-Scan traverses a fixed component ordering using random proposed update count $f_{q+1}$. For transition $q \to q+1$, it generates $f_{q+1} \sim \text{Binomial}(D, 1/Q)$, but executes at most the number of components remaining in the unused pool. Because overshoots are truncated whereas undershoots leave components unflipped, the induced mean overlap similarity lies slightly above the linear target near pool exhaustion; Proposition 2 in supplementary note 2.2.2 gives the exact mean law. At the terminal pair $(0, Q)$, whose separation is $\Delta = Q$, and for which $T_{\text{lin}}(Q) = 0$, the residual mean overlap is, for large $D$, $\mathrm{E}[S(0,Q)] \approx ((1 - 1/Q)/(2\pi D))^{1/2}$, see Remark 3 in supplementary note 2.2.3. Empirically, the fixed-order variants show a slightly larger terminal bias than Pool-Bernoulli.

#### 5.2.3 Fixed-Count and Equal-Swap Exactness

Suppose that every transition flips exactly $F = D/Q$ previously unused components, where $F$ is an integer. Then the remaining pool is depleted deterministically and, for $a + \Delta \leq Q$, $\mathrm{E}[S(a, a+\Delta)] = 1 - \Delta \cdot F/D = T_{\text{lin}}(\Delta)$, while $\text{dist}_{\text{Ham}}(\mathbf{x}_a, \mathbf{x}_{a+\Delta}) = \Delta \cdot F$. Thus, fixed-count constructions attain the target mean overlap similarity and the exact normalized Hamming similarity; see Corollary 4 in supplementary note 2.2.4. They do not necessarily preserve $|\mathbf{x}_q|$, because the split between $1 \to 0$ and $0 \to 1$ flips remains random. The Fixed-kpD constructions in section 5.3 additionally enforce $k = D/(2Q)$ flips in each direction per transition. Consequently, $|\mathbf{x}_q| = D/2$, $O(a,b) = D/2 - \Delta \cdot k$, and $S(a,b) = 1 - \Delta/Q = T_{\text{lin}}(\Delta)$ for all levels exactly, in every realization.

### 5.3 Linear-Similarity Derandomization Variants and Generator Constructions

The linear-family generators instantiate the five derandomization variants using the remaining-pool and deterministic mechanisms analyzed above. Baseline uses i.i.d. $\text{Bernoulli}(1/2)$ initialization and one of two random-count constructions: Pool-Bernoulli or random-count Perm-Scan. Fixed-pD uses the same constructions but initializes $\mathbf{x}_0$ uniformly subject to $|\mathbf{x}_0| = D/2$. Their finite-capacity corrections are in section 5.2.2.

Fixed-F uses Bernoulli initialization and flips exactly $F = D/Q$ previously unused components per transition, implemented either by Pool-Uniform or fixed-count Perm-Scan. Fixed-FpD combines the same fixed-count constructions with $|\mathbf{x}_0| = D/2$. As shown in section 5.2.3, both variants attain the target mean overlap similarity and exact normalized Hamming similarity, although $|\mathbf{x}_q|$ need not remain fixed because the numbers of $1 \to 0/0 \to 1$ flips are not constrained to be equal. A representative generator construction is illustrated in Figure 3.

Fixed-kpD additionally enforces exact balance with $k = D/(2Q)$ flips in each direction per transition. Block-Swap realizes this by complementing successive disjoint balanced blocks, whereas Float generates the levels by cyclic shifts of a contiguous half-one vector. Both preserve $|\mathbf{x}_q| = D/2$ and realize $T_{\text{lin}}(\Delta)$ exactly in every realization. The quantities $F$ and $k$ are assumed to be integers where required.

| | 1 | 2 | 3 | 4 | 5 | 6 | 7 | 8 | 9 | 10 | 11 | 12 | 13 | 14 | 15 | 16 |
|---|---|---|---|---|---|---|---|---|---|---|---|---|---|---|---|---|
| $\mathbf{x}_0$, $\lvert\mathbf{x}_0\rvert = D/2 = 8$ | 1 | 1 | 0 | 1 | 0 | 1 | 0 | 0 | 1 | 0 | 1 | 0 | 1 | 0 | 0 | 1 |
| $\mathbf{M}_0$, $\lvert\mathbf{M}_0\rvert = F = 4$ | 0 | 1 | 0 | 0 | 1 | 0 | 0 | 0 | 1 | 0 | 0 | 0 | 1 | 0 | 0 | 0 |
| $\mathbf{x}_1$, $\lvert\mathbf{x}_1\rvert = 6$ | 1 | 0 | 0 | 1 | 1 | 1 | 0 | 0 | 0 | 0 | 1 | 0 | 0 | 0 | 0 | 1 |
| $\mathbf{M}_1$, $\lvert\mathbf{M}_1\rvert = F = 4$ | 1 | 0 | 0 | 1 | 0 | 1 | 1 | 0 | 0 | 0 | 0 | 0 | 0 | 0 | 0 | 0 |
| $\mathbf{x}_2$, $\lvert\mathbf{x}_2\rvert = 4$ | 0 | 0 | 0 | 0 | 1 | 0 | 1 | 0 | 0 | 0 | 1 | 0 | 0 | 0 | 0 | 1 |

Figure 3. Linear-family Fixed-FpD Pool-Uniform generator construction. The example uses $D = 16$, $|\mathbf{x}_0| = D/2 = 8$, and two successive updates with $|\mathbf{M}_0| = |\mathbf{M}_1| = F = 4$. Blue mask components are selected for flip; orange HV components change by XOR. Red positions in $\mathbf{M}_1$ were selected in $\mathbf{M}_0$ and are no longer eligible, illustrating the without-replacement rule.

## 6 Accuracy of Approximating the Target Similarity Laws

This section summarizes exact finite-D expressions for $\mu_{\text{th}}(a,b)$, $\text{bias}_{\text{th}}(a,b)$, and $v_{\text{th}}(a,b)$ for the generators introduced earlier, relative to the target laws $T_{\text{exp}}$ and $T_{\text{lin}}$ defined in section 3. The corresponding $\text{RMSE}_{\text{th}}$ is obtained using the evaluation protocol of section 3.3 and compared with the experiments in section 7.2. Full derivations are provided in supplementary note 3, which mirrors the structure of this section.

### 6.1 Finite-D Accuracy for Exponential Similarity Family

We analyze the SC scheme and the generator constructions introduced in section 4.3. For Baseline, Fixed-pD, Fixed-F, and Fixed-FpD, the finite-D formulas are expressed in terms of two one-step parameters, $\rho_{\text{step}}$ and $\gamma_{\text{step}}$, which summarize the marginal and pair-wise behavior of a single transition. Fixed-kpD is handled separately via an overlap-moment recursion. Because $S(a,b)$, $\text{bias}_{\text{th}}(a,b)$, and $v_{\text{th}}(a,b)$ are symmetric in $(a,b)$, we assume $a \le b$, so that $\Delta = |a - b| = b - a$.

#### 6.1.1 Finite-D Formulas and Variance Summary

Let $M_q \subset \{1,\ldots,D\}$ denote the index set flipped in the level transition (step) $q \to q+1$, and use the $\pm 1$ encoding $y_q(i) := 2x_q(i) - 1$. Define $m_q(i) \in \{+1, -1\}$ by $m_q(i) = -1$ iff $i \in M_q$ and $m_q(i) = +1$ otherwise, so $y_{q+1}(i) = m_q(i) y_q(i)$. Define the one-step parameters (level and component permutation invariant) $\rho_{\text{step}} := \mathrm{E}[m_q(i)]$ and $\gamma_{\text{step}} := \mathrm{E}[m_q(i) m_q(j)]$ for $i \neq j$. $\rho_{\text{step}}$ determines the induced mean similarity profile $\mu_{\text{th}}(a,b)$, whereas $\gamma_{\text{step}}$ captures within-step dependence between distinct components and enters the finite-D variance. Independent within-step flips give $\gamma_{\text{step}} = \rho_{\text{step}}^2$; fixed-count flip selection generally does not.

**Mean similarity and bias.** Across $\Delta$ transitions, $\mathrm{E}[y_a(i)y_b(i)] = \rho_{\text{step}}^{\Delta}$. Global sign-flip symmetry at $p = 1/2$ implies $\mathrm{E}[y_q(i)] = 0$. Hence $\mu_{\text{th}}(a,b) := \mathrm{E}[S(a,b)] = 1/2 + (1/2)\rho_{\text{step}}^{\Delta}$. For Baseline, Fixed-pD, Fixed-F, and Fixed-FpD, the marginal per-component flip probability is $r$, with $r = F/D$ in the fixed-count variants. Hence $\rho_{\text{step}} = \mathrm{E}[m_q(i)] = 1 - 2r = \rho$, so $\mu_{\text{th}}(a,b) = T_{\exp}(\Delta)$ and $\text{bias}_{\text{th}}(a,b) = 0$. The complete derivation is given in supplementary note 3.2.

**Variance template.** Define $A_q := \sum_i y_q(i)$ and $C_{\text{ab}} := \sum_i (y_a(i)y_b(i))$. Then $O(a,b) = \sum_i x_a(i)x_b(i) = D/4 + (A_a + A_b + C_{\text{ab}})/4$. Under the global sign flip $\mathrm{Cov}(A_a + A_b, C_{\text{ab}}) = 0$, thus

$$\mathrm{v}_{\text{th}}(a,b) := \mathrm{Var}(S(a,b)) = (\mathrm{Var}(A_a + A_b) + \mathrm{Var}(C_{\text{ab}}))\ /\ (4D^2). \tag{6.1}$$

Derivation shows that

$$\mathrm{Var}(C_{\text{ab}}) = D(1 - \rho_{\text{step}}^{2\Delta}) + D(D-1)(\gamma_{\text{step}}^{\Delta} - \rho_{\text{step}}^{2\Delta}). \tag{6.2}$$

The variant-specific values of $\rho_{\text{step}}$, $\gamma_{\text{step}}$, and $\mathrm{Var}(A_a + A_b)$ are summarized below, their derivations and a compact tabular summary are provided in supplementary note 3.2.

For Bernoulli$(1/2)$ initialization, $y_q$ remains i.i.d. Rademacher across components, implying $\mathrm{Cov}(A_a, A_b) = D\rho_{\text{step}}^{\Delta}$, and

$$\mathrm{Var}(A_a + A_b) = 2D + 2D\rho_{\text{step}}^{\Delta}. \tag{6.3}$$

For Baseline, $\rho_{\text{step}} = 1 - 2r$ and $\gamma_{\text{step}} = \rho_{\text{step}}^2$. For Fixed-F, $\rho_{\text{step}} = 1 - 2F/D$ and $\gamma_{\text{step}} = 1 - 4F(D-F)/(D(D-1))$. Substitution into (6.1)-(6.3) gives $\mathrm{v}_{\text{th}}(a,b)$.

For $|\mathbf{x}_0| = D/2$, we have $\mathrm{E}[y_0(i)] = 0$ and $\mathrm{E}[y_0(i)y_0(j)] = -1/(D-1)$ for $i \neq j$. Propagation through the level transitions gives

$$\mathrm{Var}(A_a + A_b) = D(2 - \gamma_{\text{step}}^{a} - \gamma_{\text{step}}^{b} + 2\rho_{\text{step}}^{\Delta}(1 - \gamma_{\text{step}}^{a})). \tag{6.4}$$

Fixed-pD uses $\gamma_{\text{step}} = \rho_{\text{step}}^2$, as in Baseline, whereas Fixed-FpD uses $\gamma_{\text{step}} = 1 - 4F(D-F)/(D(D-1))$, as in Fixed-F. These provide $\mathrm{Var}(A_a + A_b)$ and $\mathrm{Var}(C_{\text{ab}})$ to obtain $\mathrm{v}_{\text{th}}(a,b)$.

#### 6.1.2 Equal-Swap, Constant Hamming-Weight (Fixed-kpD)

Fixed-kpD preserves $|\mathbf{x}_q| = D/2$ at every level. With $k := rD/2$ we have $\rho_{\text{step}} = 1 - 4k/D = 1 - 2r = \rho$, so $\mu_{\text{th}}(a,b) = T_{\exp}(\Delta)$ and $\text{bias}_{\text{th}}(a,b) = 0$. Because its equal-swap update samples from the current ones and zeros, the common $(\rho_{\text{step}}, \gamma_{\text{step}})$ variance template does not apply. The finite-D variance derivation is given in supplementary note 3.2.4.

### 6.2 Finite-D Accuracy for Linear Similarity Family

For the deterministic equal-swap constructions Fixed-kpD Block-Swap and Float, $S(a,b) = T_{\mathrm{lin}}(\Delta)$ in every realization, so $\mathrm{bias}_{\mathrm{th}}(a,b) = \mathrm{v}_{\mathrm{th}}(a,b) = 0$. For the remaining generators, let $K$ denote the number of distinct components flipped between levels $a$ and $b$. In Fixed-F and Fixed-FpD, $K = \Delta \cdot F$ is deterministic. In the random-count Baseline and Fixed-pD constructions, $K$ is random because the proposed update counts are truncated by the remaining-pool capacity. Thus, the fixed-count variants have no update-count induced terminal bias, whereas the random-count variants acquire terminal bias and an additional variance term through $\mathrm{Var}(K)$.

6.2.1 Fixed-Count Linear Variants

The Fixed-F and Fixed-FpD results apply to both Pool-Uniform and Perm-Scan. With integer $F = D/Q$, both satisfy $\mu_{\mathrm{th}}(a,b) = T_{\mathrm{lin}}(\Delta)$ and therefore have zero bias. For Fixed-F, $\mathrm{Bernoulli}(1/2)$ initialization yields $\mathrm{v}_{\mathrm{th}}(a,b) = (1 - \Delta/Q)/D$. Fixed-FpD has the same mean but a different finite-D variance because the constraint $|\mathbf{x}_0| = D/2$ induces dependence across components; the resulting hypergeometric variance is derived in supplementary note 3.3.

6.2.2 Fixed-Order Random-Count Linear Variants

For both Fixed-pD and Baseline random-count Perm-Scan constructions, $\mathrm{E}[S(a,b) \mid K] = (D - K)/D$, giving $\mu_{\mathrm{th}}(a,b) = 1 - \mathrm{E}[K]/D$, $\mathrm{bias}_{\mathrm{th}}(a,b) = \Delta/Q - \mathrm{E}[K]/D$. For Baseline, the variance can be written as $\mathrm{v}_{\mathrm{th}}(a,b) = (1 - \mathrm{E}[K]/D)/D + \mathrm{Var}(K)/D^2$. For Fixed-pD, constant-weight initialization changes the conditional-overlap variance, while the same $\mathrm{Var}(K)/D^2$ term accounts for update-count variability. Under the Binomial proposed-count model, the required moments of $K$, together with the additional cumulative-flip-count moments required for Fixed-pD, are evaluated at finite $D$. Full derivations are given in supplementary note 3.3.

# 7 Experimental Investigation

This section validates the finite-D theory and illustrates the realized similarity profiles. Section 7.1 presents representative overlap-similarity profiles for the exponential and linear families, and section 7.2 compares empirical RMSE with the theoretical predictions.

## 7.1 Examples of Similarity Profiles

Figure 4 shows representative exponential-family overlap-similarity profiles from three independent realizations for the five derandomization variants. Increasing derandomization generally produces smoother realized profiles, although the visual ordering does not exactly coincide with the RMSE ordering in section 7.2.1.

Figure 5 shows representative linear-family overlap-similarity profiles from three independent realizations. The profiles become increasingly regular across the derandomization variants, with Fixed-kpD reproducing the linear target realization-wise by construction.

Corresponding Hamming-similarity profiles and additional parameter settings are given in supplementary note 4.1.

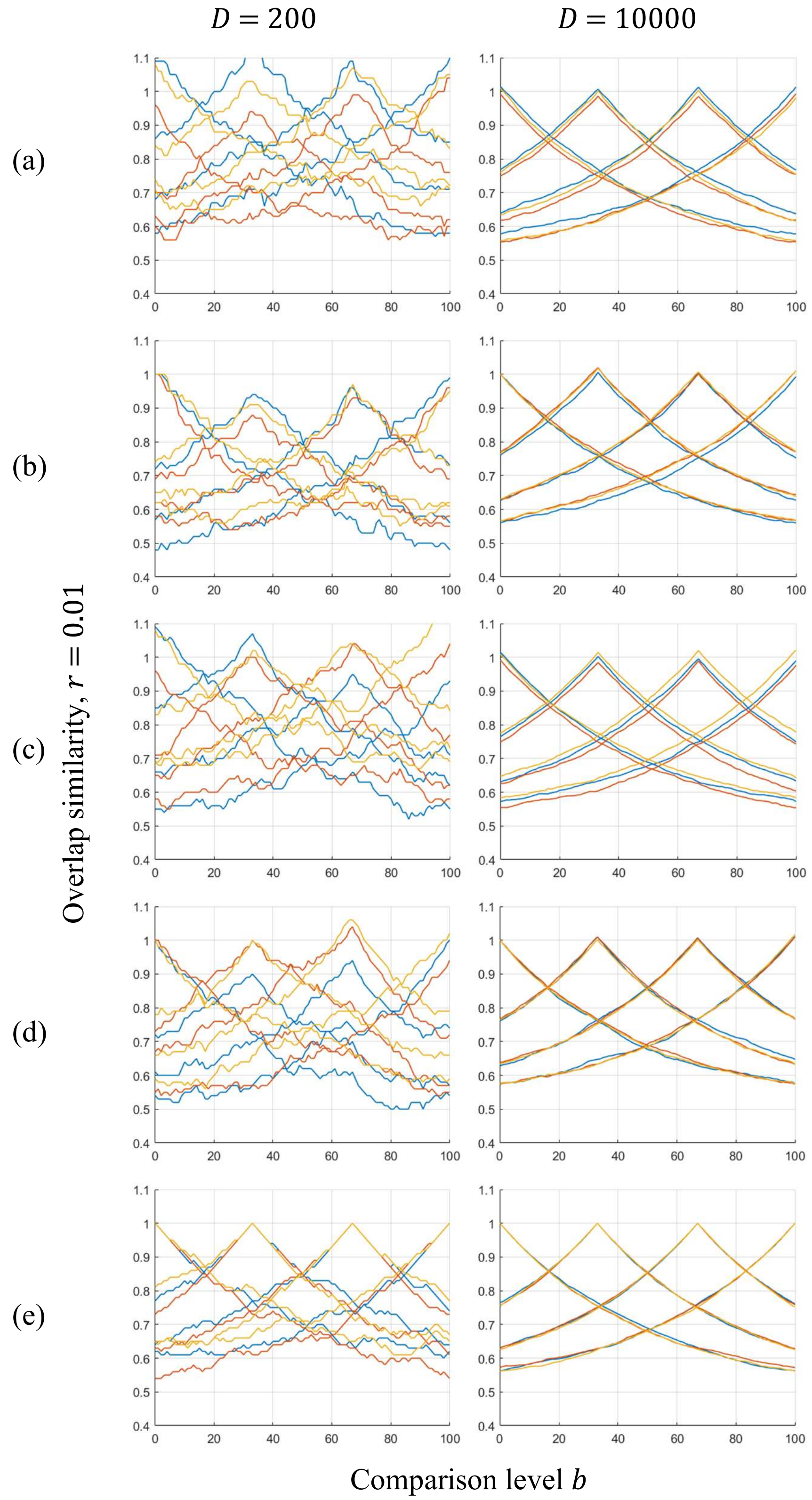

Figure 4: Exponential similarity profiles ($2\text{sim}_{\text{dot}}/D$ as a function of comparison level $b$) from three independent realizations, for anchors $a \in \{0,33,67,100\}$. Five generator constructions: Baseline (a), Fixed-pD (b), Fixed-F (c), Fixed-FpD (d), Fixed-kpD (e). $Q = 100$, $D \in \{200,10000\}$, $r = 0.01$.

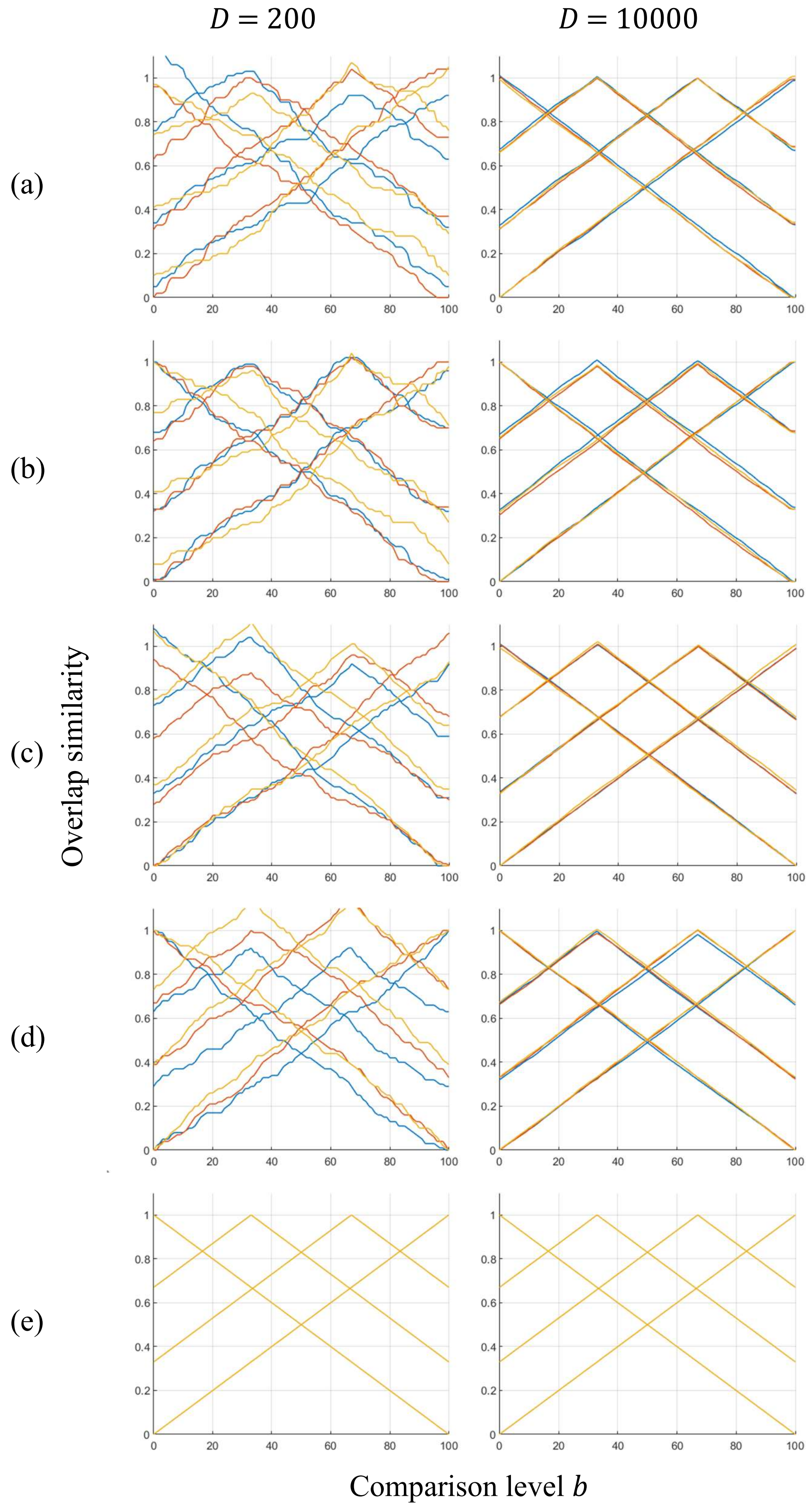

Figure 5: Linear similarity profiles ($2\mathrm{sim}_{\mathrm{dot}}/D$ as a function of comparison level $b$) from three independent realizations, for anchors $a \in \{0,33,67,100\}$. Five generator constructions: Baseline (a), Fixed-pD (b), Fixed-F (c), Fixed-FpD (d), Fixed-kpD (e). $Q = 100$, $D \in \{200,10000\}$.

### 7.2 RMSE of Similarity: Experiments versus Theory

Using the error metrics of section 3.3, we compare empirical anchor-wise $\mathrm{RMSE}_{\mathrm{emp}}(a)$ with the theoretical predictions of section 6, using the same anchor-wise averaging convention.

#### 7.2.1 Exponential Similarity Family

For the exponential similarity family, Table 3 compares empirical and theoretical anchor-wise RMSE across the five derandomization variants and flip rates $r$. Additional parameter settings are reported in supplementary note 4.2.

Table 3: Exponential similarity family: $\mathrm{RMSE}_{\mathrm{emp}}(a)$ and $\mathrm{RMSE}_{\mathrm{th}}(a)$ for anchors $a \in \{0,33,67,100\}$ and $r \in \{0.01,0.02,0.05,0.10,0.20\}$. 'Mean' averages RMSE over anchors. Parameters: $D = 200$, $Q = 100$, $10000$ realizations. Theory rows follow the empirical rows for each generator.

| Variant | Generator | $r$ | $a = 0$ | $a = 33$ | $a = 67$ | $a = 100$ | Mean |
|---|---|---|---|---|---|---|---|
| 1 | Baseline | 0.01 | 0.066729 | 0.068688 | 0.068732 | 0.066715 | 0.067716 |
| 1 | Baseline | 0.02 | 0.064383 | 0.066804 | 0.066408 | 0.064496 | 0.065523 |
| 1 | Baseline | 0.05 | 0.062688 | 0.063955 | 0.064030 | 0.062906 | 0.063395 |
| 1 | Baseline | 0.10 | 0.061918 | 0.062556 | 0.062725 | 0.061906 | 0.062276 |
| 1 | Baseline | 0.20 | 0.061407 | 0.061920 | 0.061590 | 0.061625 | 0.061636 |
| 1 theory | Baseline | 0.01 | 0.067227 | 0.068927 | 0.068927 | 0.067227 | 0.068077 |
| 1 theory | Baseline | 0.02 | 0.064815 | 0.066910 | 0.066910 | 0.064815 | 0.065863 |
| 1 theory | Baseline | 0.05 | 0.062709 | 0.063995 | 0.063995 | 0.062709 | 0.063352 |
| 1 theory | Baseline | 0.10 | 0.061963 | 0.062581 | 0.062581 | 0.061963 | 0.062272 |
| 1 theory | Baseline | 0.20 | 0.061584 | 0.061828 | 0.061828 | 0.061584 | 0.061706 |
| | | | | | | | |
| 2 | Fixed-pD | 0.01 | 0.043314 | 0.059840 | 0.063591 | 0.063186 | 0.057483 |
| 2 | Fixed-pD | 0.02 | 0.046645 | 0.063938 | 0.065462 | 0.063563 | 0.059902 |
| 2 | Fixed-pD | 0.05 | 0.048522 | 0.063254 | 0.063321 | 0.062259 | 0.059339 |
| 2 | Fixed-pD | 0.10 | 0.049257 | 0.062440 | 0.062317 | 0.061459 | 0.058868 |
| 2 | Fixed-pD | 0.20 | 0.049650 | 0.061556 | 0.061756 | 0.061401 | 0.058591 |
| 2 theory | Fixed-pD | 0.01 | 0.043422 | 0.059677 | 0.063892 | 0.063620 | 0.057653 |
| 2 theory | Fixed-pD | 0.02 | 0.046737 | 0.063847 | 0.065379 | 0.063502 | 0.059866 |
| 2 theory | Fixed-pD | 0.05 | 0.048680 | 0.063414 | 0.063482 | 0.062187 | 0.059441 |
| 2 theory | Fixed-pD | 0.10 | 0.049308 | 0.062305 | 0.062305 | 0.061685 | 0.058901 |
| 2 theory | Fixed-pD | 0.20 | 0.049612 | 0.061671 | 0.061671 | 0.061426 | 0.058595 |
| | | | | | | | |
| 3 | Fixed-F | 0.01 | 0.065236 | 0.066440 | 0.066160 | 0.065170 | 0.065751 |
| 3 | Fixed-F | 0.02 | 0.063698 | 0.065025 | 0.065177 | 0.063354 | 0.064314 |
| 3 | Fixed-F | 0.05 | 0.062030 | 0.063253 | 0.062977 | 0.062050 | 0.062577 |
| 3 | Fixed-F | 0.10 | 0.061580 | 0.062353 | 0.061908 | 0.061800 | 0.061910 |
| 3 | Fixed-F | 0.20 | 0.061404 | 0.061330 | 0.061592 | 0.061491 | 0.061455 |
| 3 theory | Fixed-F | 0.01 | 0.065080 | 0.066281 | 0.066281 | 0.065080 | 0.065681 |
| 3 theory | Fixed-F | 0.02 | 0.063595 | 0.064844 | 0.064844 | 0.063595 | 0.064220 |
| 3 theory | Fixed-F | 0.05 | 0.062190 | 0.062978 | 0.062978 | 0.062190 | 0.062584 |
| 3 theory | Fixed-F | 0.10 | 0.061686 | 0.062032 | 0.062032 | 0.061686 | 0.061859 |
| 3 theory | Fixed-F | 0.20 | 0.061427 | 0.061516 | 0.061516 | 0.061427 | 0.061472 |
| | | | | | | | |
| 4 | Fixed-FpD | 0.01 | 0.040203 | 0.056592 | 0.061208 | 0.060877 | 0.054720 |
| 4 | Fixed-FpD | 0.02 | 0.044950 | 0.061525 | 0.063206 | 0.062512 | 0.058048 |
| 4 | Fixed-FpD | 0.05 | 0.048091 | 0.062148 | 0.062821 | 0.061915 | 0.058744 |
| 4 | Fixed-FpD | 0.10 | 0.048943 | 0.061676 | 0.061805 | 0.061831 | 0.058564 |
| 4 | Fixed-FpD | 0.20 | 0.049432 | 0.061582 | 0.061711 | 0.061210 | 0.058483 |
| 4 theory | Fixed-FpD | 0.01 | 0.040036 | 0.056655 | 0.061063 | 0.061370 | 0.054781 |
| 4 theory | Fixed-FpD | 0.02 | 0.045039 | 0.061704 | 0.063273 | 0.062263 | 0.058070 |
| 4 theory | Fixed-FpD | 0.05 | 0.048013 | 0.062391 | 0.062459 | 0.061666 | 0.058632 |
| 4 theory | Fixed-FpD | 0.10 | 0.048961 | 0.061755 | 0.061755 | 0.061408 | 0.058470 |
| 4 theory | Fixed-FpD | 0.20 | 0.049418 | 0.061359 | 0.061359 | 0.061270 | 0.058352 |
| | | | | | | | |
| 5 | Fixed-kpD | 0.01 | 0.025899 | 0.019695 | 0.019470 | 0.025707 | 0.022693 |
| 5 | Fixed-kpD | 0.02 | 0.030873 | 0.026307 | 0.026179 | 0.030932 | 0.028573 |
| 5 | Fixed-kpD | 0.05 | 0.033549 | 0.031673 | 0.031762 | 0.033659 | 0.032661 |
| 5 | Fixed-kpD | 0.10 | 0.034446 | 0.033597 | 0.033652 | 0.034553 | 0.034062 |
| 5 | Fixed-kpD | 0.20 | 0.034872 | 0.034479 | 0.034534 | 0.034927 | 0.034703 |
| 5 theory | Fixed-kpD | 0.01 | 0.025729 | 0.019508 | 0.019508 | 0.025729 | 0.022619 |
| 5 theory | Fixed-kpD | 0.02 | 0.030663 | 0.026237 | 0.026237 | 0.030663 | 0.028450 |
| 5 theory | Fixed-kpD | 0.05 | 0.033551 | 0.031749 | 0.031749 | 0.033551 | 0.032650 |
| 5 theory | Fixed-kpD | 0.10 | 0.034458 | 0.033628 | 0.033628 | 0.034458 | 0.034043 |
| 5 theory | Fixed-kpD | 0.20 | 0.034893 | 0.034513 | 0.034513 | 0.034893 | 0.034703 |

Across all variants and flip rates, the theoretical predictions closely match the experimental RMSE values in both magnitude and anchor dependence. Anchor-averaged RMSE follows the ordering Baseline $\approx$ Fixed-F $>$ Fixed-pD $>$ Fixed-FpD $>$ Fixed-kpD. The reduction from Baseline and Fixed-F to the constant-weight variants reflects the removal of random initial Hamming-weight imbalance, while Fixed-kpD gives the lowest RMSE, reflecting its additional equal-swap constraint and Hamming-weight preservation.

Anchor dependence differs across variants. For Baseline and Fixed-F, interior anchors contain a larger proportion of small-separation comparisons and exhibit slightly larger RMSE than endpoint anchors. For Fixed-pD and Fixed-FpD, fixing the Hamming weight of the initial HV suppresses variability at $a = 0$, producing a noticeably smaller RMSE there. For Fixed-kpD, the per-separation error is larger at greater separations; consequently, endpoint anchors, which contain a larger proportion of large-separation comparisons, have larger RMSE than interior anchors. The anchor-to-anchor spread generally narrows as $r$ increases, consistent with the faster approach of the separation-dependent variance terms to their long-separation regime. A detailed interpretation of these trends in terms of the finite-D variance expressions of section 6 and supplementary note 3 is given in supplementary note 4.2.2.

#### 7.2.2 Linear Similarity Family

For the linear similarity family, Table 4 compares empirical and theoretical RMSE and reports bias for the random-count variants. Additional parameter settings are reported in supplementary note 4.2.3.

Table 4: Linear similarity family: $\mathrm{RMSE_{emp}}(a)$ and $\mathrm{RMSE_{th}}(a)$ for anchors $a \in \{0,33,67,100\}$. Parameters: $D = 200$, $Q = 100$, $10000$ realizations. Bias is reported for Baseline and Fixed-pD. Theory rows follow the empirical rows for each generator.

RMSE

| Variant | Generator | $a = 0$ | $a = 33$ | $a = 67$ | $a = 100$ | Mean |
|---|---|---|---|---|---|---|
| 1 | Baseline (Pool-Bernoulli) | 0.068781 | 0.069788 | 0.070744 | 0.069581 | 0.069723 |
| 1 | Baseline (Perm-Scan) | 0.068995 | 0.070054 | 0.071395 | 0.070264 | 0.070177 |
| 1 theory | Baseline (Perm-Scan) | 0.069250 | 0.070040 | 0.070765 | 0.070162 | 0.070054 |
| 2 | Fixed-pD (Pool-Bernoulli) | 0.055733 | 0.066032 | 0.066322 | 0.057171 | 0.061315 |
| 2 | Fixed-pD (Perm-Scan) | 0.055820 | 0.066006 | 0.066568 | 0.058288 | 0.061670 |
| 2 theory | Fixed-pD (Perm-Scan) | 0.055793 | 0.065745 | 0.066428 | 0.058055 | 0.061505 |
| 3 | Fixed-F (Pool-Uniform) | 0.049807 | 0.059829 | 0.059734 | 0.049916 | 0.054822 |
| 3 | Fixed-F (Perm-Scan) | 0.050224 | 0.060341 | 0.059971 | 0.049781 | 0.055079 |
| 3 theory | Fixed-F (Perm-Scan) | 0.050000 | 0.059955 | 0.059955 | 0.050000 | 0.054977 |
| 4 | Fixed-FpD (Pool-Uniform) | 0.028929 | 0.055150 | 0.055760 | 0.028929 | 0.042192 |
| 4 | Fixed-FpD (Perm-Scan) | 0.028658 | 0.054637 | 0.054938 | 0.028658 | 0.041723 |
| 4 theory | Fixed-FpD (Perm-Scan) | 0.028795 | 0.055139 | 0.055139 | 0.028795 | 0.041967 |
| 5 | Fixed-kpD (Block-Swap) | 0.000000 | 0.000000 | 0.000000 | 0.000000 | 0.000000 |
| 5 | Fixed-kpD (Float) | 0.000000 | 0.000000 | 0.000000 | 0.000000 | 0.000000 |
| 5 theory | Fixed-kpD (Float) | 0.000000 | 0.000000 | 0.000000 | 0.000000 | 0.000000 |

Bias

| Variant | Generator | $a = 0$ | $a = 33$ | $a = 67$ | $a = 100$ | Mean |
|---|---|---|---|---|---|---|
| 1 | Baseline (Pool-Bernoulli) | 0.004456 | 0.004539 | 0.004679 | 0.026600 | 0.010068 |
| 1 | Baseline (Perm-Scan) | 0.004720 | 0.004712 | 0.004735 | 0.027312 | 0.010370 |
| 1 theory | Baseline (Perm-Scan) | 0.004726 | 0.004726 | 0.004726 | 0.027148 | 0.010331 |
| 2 | Fixed-pD (Pool-Bernoulli) | 0.004493 | 0.004462 | 0.004488 | 0.025736 | 0.009795 |
| 2 | Fixed-pD (Perm-Scan) | 0.004697 | 0.004786 | 0.004767 | 0.027507 | 0.010439 |
| 2 theory | Fixed-pD (Perm-Scan) | 0.004726 | 0.004726 | 0.004726 | 0.027148 | 0.010331 |

The theoretical predictions closely match the corresponding experiments in both RMSE magnitude and anchor dependence. The anchor-averaged RMSE decreases in the order Baseline $>$ Fixed-pD $>$ Fixed-F $>$ Fixed-FpD $>$ Fixed-kpD. The fixed-count variants are essentially unbiased, whereas the random-count variants exhibit nonzero terminal bias near level $Q$. Fixed-kpD matches the linear target exactly, giving zero bias and variance.

Anchor dependence also differs across variants. Fixed-F and Fixed-FpD have lower RMSE at endpoint anchors than at interior anchors, with the endpoint reduction stronger for Fixed-FpD. For Baseline and Fixed-pD, the terminal boundary increases bias while reducing the remaining run-to-run variability; constant-weight initialization lowers the overall RMSE of Fixed-pD relative to Baseline. A detailed interpretation of these trends using the finite-D formulas of section 6 is given in supplementary note 4.2.4.

# 8 General Discussion

This section synthesizes the theoretical and empirical results into a mechanism-level account of scalar-codebook accuracy. We then use this account to derive generator-design implications and to identify limitations and directions for further work.

## 8.1 Mechanisms of Derandomization

The results show that scalar-codebook derandomization acts by constraining distinct sources of stochastic variability in the level-transition process. This motivates separating three objects: the prescribed target similarity law, the derandomization variant, and the generator construction. The target law specifies the desired dependence of similarity on level separation; the variant specifies which sources of randomness are constrained; and the generator specifies how those constraints are implemented through component selection, update rules, reuse, and capacity handling. This separation is essential because matching the induced mean similarity profile does not determine finite-D accuracy: generators with the same mean profile can retain different dependence structures and therefore different variance, RMSE, anchor dependence, and terminal behavior.

The derandomization variants clarify how distinct sources of variability are controlled. Constant-Hamming-weight initialization removes random-start imbalance in the initial HV. Fixing the update count additionally removes update-count variability, but does not eliminate fluctuations in the balance between $1 \rightarrow 0$ and $0 \rightarrow 1$ updates. Equal-swap updates impose this additional balance constraint and preserve Hamming weight at every level. These constraints are therefore complementary rather than interchangeable: each suppresses a different source of finite-D variability, and its effect on accuracy depends on which source dominates for the similarity family and parameter regime under consideration. The variants should therefore be viewed as controlling distinct constraint axes rather than as differing only in an overall degree of derandomization: Fixed-pD and Fixed-F isolate initialization and update-count control, respectively, whereas Fixed-FpD combines these controls and Fixed-kpD additionally enforces transition balance. Their relative effect on accuracy depends on the similarity family and operating regime.

The generator construction determines the dependence that remains after these constraints are imposed. Selection rules control component reuse and across-component dependence, update rules determine how selected components change state, and capacity handling determines the behavior when the eligible component pool becomes small. The distinction between the exponential and linear families illustrates the consequences. In the exponential family, with-replacement reuse produces the geometric induced mean profile, while the derandomization constraints primarily modify finite-D variability. In the linear family, without-replacement updates produce the linear dependence on separation, but the shrinking component pool introduces an additional finite-capacity effect. Random-count constructions can therefore develop terminal deviations even when their nominal update rule is calibrated to the target law.

This view also explains why different constructions within the same derandomization variant need not be statistically equivalent. Remaining-pool Bernoulli thinning and fixed-order proposed-count selection, for example, both implement random-count without-replacement updates, but they handle finite remaining capacity differently near pool exhaustion and can consequently produce different bias and variance. Taken together, the results reveal a hierarchy of mechanisms: single-component transition statistics govern the induced mean profile, dependence structures govern much of the finite-D variability, and finite-capacity effects can introduce additional systematic terminal bias. This hierarchy provides a more informative basis for generator design than mean target matching alone.

### 8.2 Accuracy and Design Implications

The results distinguish several levels of similarity-law matching. For the exponential SC constructions, fixing the marginal per-component flip probability fixes the induced mean decay, so different variants can share the same mean similarity profile while retaining different finite-D variability. For the linear fixed-count constructions, mean overlap similarity matches the target when $F = D/Q$, while normalized Hamming similarity matches it realization-wise; the deterministic equal-swap constructions achieve realization-wise matching for both similarities. Thus, mean and realization-wise target matching are distinct design properties and should not be conflated when assessing scalar-codebook accuracy.

Importantly, deviations from the target mean are themselves informative: in the linear random-count constructions they arise from the interaction between stochastic update counts and finite remaining capacity, with fixed-order truncation producing an upward terminal bias.

Finite-D accuracy therefore requires more than analysis of the mean profile. The bias–variance–RMSE decomposition reveals that similar RMSE values can arise from different error compositions: a constraint may reduce realization-to-realization variance without changing the mean, whereas finite-capacity truncation in linear random-count constructions can introduce systematic terminal bias. Derandomization can therefore change not only the magnitude of the approximation error but also the balance between its stochastic and systematic components.

The same analysis explains why accuracy depends on reference scalar level. Reference-level RMSE averages comparisons over different distributions of separations and, in the linear family, over different degrees of exposure to the terminal boundary. In the exponential family, the resulting reference-level dependence is governed mainly by the separation dependence in the finite-D variance and by whether initialization is Bernoulli or constant-Hamming-weight. In the linear family, this effect is supplemented by depletion of the remaining component pool and the associated finite-capacity corrections. The close agreement between these theoretical predictions and the simulations across the tested settings indicates that the observed anchor effects are structured consequences of the generator mechanisms rather than incidental simulation fluctuations.

The agreement between theory and simulation across the tested dimensions, quantization ranges, anchors, similarity families, variants, and constructions supports a mechanism-based design principle: generator selection should be governed by the dominant error source in the intended operating regime rather than by a universal ranking of derandomization strength. Constant-Hamming-weight initialization is effective when random-start imbalance is important; fixed-count updates are useful when update-count variability dominates; and equal-swap updates are required when Hamming-weight drift must be eliminated. For linear random-count constructions, finite-capacity effects must additionally be considered when the application emphasizes levels near the end of the quantization range. More generally, the appropriate construction is the least restrictive one that controls the error sources relevant to the intended operating regime, since additional constraints target specific stochastic mechanisms rather than providing a uniform improvement under all conditions.

For practical use, Table 5 summarizes target-law matching and finite-D accuracy across the considered derandomization variants in the exponential and linear families.

Table 5. Target-law matching and finite-D accuracy of the five presented derandomization variants. Variant definitions are given in Section 3.1; the exponential and linear generator constructions are described in Sections 4.3 and 5.3, respectively, and their finite-D accuracy is analyzed in Sections 6.1 and 6.2. The exponential-family entries refer to the SC constructions; exact linear-family statements assume the integer conditions stated in Section 5.

| Variant | Exponential similarity family | Linear similarity family |
|---|---|---|
| Baseline | Mean overlap similarity matches the exponential target, so bias is zero. Random initialization and random update masks produce nonzero overlap-similarity variance and RMSE. | Constructions: Pool-Bernoulli; random-count Perm-Scan. Near pool exhaustion, mean overlap similarity lies above the linear target, producing upward terminal bias. Overlap-similarity variance and RMSE are nonzero; Perm-Scan has a slightly larger terminal bias than Pool-Bernoulli. |
| Fixed-pD | Mean overlap similarity matches the exponential target, so bias is zero. Constant-Hamming-weight initialization removes random-start imbalance, but random update masks produce nonzero overlap-similarity variance and RMSE. | Constructions: Pool-Bernoulli; random-count Perm-Scan. Near pool exhaustion, mean overlap similarity lies above the linear target, producing upward terminal bias. Constant-Hamming-weight initialization removes random-start imbalance, but overlap-similarity variance and RMSE remain nonzero; Perm-Scan has a slightly larger terminal bias than Pool-Bernoulli. |
| Fixed-F | Mean overlap similarity matches the exponential target, so bias is zero. Fixed update counts remove update-count variability, but Bernoulli initialization and random fixed-count masks produce nonzero overlap-similarity variance and RMSE. | Constructions: Pool-Uniform; fixed-count Perm-Scan. Mean overlap similarity matches the linear target, so bias is zero, and normalized Hamming similarity matches the linear target exactly in every realization. Overlap-similarity variance and RMSE remain nonzero. |
| Fixed-FpD | Mean overlap similarity matches the exponential target, so bias is zero. Constant-Hamming-weight initialization and fixed update counts remove random-start imbalance and update-count variability, but random fixed-count masks produce nonzero overlap-similarity variance and RMSE. | Constructions: Pool-Uniform; fixed-count Perm-Scan. Mean overlap similarity matches the linear target, so bias is zero, and normalized Hamming similarity matches the linear target exactly in every realization. Constant-Hamming-weight initialization removes random-start imbalance, but overlap-similarity variance and RMSE remain nonzero. |
| Fixed-kpD | Mean overlap similarity matches the exponential target, so bias is zero. Equal-swap updates preserve Hamming weight at every level, but random selection of the swapped components produces nonzero overlap-similarity variance and RMSE. | Constructions: Block-Swap; Float. Overlap similarity and normalized Hamming similarity match the linear target exactly in every realization. Bias, variance, and RMSE are zero. |

### 8.3 Limitations and Future Directions

The present analysis focuses on dense binary scalar codebooks and on exponential and linear target-similarity families. Other HV representations and similarity laws may change both the relevant similarity statistics and the dependence induced by the generator mechanisms. An important direction is therefore to determine which elements of the proposed mechanism-level framework generalize across representations and target laws, and which require representation-specific finite-D analyses.

The theory also isolates the scalar-codebook layer rather than the complete HDC pipeline. In practical systems, scalar HVs undergo binding or permutation, aggregation, and downstream similarity or classification operations. Future work should therefore establish how scalar-level bias, variance, reference-level dependence, and boundary effects propagate through these transformations, including whether aggregation suppresses them, preserves them, or creates interactions with other sources of representation variability.

The exact finite-D random-count variance theory developed here applies to the fixed-order Perm-Scan construction. For adaptive Pool-Bernoulli, Proposition 1 gives the exact mean recursion and hence the bias, but a corresponding finite-D variance and RMSE theory for its state-dependent remaining-pool dynamics is not developed here. Extending the variance analysis to this construction would complete the theoretical comparison of the random-count mechanisms.

## 9 Conclusions

This paper developed a unified derandomization framework for dense binary scalar codebooks in hyperdimensional computing. The framework treats scalar codebook generation as a sequence of level transitions and separates three aspects that are often conflated: the target similarity law, the derandomization constraints, and the concrete generator construction. This separation makes it possible to analyze exponential and linear target-similarity families within a common transition-based view, while still explaining why generators with the same target law or the same derandomization variant can differ at finite dimension.

Derandomization in this setting is mechanism-specific control of finite-dimensional error rather than a single generic reduction of randomness. Constant Hamming-weight initialization removes random-start imbalance; fixed update counts remove update-count variability; and exact balance constraints prevent Hamming-weight drift. These constraints affect different parts of the error, so additional constraints are useful when they target the error source that dominates in the intended operating regime.

The generator-construction view further explains the remaining deviations from the target law. Selection rules, update mechanisms, dependence within a transition, and truncation or capacity handling determine bias, variance, reference-level effects, and boundary behavior. In particular, finite-capacity effects in the linear family and dependence induced by fixed-count or balanced updates cannot be inferred from the target similarity law alone. This is why mean target matching, although important, is not sufficient to characterize the finite-dimensional accuracy of a scalar codebook generator.

The finite-dimensional theory derived in this paper connects these mechanisms to induced mean similarity, bias, variance, and RMSE, and the simulations validate the predicted trends across similarity families, dimensions, quantization ranges, reference scalar levels, variants, and generator constructions. The resulting design rule is mechanism-specific: choose the target similarity law, identify whether the dominant finite-dimensional error arises from random-start imbalance, update-count variability, Hamming-weight drift, within-transition dependence, or finite-capacity effects, and match the derandomization mechanism to the dominant source of finite-dimensional error while respecting implementation constraints.

## Acknowledgments

The work of D.R. was supported in part by the Swedish Foundation for Strategic Research (SSF, grant nos. UKR22-0024, UKR24-0014).
The work of E.O. and D.R. was supported in part by the Swedish Research Council (VR grant no. 2022-04657).
The work of E.O. was supported in part by STINT, the Swedish Foundation for International Cooperation in Research and Education (STINT, grant no. MG2020-8842) and Intel Neuromorphic Research Community Project: Unsupervised learning in NLP tasks on using Vector-Symbolic representations on phasor-based associative memory.
D.K. acknowledges funding from the Swedish Strategic Research Foundation under the Future Research Leaders program (Grant No. FFL24-0111) and the Swedish Research Council under the Starting Grant program (Grant No. 2025-05421). This work was supported in part by the AFOSR under award number FA8655-25-1-7007.
V.S. acknowledges funding from European Union under the REFRESH project number CZ.10.03.01/00/22 /0000048 via the Operational Programme Just Transition; CLARA project under European Union’s HORIZON EUROPE No 101136607.
E.O. and D.R. acknowledge the computational resources provided by the National Academic Infrastructure for Supercomputing in Sweden (NAISS), partially funded by the Swedish Research Council through grant agreement no. 2022-06725, as well as help and support of Tim Ufer and Philip Gard.

# 1 Supplementary Note 1: Exponential Similarity Family

## 1.1 Subtract–Add

### 1.1.1 Derivation of Mean Between-Level Similarity (Subtract–Add)

Fix an anchor level $a$ and let $P_q := \Pr(X_{a+q} = 1, X_a = 1)$, $q \geq 0$, so $q$ is the level separation from the anchor, $q = \Delta$. In one SA transition, Degrade flips a current 1 to 0 with probability $r$; Refill then flips each current 0 after Degrade to 1 with probability α. We choose $\alpha = r/(1+r)$, because this preserves $p = 1/2$.

Let us write a recursion for $P_q$. Before the Degrade–Refill transition from level $a+q$ to $a+q+1$, we have $\Pr(X_{a+q} = 1) = 1/2$, $\Pr(X_{a+q} = 0) = 1/2$. The joint probabilities for $(X_{a+q}, X_a)$ are as follows. (1,1): $P_q$; (1,0): $1/2 - P_q$, because $\Pr(X_{a+q} = 1) = 1/2$; (0,1): $1/2 - P_q$, because $\Pr(X_a = 1) = 1/2$; (0,0): $P_q$, because the probabilities sum to 1 (so that $1 - (P_q + 1/2 - P_q + 1/2 - P_q) = P_q$).

We now compute $P_{q+1}$. There are two contributions. First, pairs with $(X_{a+q}, X_a) = (1,1)$ remain overlapping if the current 1 survives Degrade, giving $(1 - r)\, P_q$. Second, after Degrade, the event $(0,1)$ consists of the original $(0,1)$ pairs and the degraded original $(1,1)$ pairs. Its probability is $(1/2 - P_q) + rP_q = 1/2 - (1 - r)\, P_q$. Refill turns these current zeros to 1 with probability α. Thus $P_{q+1} = (1 - r)\, P_q + [1/2 - (1 - r)\, P_q]\, \alpha = (1 - r)(1 - \alpha)\, P_q + \alpha/2$. With $\alpha = r/(1+r)$, we have $1 - \alpha = 1/(1+r)$, so the decay parameter becomes $\rho_{\mathrm{SA}} := (1 - r)(1 - \alpha) = (1 - r)/(1+r)$, and the constant term is $c_{\mathrm{SA}} := \alpha/2 = r/[2(1+r)]$.

The recursion can be written as $P_{q+1} = \rho_{\mathrm{SA}}\, P_q + c_{\mathrm{SA}}$, with $P_0 = 1/2$. Unrolling gives $P_{q+1} = \rho_{\mathrm{SA}}^{q+1}\, P_0 + c_{\mathrm{SA}}\, (1 + \rho_{\mathrm{SA}} + \ldots + \rho_{\mathrm{SA}}^{q})$, where the sum of geometric progression is $(1 - \rho_{\mathrm{SA}}^{q+1})/(1 - \rho_{\mathrm{SA}})$. With $P_0 = 1/2$ and $c_{\mathrm{SA}}/(1 - \rho_{\mathrm{SA}}) = 1/4$, this yields $P_q = 1/4 + \rho_{\mathrm{SA}}^{q}/4$. Since $S(a,b) = 2\, O(a,b)/D$ and each component contributes overlap with probability $P_q$, for two levels $a$ and $b$ with separation $\Delta := |a - b|$, we have $P_\Delta = \Pr(X_a = 1, X_b = 1) = 1/4 + (1/4)\rho_{\mathrm{SA}}^{\Delta}$, hence $\mathrm{E}[S(a,b)] = T_{\mathrm{SA}}(\Delta) = 2P_\Delta = 1/2 + (1/2)\rho_{\mathrm{SA}}^{\Delta}$.

### 1.1.2 Properties

**Parameter range:** The update is well-defined for $0 \leq r \leq 1$. With $\rho_{\mathrm{SA}} = (1 - r)/(1+r)$ from the recursion above, we have $\rho_{\mathrm{SA}} \in [0,1]$, and the similarity law $T_{\mathrm{SA}}(\Delta) = 1/2 + (1/2)\, \rho_{\mathrm{SA}}^{\Delta}$ is well defined for all integer $\Delta \geq 0$.

**One-step similarity and distance:** For consecutive levels $(\Delta = 1)$, $T_{\mathrm{SA}}(1) = 1/2 + (1/2)\, \rho_{\mathrm{SA}} = 1/2 + (1/2)\, (1 - r)/(1+r) = 1/(1+r)$. The normalized Hamming distance between levels at distance $\Delta$ is $\mathrm{E}[\mathrm{dist}_{\mathrm{Ham}}(x_a, x_b)] \,/\, D = (1 - \rho_{\mathrm{SA}}^{\Delta}) \,/\, 2$, so for consecutive levels $(\Delta = 1)$ this gives $\mathrm{E}[\mathrm{dist}_{\mathrm{Ham}}(x_a, x_{a+1})] \,/\, D = (1 - \rho_{\mathrm{SA}})/2 = (1 - (1 - r)/(1+r))/2 = r/(1+r)$.

**Executed and net flips:** At $p = 1/2$, Degrade executes $r/2$ flips (1→0) in expectation and Refill executes another $r/2$ flips (0→1), so the expected number of executed flips per component per level transition is $\mathrm{E}[N_{\mathrm{exec}}] = r$. Some components flip twice in one step (1→0 in Degrade, then 0→1 in Refill). For $p = 1/2$ the double-flip probability per component is $\Pr(\mathrm{double}) = r^2 / (2(1+r))$. Each double-flip event contributes two executed flips but zero net flips, so $\mathrm{E}[N_{\mathrm{exec}}] = \Pr(\text{net flip}) + 2\Pr(\text{double})$. With $\mathrm{E}[N_{\mathrm{exec}}] = r$ and $\Pr(\mathrm{double}) = r^2/(2(1+r))$, this gives $\Pr(\text{net flip}) = r - r^2/(1+r) = r/(1+r)$ which agrees with the expression obtained from the distance formula above.

## 1.2 Single–Commit

### 1.2.1 Derivation of Mean Between-Level Similarity (Single–Commit)

It is convenient to recode components as $\{-1,+1\}$. Define $Z_q \in \{-1,+1\}$ by $Z_q = 2\,X_q - 1$, so $X_q = 1$ corresponds to $Z_q = +1$ and $X_q = 0$ corresponds to $Z_q = -1$, and $Z_q^2 = 1$. Under the SC update, each component flips with probability $r$, so $Z_{q+1} = Z_q\,V_q$, where $V_q \in \{-1,+1\}$ is independent of $Z_q$, and $V_q$s are independent across level transitions, with $\Pr(V_q = -1) = r$ and $\Pr(V_q = +1) = 1 - r$. Then $\mathrm{E}[V_q] = (1 - r) - r = 1 - 2r$ and $\mathrm{E}\left[Z_q\,Z_{q+1}\right] = \mathrm{E}\left[Z_q^2\,V_q\right] = \mathrm{E}\left[V_q\right] = 1 - 2r$ because $Z_q^2 = 1$. For two levels $a$ and $b$ with separation $\Delta := |a - b|$, iterating the update over $\Delta$ transitions gives $\mathrm{E}[Z_a\,Z_b] = (1 - 2r)^{\Delta}$.

Relating this to the overlap of binary vectors, note that since $\Pr(X_a = 1) = \Pr(X_b = 1) = 1/2$, we have $\mathrm{E}[Z_a\,Z_b] = 4\Pr(X_a = 1, X_b = 1) - 1$, $\Pr(X_a = 1, X_b = 1) = (1 + (1 - 2r)^{\Delta})/4$. Using $S(a,b) = 2\,O(a,b)/D$ at $p = 1/2$ and the fact that each component contributes overlap with probability $\Pr(X_a = 1, X_b = 1)$, the induced mean similarity is $\mathrm{E}[S(a,b)] = T_{\mathrm{SC}}(\Delta) = 1/2 + (1/2)\,(1 - 2r)^{\Delta}$.

### 1.2.2 Properties

**Parameter range:** The update is well-defined for $0 \le r \le 1$. With $\rho_{\mathrm{SC}} := 1 - 2r$ we have $\rho_{\mathrm{SC}} \in [-1,1]$, and the similarity law $T_{\mathrm{SC}}(\Delta) = 1/2 + (1/2)\,\rho_{\mathrm{SC}}^{\Delta}$ is well defined for all integer $\Delta \ge 0$. For many applications we restrict to $0 \le r \le 1/2$ so that $\rho_{\mathrm{SC}} \in [0,1]$ and the similarity decays monotonically with $\Delta$.

**One-step similarity and distance.** For $\Delta = 1$, $T_{\mathrm{SC}}(1) = 1/2 + (1/2)\,\rho_{\mathrm{SC}} = 1/2 + (1/2)\,(1 - 2r) = 1 - r$. The normalized Hamming distance follows from the component-wise XOR identity $x_a \oplus x_b = x_a + x_b - 2(x_a \wedge x_b)$. At $p = 1/2$ this gives $\mathrm{E}[\mathrm{dist}_{\mathrm{Ham}}(x_a, x_b)] / D = 1 - \mathrm{E}[\mathrm{sim}_{\mathrm{dot}}(x_a, x_b)]/(D/2) = (1 - \rho_{\mathrm{SC}}^{\Delta})/2$. For consecutive levels ($\Delta = 1$) this gives $\mathrm{E}[\mathrm{dist}_{\mathrm{Ham}}(x_a, x_{a+1})] / D = r$.

**Executed versus net flips.** Each component flips with probability $r$ and never flips twice within the same step, so executed and net flip probabilities coincide and equal $r$ per component per step (or $rD$ per vector per step).

### 1.3 Comparison of Subtract–Add and Single–Commit

At $p = 1/2$ the SC and SA updates have $\rho_{SC} = 1 - 2r$ and $\rho_{SA} = (1 - r)/(1+r)$. At $p = 1/2$ both SC and SA execute $rD$ flips per step in expectation, so comparing their decay factors ρ at the same $r$ compares the schemes at equal expected executed-flip cost. For $0 < r \le 1/2$, we have $1 - 2r < (1 - r)/(1+r)$, so SC decorrelates faster than SA at this equal-cost calibration.

Let $u$ denote the net flip probability per component per level transition, equivalently $u = \mathrm{E}[\mathrm{dist}_{\mathrm{Ham}}(x_a, x_{a+1})] / D$. If the comparison is made at fixed $u$, then SA has $u = r/(1+r)$ and hence $r = u/(1 - u)$, whereas for SC at $p = 1/2$ the net flip probability equals the flip probability $r$ itself, so setting $r = u$ parameterizes SC directly by $u$. Under this fixed-net-flip calibration, both updates have decay $\rho = 1 - 2u$ and similarity $T(\Delta) = 1/2 + (1/2)(1 - 2u)^{\Delta}$; in particular, $T(1) = 1 - u$. However, SA requires extra executed flips to achieve the same net $u$, because components can flip twice within one Degrade–Refill transition. In expectation, SC executes $uD$ flips vs $Du/(1 - u)$ for SA.

**Feasibility:** At $p = 1/2$, in this comparison we restrict SC to $0 \le r \le 1/2$ (so that $\rho_{SC} \in [0,1]$ and the similarity decays monotonically), while SA permits $0 \le r \le 1$ ($\rho_{SA} \in [0,1]$). For SC, $r = 1/2$ gives $\rho = 0$ and $T(\Delta) = 1/2$ for all $\Delta \ge 1$. For SA, $r = 1$ gives similarity $1/2$ and $\rho = 0$.

**Comparison with "scatter code"** (Smith & Stanford, 1990): For one component under the symmetric update with per-step flip probability $r = 1/D$, the number of flips over $q$ steps is $\mathrm{Binomial}(q, 1/D)$. The component differs from its start iff the flip count is odd, which occurs with probability $\Pr(\mathrm{odd}) = (1 - (1 - 2/D)^q)/2$. Hence $\mathrm{E}(\mathrm{dist}_{\mathrm{Ham}})/D = (1 - \rho^{|a-b|})/2$ with $\rho = 1 - 2r = 1 - 2/D$ and $|a - b| = q$.

**Useful limiting regimes:** Small $q \ll D$: $(1 - 2/D)^q \approx 1 - 2q/D$, so $\mathrm{E}[\mathrm{dist}_{\mathrm{Ham}}] \approx q$. For $q \sim D$: $(1 - 2/D)^q \approx e^{-2q/D}$, so $\mathrm{E}[\mathrm{dist}_{\mathrm{Ham}}]/D \approx (1 - e^{-2q/D})/2$ and $\mathrm{E}[\mathrm{dist}_{\mathrm{Ham}}] \approx (D/2)(1 - e^{-2q/D})$. As $q \to \infty$, $\mathrm{E}[\mathrm{dist}_{\mathrm{Ham}}] \to D/2$.

## 1.4 Markov Chain and Eigenvalue Perspective

Above, we derived expected similarity for SA and SC directly from explicit flip and overlap recursions. Here we give a complementary Markov-chain perspective on the same result: for each exponential-family update, the corresponding decay parameter ρ is the nontrivial eigenvalue of an associated 2×2 single-component Markov kernel. This viewpoint shows how the same decay factor obtained from the flip/overlap recursions also arises from standard spectral facts for finite Markov chains (see, e.g., (Diaconis & Stroock, 1991), (Fill, 1991), (Diaconis & Saloff-Coste, 1993), (Diaconis, 2009)).

### 1.4.1 Single-Component Kernels and Eigenvalues at $p = 1/2$

Consider a single component $X_q \in \{0,1\}$ evolving across levels as a level-homogeneous two-state Markov chain (same one-component transition rule at every level transition $q \to q+1$). The update is specified by a 2×2 transition kernel with elements $K(i,j) = \Pr\,(X_q = i \to X_{q+1} = j)$: $K = [[\Pr(0 \to 0), \Pr(0 \to 1)], [\Pr(1 \to 0), \Pr(1 \to 1)]]$. Rows are 'from' states (before the arrow), columns are 'to' states (after the arrow); row $i$ gives the transition probabilities out of state $i$, and each row sums to 1.

Let $\pi_q$ denote the probability mass function (pmf) of $X_q$, represented as a row vector $\pi_q := (\Pr(X_q = 0), \Pr(X_q = 1))$. Then the evolution is $\pi_{q+1} = \pi_q\, K$. Now specialize to $p = 1/2$, i.e., $\Pr(X_q = 1) = 1/2$ for all $q$. This holds if we start from $\pi_0 = (1/2,1/2)$ and choose parameters so that $\pi^* = (1/2,1/2)$ is stationary for $K$. This requires $\Pr(0 \to 1) = \Pr(1 \to 0)$; denote this common flip probability by θ. Then $K$ takes the symmetric "equal-flip" form $K(\theta) = [[1 - \theta, \theta], [\theta, 1 - \theta]]$, $0 \le \theta \le 1$.

To connect this to overlap/similarity, recode $\{0,1\}$ components to the centered sign variable $Z_q = 2\,X_q - 1 \in \{-1,+1\}$. Under $\pi^* = (1/2,1/2)$ one has $\mathrm{E}[Z_q] = 0$. The eigenvalues of $K(\theta)$ are $\lambda_1$ and $\lambda_2$. Since each row sums to 1, $\lambda_1 = 1$. While $\lambda_2 = 1 - 2\theta$ controls correlation decay: $\mathrm{E}[Z_a\, Z_{a+\Delta}] = \lambda_2^{\Delta}$. In our notation $\rho := \lambda_2$ and $\Pr(X_a = 1, X_b = 1) = (1 + \mathrm{E}[Z_a\, Z_b]) / 4 = (1 + \rho^{\Delta}) / 4$. Summing over components and using $S(a,b) = 2 \cdot O(a,b)/D$ at $p = 1/2$ yields the mean similarity law $\mathrm{E}[S(a,b)] = 1/2 + (1/2) \cdot \rho^{\Delta}$ corresponding to the exponential target law $T(\Delta)$ in section 4.

For the SC scheme at $p = 1/2$, we have $\theta = r$ and $\rho_{SC} = 1 - 2\,r$, in agreement with section 4.2. For SA at $p = 1/2$, one full level transition is the composition of two one-component substeps: Degrade flips $1 \to 0$ (0 stays 0) with probability $r$, Refill then flips $0 \to 1$ (1 stays 1) with probability α chosen so that $\pi^* = (1/2,1/2)$ is stationary for the composed transition. At $p = 1/2$, stationarity requires $\Pr(0 \to 1) = \Pr(1 \to 0)$.

For the composed SA step, these probabilities are α and $r(1 - \alpha)$, respectively; hence $\alpha = r(1 - \alpha)$, so $\alpha = r/(1+r)$. The resulting one-component transition probabilities for the composed SA step are $\Pr(0 \to 0) = 1 - \alpha$, $\Pr(0 \to 1) = \alpha$, $\Pr(1 \to 0) = r\,(1 - \alpha)$, $\Pr(1 \to 1) =$ $(1-r)+r\,\alpha$, so $K_{\mathrm{SA}} = [[1-\alpha,\alpha],[r\,(1-\alpha),(1-r)+r\,\alpha]]$. Substituting $\alpha = r/(1+r)$ gives $K_{\mathrm{SA}} = [[1/(1+r),r/(1+r)],\ [r/(1+r),1/(1+r)]]$. This matrix has eigenvalues $\lambda_1 = 1$, $\lambda_2 = (1-r)/(1+r) = \rho_{\mathrm{SA}}$, matching section 4.2.

1.4.2 From One Component to $D$ Components

For SC and SA with independently updated components, the full chain on $\{0,1\}^D$ is the $D$-fold product of the one-component chain ($K_{\mathrm{SC}}$ or $K_{\mathrm{SA}}$); consequently, the $D$-dimensional eigenvalues are products of the one-component eigenvalues. The overlap $O(a,b) = \mathrm{sim}_{\mathrm{dot}}(x_a,x_b)$ is the sum of $D$ component-wise overlaps. Linearity of expectation gives $\mathrm{E}[O(a,b)] = D\ \Pr(X_a(i) = 1, X_b(i) = 1)$. Since $\Pr(X_a(i) = 1, X_b(i) = 1) = (1+\rho^{\Delta})/4$, we obtain $\mathrm{E}[S(a,b)] = 1/2+(1/2)\rho^{\Delta}$. Thus, both SC and SA follow $T(\Delta) = 1/2+(1/2)\,\rho^{\Delta}$, with $\rho_{\mathrm{SC}} = 1-2\,r$ for SC and $\rho_{\mathrm{SA}} = (1-r)/(1+r)$ for SA.

1.4.3 Derandomization Variants

For Baseline, the component-wise SC and SA update schemes are treated above. The fixed-count SC variants of section 4.3, including Fixed-F, Fixed-FpD and Fixed-kpD (equal-swap), define Markov chains on $\{0,1\}^D$, but they are not product chains (i.e., the $D$ components are not updated independently at each step). Each step flips a constrained number of components (and, for equal-swap, preserves $|x_q|$), which couples components within a step. Nevertheless, for each fixed component, the marginal flip probability conditional on its current value is $r$. Hence the marginal single-component transition probabilities satisfy $\Pr(0\to1) = \Pr(1\to0) = r$ and the induced single-component kernel is $K(r) = [[1-r,r],[r,1-r]]$ with nontrivial eigenvalue $\rho = 1-2r$. Although components are coupled within a step, this coupling affects their joint behavior (and thus fluctuations), but it does not change the one-step transition probabilities of a single component; therefore, the mean overlap/similarity still uses the same $\rho$. Consequently, in the $p = 1/2$ setting the fixed-count SC variants share the same induced mean similarity law $\mathrm{E}[S(a,b)] = 1/2+(1/2)\rho^{\Delta}$, while they differ in finite-$D$ fluctuation properties because the per-step flips are constrained rather than independent.

1.4.4 Related Markov-Chain Literature and Scope

Our sequence of derandomization variants begins with the Baseline variants. Baseline SC (component-wise random flips) and Baseline SA (component-wise Degrade-Refill) both induce, at the component level, a two-state Markov kernel applied independently across components. Consequently, the full $D$-dimensional evolution (up to the chosen initial law) is a product chain of identical two-state component chains. In this setting, overlap/similarity with a fixed anchor reduces to a count statistic: how many components are 1 both in the anchor and in the level reached after $\Delta$ steps.

When the component kernel is symmetric/balanced, this count process is closely related to classical Ehrenfest-type chains; general cutoff/mixing results for Ehrenfest chains and product chains (e.g., (Chen et al., 2012), (Chen & Kumagai, 2018)) can therefore serve as optional background if one wishes to discuss worst-case convergence behavior of these baselines. For fixed-count SC derandomization variants, the evolution coincides with classical random walks on discrete state spaces. In the Fixed-F setting (flip exactly $F$ components per level), the evolution is a random walk on the $D$-dimensional hypercube. Related asymptotic and spectral analyses include (Diaconis et al., 1990) for single-bit flips and (Nestoridi, 2017) for fixed-count walks. In the Fixed-kpD / equal-swap setting (swap exactly $k$ 1's with $k$ 0's per level), the evolution is a $k$-swap Bernoulli–Laplace urn model on the constant Hamming-weight state space of $D$-bit vectors with $D/2$ ones. Equivalently, it is a $k$-swap walk on the Johnson scheme, with the $k = 1$ case corresponding to the standard random walk on the Johnson graph. This connects our equal-swap dynamics to the classical $k = 1$ Bernoulli–Laplace analysis of (Diaconis & Shahshahani, 1987) and (Donnelly et al., 1994), and to later $k$-swap regimes such as $k = o(D)$ and $k = \Theta(D)$ in (Eskenazis & Nestoridi, 2020) and (Alameda et al., 2024). This is consistent with our one-component viewpoint: these classical analyses diagonalize the full transition operator, which has many eigenvalues. Our mean-similarity calculation uses only the eigenvalue associated with the overlap; this eigenvalue corresponds to the decay factor ρ.

The cited cutoff/mixing literature for product/Ehrenfest chains (baseline) or for hypercube/Bernoulli–Laplace walks (fixed-count SC derandomizations) primarily targets worst-case convergence to stationarity in global metrics (total variation or separation) and often identifies sharp transition windows ("cutoff"), typically as a function of an adversarial initial configuration. Our focus is different: we analyze average-case similarity dynamics induced by encoding (mean similarity profile and variance/RMSE), typically from randomly generated anchors, or from anchors produced by the same codebook-generation process, rather than from adversarial worst-case starts. Consequently, exponential mean similarity profiles do not conflict with cutoff phenomena: cutoff is a global statement about the full configuration distribution, while our profiles track a low-dimensional observable (overlap/similarity) governed by low-order spectral structure (the decay factor ρ), which can relax smoothly even when worst-case distance to stationarity remains large.

#### 1.4.5 Summary

A one-component two-state 2×2 Markov kernel matrix with transition probabilities $g = \Pr(0{\to}1)$ and $h = \Pr(1{\to}0)$ has the second (nontrivial) eigenvalue $1 - g - h$. The target density $p$ enters through the stationary-balance condition $(1 - p)g = ph$. In this paper we specialize to $p = 1/2$. For both the SC kernel and the composed one-step SA kernel, after choosing the SA Refill probability to preserve density $1/2$, the off-diagonal probabilities are equal: $g = h = \theta$ and the nontrivial eigenvalue equals the one-step decay parameter $\rho_{\text{step}}$ used in section 6.1.1.

For component-wise SC/SA, this eigenvalue is ρ derived from the one-step similarity recursion, with $\rho_{SC} = 1 - 2r$ and $\rho_{SA} = (1 - r)/(1{+}r)$. For fixed-count and equal-swap SC variants, component selections are coupled within a step, but each individual component still has flip probability r; therefore, the mean overlap/similarity evolves with the same decay factor $\rho_{SC} = 1 - 2r$, while the coupling affects finite-D variance.

Thus, the Markov-kernel viewpoint reinterprets the same decay parameter already used in sections 4.2 and 6.1.1.

# 2 Supplementary Note 2: Linear Similarity Family Mean Laws and Finite-Capacity Effects

## 2.1 Proof of Lemma 5.1

**Proof.** Step 1 (pool depletion ⇒ agreement probability). This step uses only the w.o.r. requirement and the independence of $J$ from the generator randomness. Under w.o.r. dynamics, the component at index $J$ differs between levels $a$ and $b$ if and only if $J \in U_a$ and $J \notin U_b$, since $U_b \subseteq U_a$ (the pool is shrinking in $q$). Hence $\Pr(x_a(J) \neq x_b(J)) = \Pr(J \in U_a) - \Pr(J \in U_b)$. Conditioning on $U_q$ and using that $J$ is uniform and independent gives $\Pr(J \in U_q \mid U_q) = |U_q|/D$, hence $\Pr(J \in U_q) = \mathrm{E}[|U_q|]/D$. Therefore $\Pr(x_a(J) \neq x_b(J)) = (\mathrm{E}[|U_a|] - \mathrm{E}[|U_b|])/D$ and $\Pr(x_a(J) = x_b(J)) = 1 - (\mathrm{E}[|U_a|] - \mathrm{E}[|U_b|])/D$. This proves (i).

Step 2 (overlap-normalized similarity ⇒ agreement probability at $p = 1/2$). This step does not use the w.o.r. requirement; it uses $p = 1/2$ complement-invariant initialization, and value-independent selection. By definition of $O(a,b)$ and uniform $J$, $\mathrm{E}[O(a,b)] = D \Pr(x_a(J) = 1, x_b(J) = 1)$, hence $\mathrm{E}[S(a,b)] = (2/D)\, \mathrm{E}[O(a,b)] = 2\Pr(x_a(J) = 1, x_b(J) = 1)$.

For any fixed $\mathbf{u}, \mathbf{m} \in \{0,1\}^D$: $1 - (\mathbf{u} \oplus \mathbf{m}) = (1 - \mathbf{u}) \oplus \mathbf{m}$. Consider $\mathbf{m}^{(q)} = \mathbf{m}_0 \oplus \dots \oplus \mathbf{m}_{q-1}$. If $\mathbf{x}_0 = \mathbf{u}$ and $\mathbf{x}_q = \mathbf{x}_0 \oplus \mathbf{m}^{(q)}$, then $1 - \mathbf{x}_q = (1 - \mathbf{x}_0) \oplus \mathbf{m}^{(q)}$. Therefore, with the same masks, replacing any fixed $\mathbf{x}_0$ by its component-wise complement $1 - \mathbf{x}_0$ complements every level $q \leq b$: $\mathbf{x}_q \to 1 - \mathbf{x}_q$. In particular, $(x_a(J), x_b(J))$ maps to $(1 - x_a(J), 1 - x_b(J))$. By Assumption (A), $\mathbf{x}_0$ and $1 - \mathbf{x}_0$ have the same distribution. Let $\mathbf{M}$ denote the entire realized mask sequence up to level $b$. By Assumption (B), the distribution of $\mathbf{M}$ is unchanged when $\mathbf{x}_0$ is replaced by $1 - \mathbf{x}_0$. Together, these imply that the joint pairs $(\mathbf{x}_0, \mathbf{M})$ and $(1 - \mathbf{x}_0, \mathbf{M})$ have the same distribution.

**Condition on M:** Since the mapping $x_0 \to 1 - x_0$ deterministically maps $(x_a(J), x_b(J))$ to $(1 - x_a(J), 1 - x_b(J))$ under the same $\mathbf{M}$, the conditional distribution of $(x_a(J), x_b(J))$ given $\mathbf{M}$ is invariant under complementation. Hence, for every realized $\mathbf{M}$, $\Pr(x_a(J) = 1, x_b(J) = 1 \mid \mathbf{M}) = \Pr(x_a(J) = 0, x_b(J) = 0 \mid \mathbf{M})$. Taking expectations over $\mathbf{M}$ gives unconditional probabilities: $\Pr(x_a(J) = 1, x_b(J) = 1) = \Pr(x_a(J) = 0, x_b(J) = 0)$. Consequently, $\mathrm{E}[S(a,b)] = 2\Pr(x_a(J) = 1, x_b(J) = 1) =$
$\Pr(x_a(J) = 1, x_b(J) = 1) + \Pr(x_a(J) = 0, x_b(J) = 0) = \Pr(x_a(J) = x_b(J))$. This proves (ii). ■

## 2.2 Random-Count Remaining-Pool Constructions

### 2.2.1 Proposition 1: Adaptive Thinning on the Remaining Pool

For some remaining-pool w.o.r. generator constructions, the selection mechanism can be expressed as adaptive Bernoulli thinning of $U_q$, with capacity capping near exhaustion. Let $F_{q+1}$ denote the executed flip count in transition $q \to q+1$. It is random but capped by the current pool size $|U_q|$. Conditional on $U_q$ with $|U_q| > 0$, each component in $U_q$ is selected independently with inclusion probability $\min(1,F/|U_q|)$, where $F = D/Q$ is the nominal target flip count per level. If $|U_q| = 0$, no component is selected. Thus $F_{q+1} \mid U_q \sim \mathrm{Bin}(|U_q|,\min(1,F/|U_q|))$, with $|U_{q+1}| = |U_q| - F_{q+1}$. Hence $\mathrm{E}[F_{q+1} \mid U_q] = \min(F,|U_q|)$, and taking expectations gives the exact recursion $\mathrm{E}[|U_{q+1}|] = \mathrm{E}[|U_q|] - \mathrm{E}[\min(F,|U_q|)] = \mathrm{E}[|U_q|] - F + \mathrm{E}[\max(F - |U_q|,0)]$.

By Lemma 5.1, the mean similarity between levels $a$ and $a+\Delta$ is $\mathrm{E}[S(a,a+\Delta)] = 1 - (\mathrm{E}[|U_a|] - \mathrm{E}[|U_{a+\Delta}|])/D = 1 - \Delta F/D + (1/D)\sum_{q=a}^{a+\Delta-1} \mathrm{E}[\max(F - |U_q|,0)]$. Thus, deviations of $\mathrm{E}[S(a,a+\Delta)]$ from the target law $T_{\mathrm{lin}}(\Delta)$ arise from these finite-capacity correction terms, which are negligible away from exhaustion and become appreciable as the probability that the remaining pool falls below $F$ increases.

### 2.2.2 Proposition 2: Fixed-Order Proposed-Count Selection with Truncation

At transitions $q \to q+1$, let $f_{q+1}$ be i.i.d. proposed counts with $f_{q+1} \sim \mathrm{Bin}(D,1/Q)$. The generator proposes $f_{q+1}$ flips but executes only $\min(f_{q+1},|U_q|)$, flipping the next previously unused components along a fixed ordering. Define $C_q := \sum_{t=1}^{q} f_t \sim \mathrm{Bin}(Dq,1/Q)$, so $|U_q| = D - \min(C_q, D) = \max(D - C_q,0)$ and $\mathrm{E}[|U_q|] = \mathrm{E}[\max(D - C_q,0)]$. By Lemma 5.1, $\mathrm{E}[S(a,a+\Delta)] = 1 - (\mathrm{E}[|U_a|] - \mathrm{E}[|U_{a+\Delta}|])/D = 1 - \Delta F/D + (1/D)\,(\mathrm{E}[\max(C_{a+\Delta} - D,0)] - \mathrm{E}[\max(C_a - D,0)])$. Thus, deviations from $T_{\mathrm{lin}}(\Delta) = 1 - \Delta F/D$ arise from the overshoot terms created by the one-sided truncation $C_q \to \min(C_q,D)$. These terms are negligible away from pool exhaustion and become appreciable near the terminal boundary, where they produce the upward terminal bias.

### 2.2.3 Remark 3: Terminal Bias Scale

At $q = Q$ in Proposition 2, $|U_Q| = \max(D - C_Q,0)$, so $\mathrm{E}[|U_Q|] > 0$ even though $D - \mathrm{E}[C_Q] = 0$. By Lemma 5.1, $\mathrm{E}[S(0,Q)] = \mathrm{E}[|U_Q|]/D = (1/D)\,\mathrm{E}[\max(D - C_Q,0)]$. Since $C_Q \sim \mathrm{Bin}(DQ,1/Q)$, it has mean $D$ and variance $D(1 - 1/Q)$. By the de Moivre–Laplace approximation for a binomial with large $DQ$, we have $D - C_Q \approx Z \sim \mathrm{Norm}(0,D(1 - 1/Q))$. The standard identity $\mathrm{E}[\max(Z,0)] = \sigma/(2\pi)^{1/2}$ therefore gives $\mathrm{E}[S(0,Q)] \approx ((1 - 1/Q)/(2\pi D))^{1/2}$. Thus, one-sided truncation induces a small positive terminal mean similarity even though $T_{\mathrm{lin}}(Q) = 0$: overshoots are clipped at zero remaining capacity while undershoots leave a positive residual pool.

Empirically, fixed-order variants (Proposition 2) show a slightly larger terminal bias than the adaptive Bernoulli-thinning variants (Proposition 1): the proposed per-level counts are i.i.d. and can overshoot the remaining pool near exhaustion, whereas Bernoulli-thinning on $U_q$ adapts via $\min(1, F/|U_q|)$ and empties the remaining pool in the next transition whenever $|U_q| < F$, since then $\min(1, F/|U_q|) = 1$ and therefore $F_{q+1} = |U_q|$.

### 2.2.4 Corollary 4: Fixed-Count W.O.R. Constructions

Assume that each level transition flips exactly $F$ previously unused components, and that the pool is exhausted exactly at level $Q$, i.e., $F = D/Q$ is an integer. Then $|U_q| = D - qF$ deterministically. Hence, for any anchor $a$ and separation $\Delta$ with $a+\Delta \le Q$, $\mathrm{E}[S(a, a+\Delta)] = 1 - \Delta \cdot F/D = T_{\text{lin}}(\Delta)$. Moreover, $\mathrm{dist}_{\text{Ham}}(\mathbf{x}_a, \mathbf{x}_{a+\Delta}) = \Delta \cdot F$, so $\mathrm{sim}_{\text{ham}} = T_{\text{lin}}(\Delta)$ exactly.

# 3 Supplementary Note 3: Accuracy of Approximating the Target Similarity Laws

This section derives exact finite-dimensional (finite-D) expressions characterizing how closely each generator matches the target law $T(\Delta)$. For each similarity family (Exponential, Linear) and each derandomization variant (Baseline, Fixed-pD, Fixed-F, Fixed-FpD, Fixed-kpD), we provide $\mu_{\text{th}}(a,b)$, $\text{bias}_{\text{th}}(a,b)$ and $\text{v}_{\text{th}}(a,b)$. $\text{RMSE}_{\text{th}}$ is then obtained from these quantities using the paper's evaluation protocol of section 3.3, and compared to experiments in section 7.2. We start with finite-N properties of empirical error estimates.

## 3.1 Finite-N Properties of Empirical Error Estimates

For each fixed pair $(a,b)$, $E[\mu_{\text{emp}}(a,b)] = \mu(a,b)$, $\text{Var}(\mu_{\text{emp}}(a,b)) = \text{v}(a,b)/N$, $\text{E}[\text{MSE}_{\text{emp}}(a,b)] = \text{MSE}(a,b)$. The empirical bias and variance have finite-$N$corrections. Since $\text{bias}_{\text{emp}}(a,b) := \mu_{\text{emp}}(a,b) - T(a,b)$, we have $\text{E}[\text{bias}^2_{\text{emp}}(a,b)] = \text{bias}^2(a,b)+\text{v}(a,b)/N$. Since $\text{v}_{\text{emp}}(a,b)$ is normalized by $1/N$, we have $\text{E}[\text{v}_{\text{emp}}(a,b)] = (1 - 1/N)\cdot\text{v}(a,b)$. Averaging these identities over $b$ gives $\text{E}[\text{BiasRMS}^2_{\text{emp}}(a)] = \text{BiasRMS}^2(a)+\text{VarRMS}^2(a)/N$, $\text{E}[\text{VarRMS}^2_{\text{emp}}(a)] = (1 - 1/N)\ \text{VarRMS}^2(a)$, $\text{E}[\text{RMSE}^2_{\text{emp}}(a)] = \text{RMSE}^2(a)$.

## 3.2 Finite-D Accuracy for Exponential Similarity Family

We analyze the SC scheme from section 4.2 and its generator constructions from section 4.3. Because $S(a,b)$, $\text{bias}_{\text{th}}(a,b)$, and $\text{v}_{\text{th}}(a,b)$ are symmetric in $(a,b)$, we assume $a \le b$ without loss of generality and write $\Delta := b - a$ throughout this section. For the Baseline, Fixed-pD, Fixed-F, Fixed-FpD derandomization variants and their generators, the theory can be expressed in terms of two one-step generator parameters $\rho_{\text{step}}$ and $\gamma_{\text{step}}$ (defined below) that summarize the randomness of a single transition $q \to q+1$.

Let $M_q \subset \{1, \dots, D\}$ be the flip-index set at transition $q \to q+1$, and use the $\pm 1$ encoding $y_q(i) := 2x_q(i) - 1$, so that $x_q(i) = (1+y_q(i))/2$. Define the per-component multiplier $m_q(i) \in \{+1, -1\}$ by $m_q(i) = -1$ iff $i \in M_q$ and $m_q(i) = +1$ otherwise, so

$$y_{q+1}(i) = m_q(i)\cdot y_q(i). \tag{3.1}$$

Under level-homogeneity and exchangeability, the following generator one-step parameters do not depend on $q$, $i$, or the particular pair $(i,j)$, $j \ne i$: $\rho_{\text{step}} := \text{E}[m_q(i)]$ (mean one-step sign multiplier), $\gamma_{\text{step}} := \text{E}[m_q(i)\cdot m_q(j)]$ for $i \ne j$ (mean product of the two multipliers for two distinct components in the same step). If the mask decisions are independent across components within a step, then $\gamma_{\text{step}} = \rho^2_{\text{step}}$; fixed-count selection generally yields $\gamma_{\text{step}} \ne \rho^2_{\text{step}}$. These one-step parameters determine the derived quantities: $\rho_{\text{step}}$ determines the mean similarity profile and hence $\mathrm{bias}_{\mathrm{th}}(a,b)$, while $(\rho_{\mathrm{step}},\gamma_{\mathrm{step}})$ determine $\mathrm{v}_{\mathrm{th}}(a,b)$. The target exponential law uses parameter $\rho$ in $T_{\exp}(\Delta)$. However, we show that $\rho_{\mathrm{step}} = \rho$ and therefore $\mathrm{bias}_{\mathrm{th}}(a,b) = 0$. For Fixed-kpD (equal-swap), $\mathrm{v}_{\mathrm{th}}(a,b)$ is computed via a dedicated overlap-moment recursion that does not use $\gamma_{\mathrm{step}}$ explicitly.

3.2.1 Common Derivation for Baseline, Fixed-pD, Fixed-F, and Fixed-FpD

**Mean similarity and bias:** Iterate (3.1) across $\Delta$ transitions. Since the masks are sampled independently across steps and their distribution is level-homogeneous, $y_b(i) = y_a(i) \prod_{t=a}^{b-1} m_t(i)$, hence $y_a(i)\cdot y_b(i) = \prod_{t=a}^{b-1} m_t(i)$, because $y_a^2(i) = 1$. Taking expectation over realizations gives

$$\mathrm{E}[y_a(i)\cdot y_b(i)] = (\mathrm{E}[m_q(i)])^{\Delta} = \rho_{\mathrm{step}}^{\Delta}. \tag{3.2}$$

Component-wise,

$$x_a(i)\cdot x_b(i) = (1+y_a(i))(1+y_b(i))/4 = \\ 1/4+(y_a(i)+y_b(i))/4+(y_a(i)\cdot y_b(i))/4. \tag{3.3}$$

At $p = 1/2$ we have $\mathrm{E}[y_a(i)] = \mathrm{E}[y_b(i)] = 0$, hence $\mathrm{E}[x_a(i)\cdot x_b(i)] = 1/4+(1/4)\mathrm{E}[y_a(i)\cdot y_b(i)] = 1/4+(1/4)\rho_{\mathrm{step}}^{\Delta}$. Since $O(a,b) = \sum_i (x_a(i)\, x_b(i))$, exchangeability gives $\mathrm{E}[O(a,b)] = D\cdot\mathrm{E}[x_a(1)\, x_b(1)]$, and using $S(a,b) = 2\, O(a,b)/D$, we obtain $\mu_{\mathrm{th}}(a,b) := \mathrm{E}[S(a,b)] = 2\cdot\mathrm{E}[x_a(1)\, x_b(1)] = 1/2+(1/2)\rho_{\mathrm{step}}^{\Delta}$. The corresponding bias is $\mathrm{bias}_{\mathrm{th}}(a,b) := \mu_{\mathrm{th}}(a,b) - T_{\exp}(\Delta)$.

**Finite-$D$ variance derivation for** $\mathrm{v}_{\mathrm{th}}(a,b)$: Define $W_i := y_a(i)\cdot y_b(i)$, $A_q := \sum_{i=1}^{D} y_q(i)$, and $C_{\mathrm{ab}} := \sum_{i=1}^{D} W_i$. Summing $x_a(i)\, x_b(i)$ over $i$ gives the decomposition via (3.3): $O(a,b) = \sum_{i=1}^{D} x_a(i)\, x_b(i) = D/4+(A_a+A_b+C_{\mathrm{ab}})/4$, so $\mathrm{Var}(O(a,b)) = (1/16)\cdot\mathrm{Var}(A_a+A_b+C_{\mathrm{ab}})$. At $p = 1/2$, the distribution over realizations of the full sign sequence is invariant under the global sign flip $y \to -y$ for all levels (the initialization is sign-symmetric and the update masks do not depend on current values). Under this flip, $A_a+A_b$ changes sign while $C_{\mathrm{ab}}$ does not. Therefore $\mathrm{Cov}(A_a+A_b,C_{\mathrm{ab}}) = 0$ and $\mathrm{Var}(O(a,b)) = (1/16)\cdot(\mathrm{Var}(A_a+A_b)+\mathrm{Var}(C_{\mathrm{ab}}))$. Finally, since $S(a,b) = 2\cdot O(a,b)/D$ for $p = 1/2$,

$$\mathrm{v}_{\mathrm{th}}(a,b) := \mathrm{Var}(S(a,b)) = 4\cdot\mathrm{Var}(O(a,b))/D^2 = \\ (\mathrm{Var}(A_a+A_b)+\mathrm{Var}(C_{\mathrm{ab}}))\ /\ (4\cdot D^2). \tag{3.4}$$

The term $\mathrm{Var}(C_{\mathrm{ab}})$ depends only on $\rho_{\mathrm{step}}$ and $\gamma_{\mathrm{step}}$. Indeed, for $i \neq j$, $W_i\, W_j = \prod_{t=a}^{b-1} (m_t(i)\cdot m_t(j))$, so $\mathrm{E}[W_i\, W_j] = \gamma_{\mathrm{step}}^{\Delta}$. Using exchangeability of the $W_i$, $\mathrm{Var}(C_{\mathrm{ab}}) = D\cdot\mathrm{Var}(W_1)+D\cdot(D-1)\cdot\mathrm{Cov}(W_1,W_2) = D\cdot(\mathrm{E}[W_1^2] - (\mathrm{E}[W_1])^2)+D\cdot(D-1)\cdot(\mathrm{E}[W_1\, W_2] - (\mathrm{E}[W_1])^2)$. We have $W_i^2 = 1$, so $\mathrm{E}[W_1^2] = 1$, $\mathrm{E}[W_1] = \rho_{\mathrm{step}}^{\Delta}$ by (3.2) and $\mathrm{E}[W_1\, W_2] = \gamma_{\mathrm{step}}^{\Delta}$. Thus

$$\mathrm{Var}(C_{\mathrm{ab}}) = D\cdot(1 - \rho_{\mathrm{step}}^{2\Delta}) + D(D-1)(\gamma_{\mathrm{step}}^{\Delta} - \rho_{\mathrm{step}}^{2\Delta}). \quad (3.5)$$

The remaining term $\mathrm{Var}(A_a + A_b)$ depends on the initialization; we evaluate it next.

### 3.2.2 Bernoulli-Start Derandomization Variants (Baseline, Fixed-F)

We have $y_0(i)$ i.i.d. Rademacher, so $A_0 = \sum_{i=1}^{D} y_0(i)$ has $\mathrm{E}[A_0] = 0$ and $\mathrm{Var}(A_0) = D$. For any $q$, each component evolves as $y_q(i) = y_0(i) \prod_{t=0}^{q-1} m_t(i)$. Since the masks are independent of $y_0$ and the update is component-wise sign multiplication, $y_q$ is a component-wise sign-flip of the i.i.d. Rademacher vector $y_0$. Hence each $y_q(i)$ is Rademacher and remains independent across $i$. Therefore $\mathrm{E}[A_q] = 0$ and $\mathrm{Var}(A_q) = D$ for all $q$.

The cross-level covariance of $A_a$ and $A_b$ is also controlled by $\rho_{\mathrm{step}}$. Indeed, since $\mathrm{E}[A_a] = \mathrm{E}[A_b] = 0$, we have $\mathrm{Cov}(A_a, A_b) = \mathrm{E}[A_a A_b] - \mathrm{E}[A_a]\,\mathrm{E}[A_b] = \mathrm{E}[A_a A_b]$. Expanding, $\mathrm{Cov}(A_a, A_b) = \sum_{i=1}^{D} \sum_{j=1}^{D} \mathrm{E}[y_a(i)\cdot y_b(j)]$. If $i \neq j$, $y_a(i) = y_0(i)\cdot\prod_{t=0}^{a-1} m_t(i)$ and $y_b(j) = y_0(j)\cdot\prod_{t=0}^{b-1} m_t(j)$. Since the masks are independent of $y_0$ and $\mathrm{E}[y_0(i)\cdot y_0(j)] = 0$ for $i \neq j$ under the Bernoulli start, we get $\mathrm{E}[y_a(i)\cdot y_b(j)] = 0$. If $i = j$, (3.2) gives $\mathrm{E}[y_a(i)\cdot y_b(i)] = \rho_{\mathrm{step}}^{\Delta}$. Summing the $D$ diagonal terms gives $\mathrm{Cov}(A_a, A_b) = D\cdot\rho_{\mathrm{step}}^{\Delta}$. Then, $\mathrm{Var}(A_a + A_b) = \mathrm{Var}(A_a) + \mathrm{Var}(A_b) + 2\cdot\mathrm{Cov}(A_a, A_b) = 2D + 2D\cdot\rho_{\mathrm{step}}^{\Delta}$. Substituting this and (3.5) into (3.4) yields $\mathrm{v_{th}}(a,b)$ for the Bernoulli-start generators once $(\rho_{\mathrm{step}}, \gamma_{\mathrm{step}})$ are specified.

**Baseline (Bernoulli start):** For a single component $i$, $m_q(i) = +1$ with probability $1 - r$ and $m_q(i) = -1$ with probability $r$, so $\rho_{\mathrm{step}} = \mathrm{E}[m_q(i)] = (1 - r)\cdot(+1) + r\cdot(-1) = 1 - 2r$. We obtain $\rho_{\mathrm{step}} = \rho$, therefore $\mu_{\mathrm{th}}(a,b) = T_{\mathrm{exp}}(\Delta)$ and $\mathrm{bias_{th}}(a,b) = 0$. For $i \neq j$, Baseline mask entries are independent across components, so $\gamma_{\mathrm{step}} = \mathrm{E}[m_q(i)\cdot m_q(j)] = \mathrm{E}[m_q(i)]\cdot\mathrm{E}[m_q(j)] = \rho_{\mathrm{step}}^{2}$. To obtain $\mathrm{v_{th}}(a,b)$, we use (3.4) with $\rho_{\mathrm{step}}$ and $\gamma_{\mathrm{step}} = \rho_{\mathrm{step}}^{2}$ to evaluate $\mathrm{Var}(C_{\mathrm{ab}})$ and $\mathrm{Var}(A_a + A_b)$.

**Fixed-F (Bernoulli start):** The mask flips exactly $F$ of $D$ components uniformly. Hence for any $i$, $\Pr(m_q(i) = -1) = F/D$ and $\Pr(m_q(i) = +1) = 1 - F/D$, so $\rho_{\mathrm{step}} = \mathrm{E}[m_q(i)] = (1 - F/D) - (F/D) = 1 - 2F/D$. For $i \neq j$, $m_q(i)\cdot m_q(j) = -1$ iff exactly one of $\{i,j\}$ is flipped, otherwise $m_q(i)\cdot m_q(j) = 1$. Therefore $\mathrm{E}[m_q(i)\, m_q(j)] = (+1)\cdot\Pr(m_q(i)\cdot m_q(j) = 1) + (-1)\cdot\Pr(m_q(i)\cdot m_q(j) = -1) = 1 - 2\cdot\Pr(m_q(i)\cdot m_q(j) = -1)$. Compute $\Pr(m_q(i)\cdot m_q(j) = -1)$. This occurs iff exactly one of $\{i,j\}$ is flipped. $\Pr(i \text{ flipped}, j \text{ not}) = \Pr(i \text{ flipped})\cdot\Pr(j \text{ not} \mid i \text{ flipped}) = (F/D)\cdot((D - F)/(D - 1))$. By symmetry, $\Pr(i \text{ not}, j \text{ flipped})$ is the same. Hence $\Pr(m_q(i)\cdot m_q(j) = -1) = 2\cdot(F/D)\cdot((D - F)/(D - 1)) = 2F(D - F)/(D(D - 1))$. Thus $\gamma_{\mathrm{step}} = 1 - 4F(D - F)/(D(D - 1))$.

Now, we specialize $\mu_{\mathrm{th}}(a,b)$ and $\mathrm{bias_{th}}(a,b)$. With $F = rD$, we get $\rho_{\mathrm{step}} = 1 - 2F/D = 1 - 2r = \rho$, hence $\mu_{\mathrm{th}}(a,b) = T_{\mathrm{exp}}(\Delta)$ and $\mathrm{bias_{th}}(a,b) = 0$. To obtain $\mathrm{v_{th}}(a,b)$, we use (3.4) with these $\rho_{\mathrm{step}}$ and $\gamma_{\mathrm{step}}$ to evaluate $\mathrm{Var}(C_{\mathrm{ab}})$ and the Bernoulli-start case for $\mathrm{Var}(A_a + A_b)$.

### 3.2.3 Constant Hamming-Weight Initialization Variants (Fixed-pD, Fixed-FpD)

Here $x_0$ has exactly $D/2$ ones, so $\mathrm{E}[y_0(i)] = 0$. For $i \neq j$ we compute $\mathrm{E}[y_0(i) \cdot y_0(j)]$ as follows: $0 = \mathrm{E}[(\sum_i y_0(i))^2] = \mathrm{E}[\sum_i (y_0(i))^2 + \sum_{i \neq j} y_0(i)\, y_0(j)] = D + D(D-1) \cdot \mathrm{E}[y_0(i)\, y_0(j)]$, hence $\mathrm{E}[y_0(i)\, y_0(j)] = -1/(D-1)$ for $i \neq j$.

**Same-level correlations:** For $i \neq j$, using $y_q(i) = y_0(i) \prod_{t=0}^{q-1} m_t(i)$ and independence across level transitions,

$$\mathrm{E}[y_q(i) y_q(j)] = (\mathrm{E}[m_q(i) m_q(j)])^q\, \mathrm{E}[y_0(i) y_0(j)] = -\gamma_{\text{step}}^q / (D-1). \tag{3.6}$$

Since $\mathrm{E}[y_q(i)] = 0$, we have $\mathrm{Cov}(y_q(i), y_q(j)) = \mathrm{E}[y_q(i)\, y_q(j)]$ for $i \neq j$ and $\mathrm{Var}(y_q(i)) = 1$. Therefore $\mathrm{Var}(A_q) = \mathrm{Var}(\sum_{i=1}^{D} y_q(i)) = \sum_i \mathrm{Var}(y_q(i)) + \sum_{i \neq j} \mathrm{Cov}(y_q(i), y_q(j)) = D + D(D-1) \cdot (-\gamma_{\text{step}}^q / (D-1)) = D \cdot (1 - \gamma_{\text{step}}^q)$.

**Cross-level covariance:** Since $\mathrm{E}[A_a] = \mathrm{E}[A_b] = 0$, we have $\mathrm{Cov}(A_b, A_a) = \mathrm{E}[A_b\, A_a] = \sum_{i=1}^{D} \sum_{j=1}^{D} \mathrm{E}[y_b(i) \cdot y_a(j)]$. If $i = j$, (3.2) gives $\mathrm{E}[y_b(i) \cdot y_a(i)] = \rho_{\text{step}}^{\Delta}$. If $i \neq j$, write $y_b(i) = y_a(i) \prod_{t=a}^{b-1} m_t(i)$. The masks in level transitions $t = a, \dots, b-1$ are independent of the past, hence $\mathrm{E}[y_b(i) \cdot y_a(j)] = \mathrm{E}[\prod_{t=a}^{b-1} m_t(i)]\, \mathrm{E}[y_a(i) \cdot y_a(j)] = \rho_{\text{step}}^{\Delta}\, \mathrm{E}[y_a(i) \cdot y_a(j)]$. Using (3.6) at $q = a$ gives $\mathrm{E}[y_a(i) \cdot y_a(j)] = -\gamma_{\text{step}}^a / (D-1)$, so $\mathrm{E}[y_b(i) \cdot y_a(j)] = -\rho_{\text{step}}^{\Delta} \cdot \gamma_{\text{step}}^a / (D-1)$ for $i \neq j$. Therefore $\mathrm{Cov}(A_b, A_a) = D \cdot \rho_{\text{step}}^{\Delta} \cdot (1 - \gamma_{\text{step}}^a)$, hence

$$\mathrm{Var}(A_a + A_b) = D(2 - \gamma_{\text{step}}^a - \gamma_{\text{step}}^b + 2\rho_{\text{step}}^{\Delta}(1 - \gamma_{\text{step}}^a)). \tag{3.7}$$

**Fixed-pD:** The per-step mask distribution is the same as in Baseline, hence $\rho_{\text{step}} = 1 - 2r = \rho$ and $\gamma_{\text{step}} = \rho_{\text{step}}^2$. Because $\rho_{\text{step}} = \rho$, we get $\mathrm{bias}_{\text{th}}(a,b) = 0$. The $v_{\text{th}}(a,b)$ is obtained by evaluating (3.4) with these $(\rho_{\text{step}}, \gamma_{\text{step}})$ and the constant Hamming-weight start case for $\mathrm{Var}(A_a + A_b)$ (3.7).

**Fixed-FpD:** The per-step mask distribution is the same fixed-count mask as in Fixed-$F$, hence $\rho_{\text{step}} = 1 - 2F/D$ and $\gamma_{\text{step}} = 1 - 4 \cdot F(D-F)/(D(D-1))$. We have $\rho_{\text{step}} = 1 - 2F/D = 1 - 2r = \rho$, hence $\mathrm{bias}_{\text{th}}(a,b) = 0$. The $v_{\text{th}}(a,b)$ is obtained by evaluating (3.4) with these $(\rho_{\text{step}}, \gamma_{\text{step}})$ and the constant Hamming-weight-start expression (3.7) for $\mathrm{Var}(A_a + A_b)$.

Table: Variance summary (Exponential family, $p = 1/2$). Variance uses the common template $v_{\text{th}}(a,b) = (\mathrm{Var}(A_a + A_b) + \mathrm{Var}(C_{\text{ab}}))/(4\, D^2)$ with $\mathrm{Var}(C_{\text{ab}})$ given by (6.2), whereas $\mathrm{Var}(A_a + A_b)$ differs between Bernoulli and constant Hamming-weight initializations. All terms are instantiated using the $\rho_{\text{step}}$ and $\gamma_{\text{step}}$ listed in the table.

| | Baseline | Fixed-pD | Fixed-F | Fixed-FpD |
|---|---|---|---|---|
| Start | Bernoulli$(1/2)$ | $\lvert x_0 \rvert = D/2$ | Bernoulli$(1/2)$ | $\lvert x_0 \rvert = D/2$ |
| Mask per step | Bernoulli$(r)$ | Bernoulli$(r)$ | exactly $F = rD$ | exactly $F = rD$ |

| | Baseline | Fixed-pD | Fixed-F | Fixed-FpD |
|---|---|---|---|---|
| $\rho_{\text{step}}$ | $1-2r$ | $1-2r$ | $1-2F/D$ | $1-2F/D$ |
| $\gamma_{\text{step}}$ | $\rho_{\text{step}}^2$ | $\rho_{\text{step}}^2$ | $1-4F(D-F)/(D(D-1))$ | $1-4F(D-F)/(D(D-1))$ |
| $\text{Var}(A_a+A_b)$ | (6.3) | (6.4) | (6.3) | (6.4) |

### 3.2.4 Constant Hamming-Weight Initialization and Fixed Swap Count (Fixed-kpD)

**Mean and bias:** Fix anchor level $a$. For $d = 0,1,\dots,b-a$ consider the overlap $O(a,a+d)$ (so at $d = b-a$ we have $a+d = b$). For one equal-swap step (from $d$ to $d+1$), condition on $O(a,a+d) = m$. At both levels $a$ and $a+d$ there are exactly $D/2$ ones. Among the $D/2$ components that are 1 at level $a$, exactly $m$ are still 1 at level $a+d$, and the remaining $D/2 - m$ are 0 at level $a+d$. One equal-swap step transforms $x_{a+d}$ into $x_{a+d+1}$ by flipping $k$ ones to zeros and $k$ zeros to ones. Define: $X$ = # of selected "1→0" flips that fall inside the level-$a$ ones; $X \mid m \sim \text{Hypergeometric}(D/2,m,k)$; $Y$ = # of selected "0→1" flips that fall inside the level-$a$ ones; $Y \mid m \sim \text{Hypergeometric}(D/2,D/2-m,k)$. Given $m$, the $k$ selected ones and the $k$ selected zeros are sampled independently from the disjoint sets. Hence $X$ and $Y$ are independent given $m$, and $O(a,a+d+1) = m - X + Y$. Taking conditional expectations $\text{E}_{\text{step}}$ over the within-step randomness in this one equal-swap transition (i.e., over the random choice of the $k$ ones and $k$ zeros selected to flip at level $a+d$), $\text{E}_{\text{step}}[O(a,a+d+1) \mid O(a,a+d) = m] = m - \text{E}_{\text{step}}[X \mid m] + \text{E}_{\text{step}}[Y \mid m]$, where $\text{E}_{\text{step}}[X \mid m] = k\cdot(m/(D/2))$ and $\text{E}_{\text{step}}[Y \mid m] = k\cdot((D/2-m)/(D/2))$. Therefore $\text{E}_{\text{step}}[O(a,a+d+1) \mid O(a,a+d) = m] = (1-4k/D)\cdot m + k = \rho_{\text{step}}\cdot m + k$, where $\rho_{\text{step}} = 1 - 4k/D$. Hence $k = (D/4)\cdot(1-\rho_{\text{step}})$, which is used in final formulas.

Applying the law of total expectation to the random variable $O(a,a+d)$ (with outer $\text{E}[\cdot]$ over generator realizations and $\text{E}_{\text{step}}$ over the within-transition selection randomness) gives: $\text{E}[O(a,a+d+1)] = \text{E}[\text{E}_{\text{step}}[O(a,a+d+1) \mid O(a,a+d)]] = \rho_{\text{step}}\cdot\text{E}[O(a,a+d)] + k$, with base case $O(a,a) = D/2$. This yields the closed form $\text{E}[O(a,a+d)] = D/4 + (D/4)\cdot\rho_{\text{step}}^d$. Therefore, $\mu_{\text{th}}(a,b) := \text{E}[S(a,b)] = 2\cdot\text{E}[O(a,b)]/D = 1/2 + (1/2)\cdot\rho_{\text{step}}^{\Delta}$. With $k := rD/2$, we have $\rho_{\text{step}} = 1 - 4k/D = 1 - 2r = \rho$, hence $\text{bias}_{\text{th}}(a,b) = \mu_{\text{th}}(a,b) - T_{\text{exp}}(\Delta) = 0$.

Finite-D variance $\text{v}_{\text{th}}(a,b)$**:** Since $\text{Var}(O(a,b)) = \text{E}[O^2(a,b)] - (\text{E}[O(a,b)])^2$, we use the closed form for $\text{E}[O(a,a+d)]$ above and compute $\text{E}[O^2\,(a,a+d)]$ by iterating the one-step update for $d = 0,\dots,b-a-1$; this yields $\text{E}[O^2(a,a+(b-a))] = \text{E}[O^2(a,b)]$ and hence $\text{Var}(O(a,b))$. Then

$$\text{v}_{\text{th}}(a,b) := 4\text{Var}(O(a,b))/D^2. \tag{3.8}$$

We already have $\text{E}[O(a,a+d)]$ in closed form above; we obtain $\text{E}[O^2(a,a+d)]$ by a one-step recursion. Start from the conditional identity (for a single step, conditional on $O(a,a+d)$):

$\mathrm{E_{step}}[O^2(a,a+d+1) \mid O(a,a+d)] =$
$\mathrm{Var_{step}}\big(O(a,a+d+1) \mid O(a,a+d)\big)+\big(\mathrm{E_{step}}[O(a,a+d+1) \mid O(a,a+d)]\big)^2$. Taking E of both sides and using the law of total expectation gives $\mathrm{E}[O^2(a,a+d+1)] = \mathrm{E}[\mathrm{E_{step}}[O^2(a,a+d+1) \mid O(a,a+d)]] = \mathrm{E}[\mathrm{Var_{step}}(O(a,a+d+1) \mid O(a,a+d))]+\mathrm{E}[(\mathrm{E_{step}}[O(a,a+d+1) \mid O(a,a+d)])^2]$.

For the first term, use the hypergeometric variance. Let $c := ((D/2) - k)/((D/2) - 1)$. Then $\mathrm{Var_{step}}(X \mid m) = k\cdot(m/(D/2))\cdot(1 - m/(D/2))\cdot c$, and $\mathrm{Var_{step}}(Y \mid m)$ has the same value, so $\mathrm{Var_{step}}(O(a,a+d+1) \mid O(a,a+d) = m) = \mathrm{Var_{step}}(X \mid m)+\mathrm{Var_{step}}(Y \mid m) = 2kc(m/(D/2))(1 - m/(D/2))$. Substituting $m = O(a,a+d)$ and averaging over realizations gives $\mathrm{E}[\mathrm{Var_{step}}(O(a,a+d+1) \mid O(a,a+d))] = 2kc((2/D)\cdot\mathrm{E}[O(a,a+d)] - (4/D^2)\cdot\mathrm{E}[O^2(a,a+d)])$.
For the second term, use the conditional mean derived above, $\mathrm{E_{step}}[O(a,a+d+1) \mid O(a,a+d)] = \rho_{\mathrm{step}}\cdot O(a,a+d)+k$: $\mathrm{E}[(\mathrm{E_{step}}[O(a,a+d+1) \mid O(a,a+d)])^2] =$
$\mathrm{E}[(\rho_{\mathrm{step}}\cdot O(a,a+d)+k)^2] = \rho_{\mathrm{step}}^2\mathrm{E}[O^2(a,a+d)]+2\rho_{\mathrm{step}}k\cdot\mathrm{E}[O(a,a+d)]+k^2$.
Adding the two terms yields the one-step recursion for $\mathrm{E}[O^2(a,a+d+1)]$ in terms of $\mathrm{E}[O(a,a+d)]$ and $\mathrm{E}[O^2(a,a+d)]$. Iterating this recursion for $d = 0,\dots,b - a - 1$ (starting from $\mathrm{E}[O(a,a)] = D/2$ and $\mathrm{E}[O^2(a,a)] = (D/2)^2$) gives $\mathrm{E}[O^2(a,a+(b - a))] = \mathrm{E}[O^2(a,b)]$, and then $\mathrm{v_{th}}(a,b)$ follows from (3.8).
**Optional closed form for** $\mathrm{E}[O^2(a,a+d)]$: Define $A := \rho_{\mathrm{step}}^2 - (8kc)/D^2$ and $B := 2\rho_{\mathrm{step}}k+(4kc)/D$. Then the recursion above can be written as $\mathrm{E}[O^2(a,a+d+1)] = A\cdot\mathrm{E}[O^2(a,a+d)]+B\cdot\mathrm{E}[O(a,a+d)]+k^2$, with $\mathrm{E}[O(a,a+d)] = D/4+(D/4)\cdot\rho_{\mathrm{step}}^d$ and $\mathrm{E}[O^2(a,a)] = (D/2)^2$. The closed form is $\mathrm{E}[O^2(a,a+d)] = A^d\cdot\mathrm{E}[O^2(a,a)]+(BD/4+k^2)(A^d - 1)/(A - 1)+(BD/4)\cdot(A^d - \rho_{\mathrm{step}}^d)/(A - \rho_{\mathrm{step}})$, provided $A \neq 1$ and $A \neq \rho_{\mathrm{step}}$; otherwise use the corresponding limit (or iterate the one-step recursion, which is numerically stable for $b - a \leq Q$).

### 3.3 Finite-D Accuracy for Linear Similarity Family

We use the evaluation protocol and error metrics of section 3.3. As above, assume $a \leq b$ and write $\Delta = b - a$. We derive $\mathrm{bias_{th}}(a,b)$ and $\mathrm{v_{th}}(a,b)$ for each linear derandomization variant, except for the deterministic equal-swap constructions Fixed-kpD Block-Swap and Float, for which $S(a,b) = T_{\mathrm{lin}}(a,b)$ exactly. Therefore $\mathrm{bias_{th}}(a,b) = 0$ and $\mathrm{v_{th}}(a,b) = 0$, and hence $\mathrm{BiasRMS_{th}}(a) = \mathrm{VarRMS_{th}}(a) = \mathrm{RMSE_{th}}(a) = 0$.

#### 3.3.1 Derandomization Variant Fixed-F (Linear)

This derivation applies to the Fixed-F generators Fixed-F (Pool-Uniform) and Fixed-F (Perm-Scan). Let $K$ be the number of distinct components flipped between levels $a$ and $b$; here $K = \Delta\cdot F$ exactly (for $\Delta \leq Q$). These components contribute 0 overlap because $x_b(i) = 1 -$ $x_a(i)$. On the remaining $D - K$ components, we have $x_b(i) = x_a(i)$, so the overlap equals the count of ones in $x_a$ over those indices. Although such components may have been flipped at earlier levels $q < a$, for each $i$ the value $x_a(i)$ equals either $x_0(i)$ or $1 - x_0(i)$ depending on the flip parity up to level $a$; since $\{x_0(i)\}$ are i.i.d. $\mathrm{Bernoulli}(1/2)$ and the update schedule is independent of $x_0$, it follows that $\{x_a(i)\}$ are i.i.d. $\mathrm{Bernoulli}(1/2)$; in particular $\mathrm{E}[x_a(i)] = 1/2$ for every $i$.

**Mean and bias:** From $\mathrm{E}[x_a(i)] = 1/2$ it follows that $\mathrm{E}[O(a,b)] = (D - K)/2$ and hence $\mu_{\mathrm{th}}(a,b) := \mathrm{E}[S(a,b)] = 2{\cdot}\mathrm{E}[O(a,b)]/D = 1 - K/D = 1 - \Delta{\cdot}F/D$. With $F = D/Q$, this equals $1 - \Delta/Q$, matching $T_{\mathrm{lin}}(\Delta)$, so $\mathrm{bias}_{\mathrm{th}}(a,b) = 0$.

**Finite-$D$ variance:** Restricting $\{x_a(i)\}$ to the unchanged index set (of size $D - K$) preserves i.i.d. $\mathrm{Bernoulli}(1/2)$, hence $O(a,b)$ is $\mathrm{Binomial}(D - K, 1/2)$ and $\mathrm{Var}(O(a,b)) = (D - K)/4$. Thus $\mathrm{v}_{\mathrm{th}}(a,b) := \mathrm{Var}(S(a,b)) = 4{\cdot}\mathrm{Var}(O)/D^2 = (D - K)/D^2 = (1 - \Delta{\cdot}F/D)/D = (1 - \Delta/Q)/D$.

### 3.3.2 Derandomization Variant Fixed-FpD (Linear)

This derivation applies to the Fixed-FpD generators Fixed-FpD (Pool-Uniform) and Fixed-FpD (Perm-Scan).

**Mean and bias.** The mean is identical to Fixed-F: $\mu_{\mathrm{th}}(a,b) = 1 - \Delta/Q$, and so $\mathrm{bias}_{\mathrm{th}}(a,b) = 0$.

**Finite-$D$ variance.** This differs from Fixed-F because $|x_0| = D/2$ induces dependence across coordinates. Let $M_q$ denote the set of component indices flipped in the transition $q \rightarrow q+1$ with $|M_q| = F$. Define the cumulative flipped set for any level $q$ as $B_q := \bigcup_{t=0}^{q-1} M_t$. Under the Linear no-reuse construction (disjoint per-level masks), $B_a \subset B_b$ and $|B_q| = qF$, so $bF \le D$. Define the following counts of initial ones: $X_a := \#\{i \in B_a : x_0(i) = 1\}$ (number of $x_0$ ones inside $B_a$); $Y_b := \#\{i \in B_b : x_0(i) = 1\}$ (number of $x_0$ ones inside $B_b$). Partition components into three disjoint sets:

(i) $i \in B_b \setminus B_a$: these components are flipped exactly once between levels $a$ and $b$, hence $x_b(i) = 1 - x_a(i)$. Therefore, they contribute 0 to overlap.

(ii) $i \in B_a$: these components were flipped before level $a$, hence $x_a(i) = x_b(i) = 1 - x_0(i)$. They contribute the number of zeros of $x_0$ inside $B_a$, which is $|B_a| - X_a$.

(iii) $i \notin B_b$: these components have not been flipped before level $b$, hence $x_a(i) = x_b(i) = x_0(i)$. They contribute the number of ones of $x_0$ outside $B_b$, which is $D/2 - Y_b$.

Summing the three contributions gives: $O(a,b) = (D/2 - Y_b) + (|B_a| - X_a) = D/2 + aF - (X_a + Y_b)$. Thus, $\mathrm{Var}(O(a,b)) = \mathrm{Var}(X_a + Y_b)$ and

$$\mathrm{v}_{\mathrm{th}}(a,b) := \mathrm{Var}(S(a,b)) = 4\mathrm{Var}(O(a,b))/D^2 = 4\mathrm{Var}(X_a + Y_b)/D^2 \qquad (3.9)$$

We now compute $\mathrm{Var}(X_a + Y_b)$. Using the law of total variance and conditioning on $Y_b$ gives $\mathrm{Var}(X_a + Y_b) = \mathrm{E}[\mathrm{Var}(X_a + Y_b \mid Y_b)] + \mathrm{Var}(\mathrm{E}[X_a + Y_b \mid Y_b])$. Because $Y_b$ is fixed under this conditioning, $\mathrm{Var}(X_a + Y_b \mid Y_b) = \mathrm{Var}(X_a \mid Y_b)$ and $\mathrm{E}[X_a + Y_b \mid Y_b] = \mathrm{E}[X_a \mid Y_b] + Y_b$. Therefore $\mathrm{Var}(X_a + Y_b) = \mathrm{E}[\mathrm{Var}(X_a \mid Y_b)] + \mathrm{Var}(\mathrm{E}[X_a \mid Y_b] + Y_b)$.

**Step 1:** $Y_b \sim \text{Hypergeometric}(D, D/2, bF)$, so $\text{Var}(Y_b) = (bF/4)\cdot(D - bF)/(D - 1)$.

**Step 2:** Conditional moments of $X_a$ given $Y_b$.

The set $B_a$ is the union of the first $a$ masks and is a size-$aF$ subset of $B_b$. Conditional on $(B_b, Y_b)$, the $Y_b$ ones are uniformly distributed over the $bF$ positions of $B_b$. So $X_a \mid (B_b, Y_b) \sim \text{Hypergeometric}(|B_b| = bF, Y_b, |B_a| = aF)$. Hence $\text{E}[X_a \mid (B_b, Y_b)] = aF\,(Y_b/bF)$, $\text{Var}(X_a \mid (B_b, Y_b)) = aF(Y_b/bF)(1 - Y_b/bF)(bF - aF)/(bF - 1)$. These expressions depend only on $Y_b$, $aF$, and $bF$ (not on the particular indices in $B_b$), therefore they also give $\text{E}[X_a \mid Y_b]$ and $\text{Var}(X_a \mid Y_b)$. Using $c := aF\,(bF - aF)/(bF - 1)$, we have $\text{Var}(X_a \mid Y_b) = c\,(Y_b/bF)(1 - Y_b/bF)$. Since $\text{E}[Y_b] = bF/2$, we have $\text{E}[(Y_b/bF)(1 - Y_b/bF)] = 1/4 - \text{Var}(Y_b)/(bF)^2$. Hence $\text{E}[\text{Var}(X_a \mid Y_b)] = c\,(1/4 - \text{Var}(Y_b)/(bF)^2)$. Also $\text{E}[X_a \mid Y_b] + Y_b = (aF/bF)\,Y_b + Y_b = (1 + aF/bF)Y_b$, so $\text{Var}(\text{E}[X_a \mid Y_b] + Y_b) = (1 + aF/bF)^2\,\text{Var}(Y_b)$. Therefore, $\text{Var}(X_a + Y_b) = c\,(1/4 - \text{Var}(Y_b)/(bF)^2) + (1 + aF/bF)^2\,\text{Var}(Y_b)$. Substituting $c$ and $\text{Var}(Y_b)$ into (3.9) gives

$$\text{v}_{\text{th}}(a,b) = \frac{D(3aF + bF) - (aF + bF)^2}{D^2(D - 1)}.$$

### 3.3.3 Derandomization Variant Fixed-pD (Linear)

This derivation applies to the Fixed-pD generator Perm-Scan. The key difference from Fixed-F and Fixed-FpD is that the per-level requested flip count $f_q$ is random rather than fixed, with $\text{E}[f_q] = D/Q$.

**Mean and bias**: Let $C_q := f_0 + \ldots + f_{q-1}$. Under the Binomial proposed-count mechanism (Proposition 2), $C_q \sim \text{Binomial}(qD, 1/Q)$, and the cumulative number of distinct flipped components by level $q$ is exactly $|B_q| := \min(D, C_q)$. Let $K$ be the number of newly flipped distinct components between levels $a$ and $b$: $K := |B_b| - |B_a|$. Hence $\text{E}[K] = \text{E}[|B_b|] - \text{E}[|B_a|]$. As in Fixed-F derivation, $\text{E}[S(a,b) \mid K] = 1 - K/D$. Therefore $\mu_{\text{th}}(a,b) = \text{E}[\text{E}[S(a,b)|K]] = \text{E}[1 - K/D] = 1 - \text{E}[K]/D$ and $\text{bias}_{\text{th}}(a,b) = \mu_{\text{th}}(a,b) - T_{\text{lin}}(\Delta) = \Delta/Q - \text{E}[K]/D$.

The required moments are evaluated exactly from $C_q \sim \text{Binomial}(qD, 1/Q)$ and $|B_q| := \min(D, C_q)$. In particular, $\text{E}[|B_q|^n] = \sum_{j=0}^{D-1} j^n \Pr(C_q = j) + D^n \Pr(C_q \geq D), n = 1,2$ and $\text{E}[|B_a||B_b|] = D^2 \Pr(C_a \geq D) + \sum_{j=0}^{D-1} j \Pr(C_a = j)(j + \text{E}[\min(D - j, \sum_{t=a}^{b-1} f_t)$.

**Finite-D variance**: Start from the law of total variance with respect to $K$: $\text{Var}(S(a,b)) = \text{E}[\text{Var}(S(a,b) \mid K)] + \text{Var}(\text{E}[S(a,b) \mid K])$.

For the first term, condition on $|B_a|$ and $|B_b|$. Since $B_a \subset B_b$ and $K := |B_b| - |B_a|$, the Fixed-FpD calculation applies with $aF$ and $bF$ replaced by $|B_a|$ and $|B_b|$, respectively:

$$\text{Var}(S(a,b) \mid |B_a|, |B_b|) = \frac{D(3|B_a| + |B_b|) - (|B_a| + |B_b|)^2}{D^2(D - 1)}.$$

Because $\mathrm{E}[S(a,b) \mid |B_a|, |B_b|] = 1 - K/D$ depends on these two quantities only through $K$, conditioning further on $|B_a|$ and $|B_b|$ introduces no additional conditional-mean variance. Hence, after averaging over $K$,

$$\mathrm{E}[\mathrm{Var}(S(a,b) \mid K)] = \frac{D(3E[|B_a|] + E[|B_b|]) - E[|B_a|^2] - 2\mathrm{E}[|B_a||B_b|] - \mathrm{E}[|B_b|^2]}{D^2(D-1)}.$$

For the second term, since $\mathrm{E}[S(a,b) \mid K] = 1 - K/D$, we have $\mathrm{Var}(\mathrm{E}[S(a,b) \mid K]) = \mathrm{Var}(1 - K/D) = \mathrm{Var}(K)/D^2$, where $\mathrm{Var}(K) = \mathrm{E}[|B_b|^2] + \mathrm{E}[|B_a|^2] - 2\,\mathrm{E}[|B_b| \cdot |B_a|] - (\mathrm{E}[K])^2)$.

Finally, $\mathrm{v_{th}}(a,b) = \mathrm{E}[\mathrm{Var}(\mathrm{S(a,b)} \mid K)] + \mathrm{Var}(K)/D^2$.

3.3.4 Derandomization Variant Baseline (Linear)

Generators covered: Baseline (Perm-Scan).

**Mean and bias:** As in the Fixed-pD derivation, $\mathrm{bias_{th}}(a,b) = \Delta/Q - \mathrm{E}[K]/D$, where the required expectation is evaluated exactly from the capped-binomial moments defined in the Fixed-pD section.

**Finite-D variance:** Unlike Fixed-pD, the i.i.d. initialization of $\mathbf{x}_0$ in Baseline implies that, conditional on $K$, the $D - K$ unchanged components are independent $\mathrm{Bernoulli}(1/2)$; hence $O(a,b)$ is $\mathrm{Binomial}(D - K, 1/2)$. So $\mathrm{E}[S(a,b) \mid K] = (D - K)/D$ and $\mathrm{Var}(S(a,b) \mid K) = 4 \cdot \mathrm{Var}(O(a,b)|K)/D^2 = 4 \cdot (D - K) \cdot (1/2)\ (1/2)\ /\ D^2 = (D - K)/D^2$. Apply the law of total variance over realizations of $K$: $\mathrm{Var}(S(a,b)) = \mathrm{E}[\mathrm{Var}(S(a,b)|K)] + \mathrm{Var}(\mathrm{E}[S(a,b)|K]) = \mathrm{E}[(D - K)/D^2] + \mathrm{Var}(1 - K/D) = (1 - \mathrm{E}[K]/D)/D + \mathrm{Var}(K)/D^2$. Here $\mathrm{E}[K]$ and $\mathrm{Var}(K)$ are evaluated as in the Fixed-pD section.

# 4 Supplementary Note 4: Experiments

## 4.1 Supplementary Similarity Profiles

This section complements section 7.1 with supplementary overlap- and Hamming-similarity profiles for the same generator families, using additional parameter settings and/or the same generators and seeds as in the main-text examples.

### 4.1.1 Exponential Similarity Family

Figure S1 shows normalized Hamming similarity for the same exponential-family generators, seeds, anchors, and parameter settings as Figure 4 in section 7.1 ($r = 0.01$). Note that $\mathrm{sim}_{\mathrm{ham}}(a,a) = 1$. We observe identical per-seed similarity curves for Baseline and Fixed-pD, and also for Fixed-F and Fixed-FpD. This per-seed equivalence occurs because these variant pairs use the same flip schedule between levels. Initialization constraints can change overlap similarity, because overlap depends on where the 1s are. However, for the same update-count schedule, they do not change the number of components that differ between two levels, so the Hamming similarity is unchanged. For Fixed-F and Fixed-FpD, deviations from the target law are visibly smaller, because the one-step Hamming distance is exactly $F$ by construction; in Baseline and Fixed-pD the Bernoulli flip mask makes the per-step flip count random, which adds visible per-seed fluctuations around the target law. Fixed-kpD behaves most regularly, and coincides with the similarity curves for $2\mathrm{sim}_{\mathrm{dot}}/D$. Figures S2, S3 show the overlap-similarity and normalized Hamming-similarity profiles, respectively, at $r = 0.1$.

### 4.1.2 Linear Similarity Family

Figure S4 shows normalized Hamming similarity for the same linear-family generators, seeds, anchors, and parameter settings as Figure 5 in the main text. Baseline and Fixed-pD share the same per-seed curves, while Fixed-F, Fixed-FpD, and Fixed-kpD share the exact linear curves. Fixed-F, Fixed-FpD, and Fixed-kpD give exact linear Hamming profiles because each transition flips exactly $F$ previously unused components.

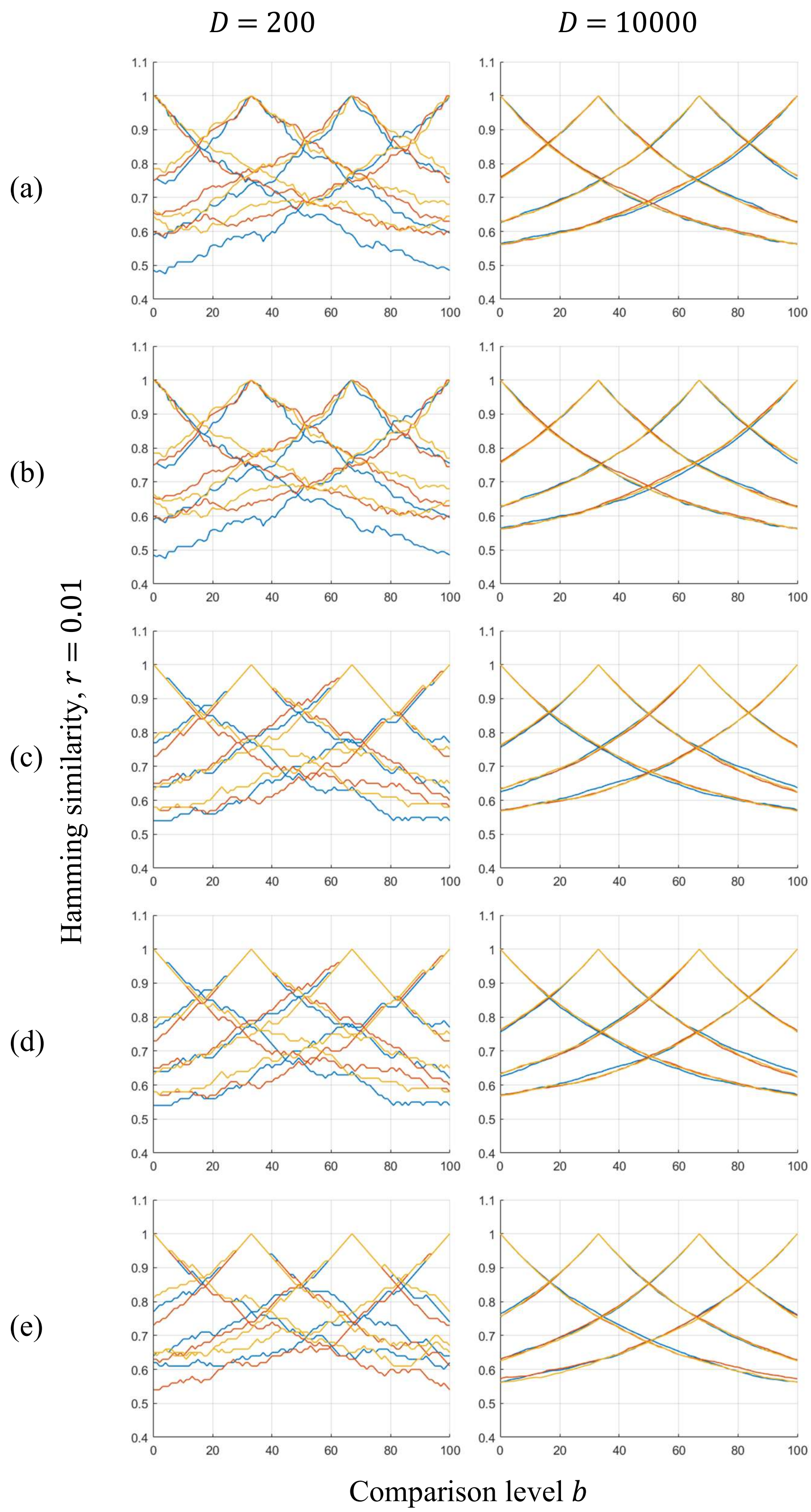

Figure S1: Exponential-family similarity profiles ($\mathrm{sim}_{\mathrm{ham}}$ as a function of comparison level $b$) from 3 realizations (seeds), starting at anchors $(0,33,67,100)$. Five generator constructions: Baseline (a), Fixed-pD (b), Fixed-F (c), Fixed-FpD (d), Fixed-kpD (e). $Q = 100$, $D \in \{200,10000\}$, $r = 0.01$.

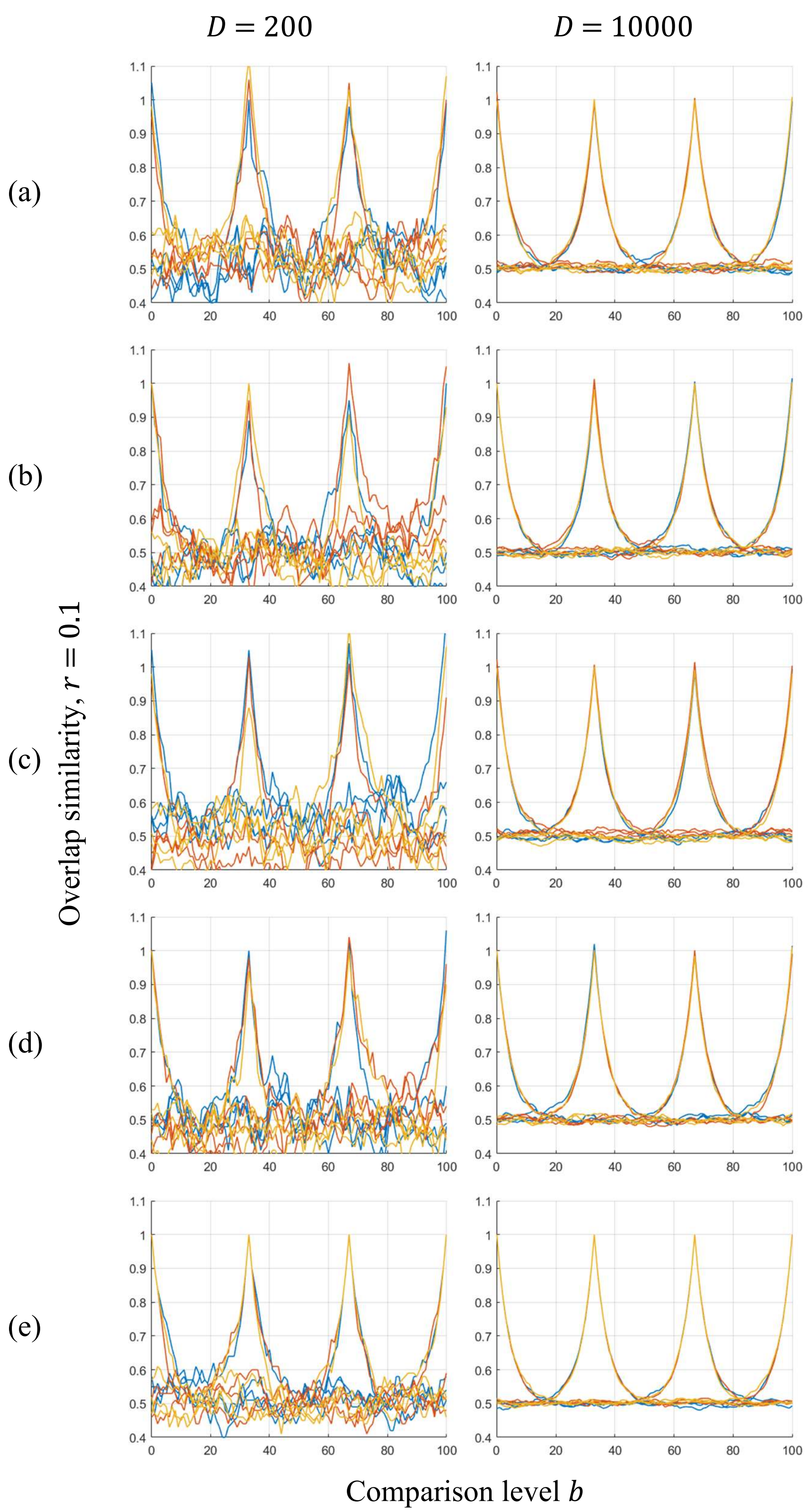

Figure S2: Exponential-family similarity profiles ($2\,\mathrm{sim}_{\mathrm{dot}}/D$ as a function of comparison level $b$) from 3 realizations (seeds), starting at anchors $(0,33,67,100)$. Five generator constructions: Baseline (a), Fixed-pD (b), Fixed-F (c), Fixed-FpD (d), Fixed-kpD (e). $Q = 100$, $D \in \{200,10000\}$, $r = 0.1$.

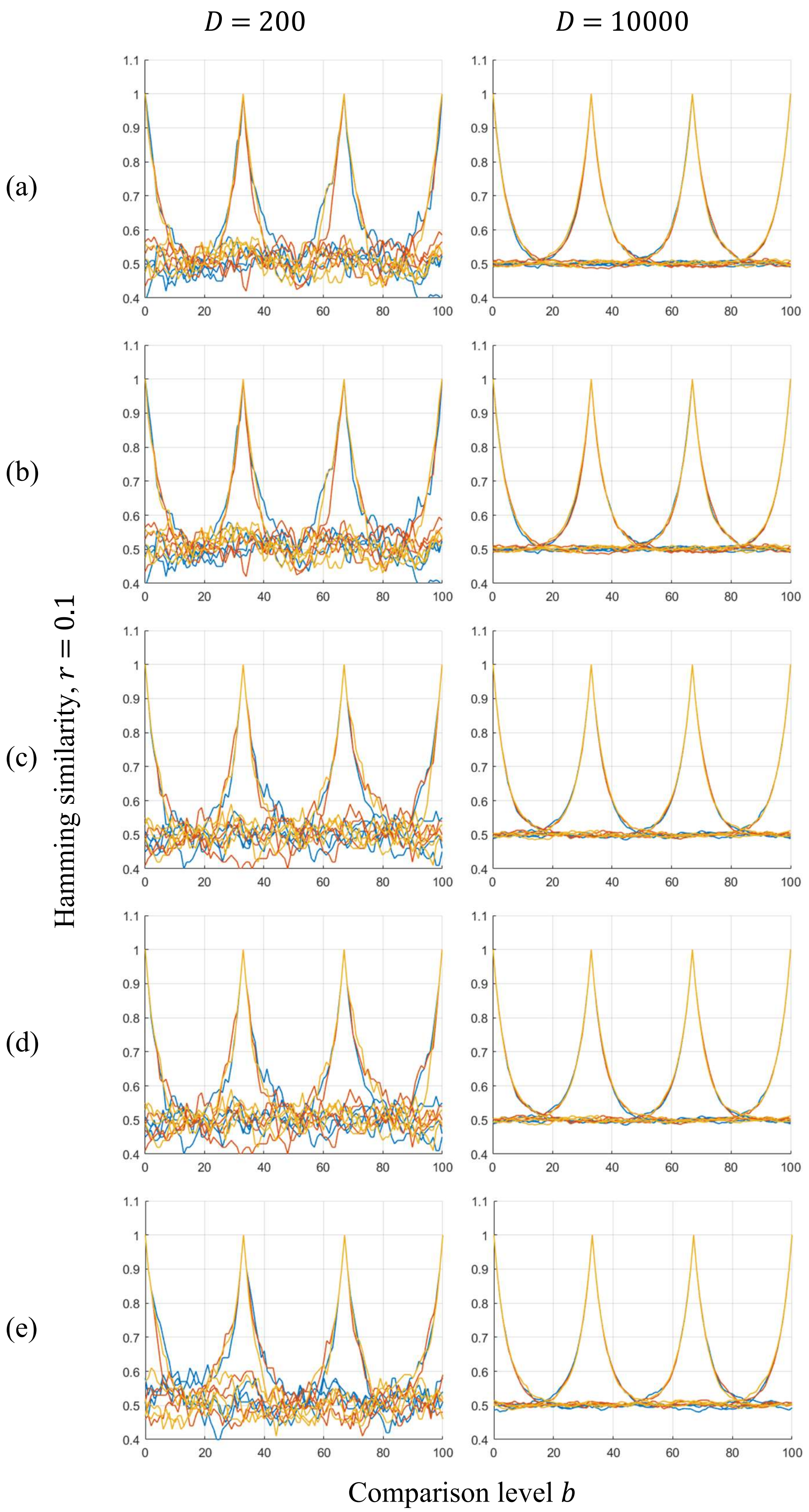

Figure S3: Exponential-family similarity profiles ($\mathrm{sim}_{\mathrm{ham}}$ as a function of comparison level $b$) from 3 realizations (seeds), starting at anchors $(0,33,67,100)$. Five generator constructions: Baseline (a), Fixed-pD (b), Fixed-F (c), Fixed-FpD (d), Fixed-kpD (e). $Q = 100$, $D \in \{200,10000\}$, $r = 0.1$.

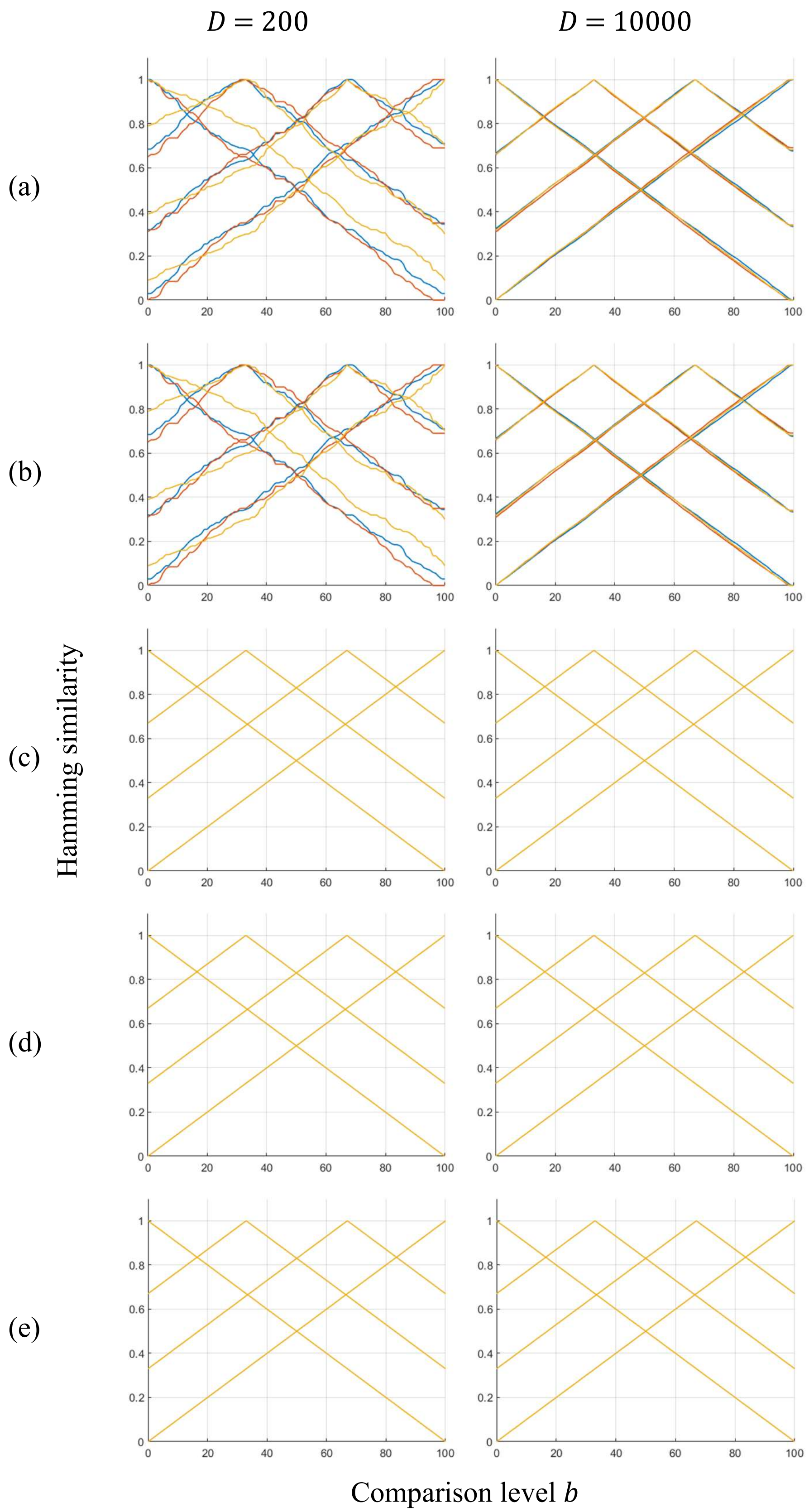

Figure S4: Linear-family similarity profiles ($\mathrm{sim}_{\mathrm{ham}}$ as a function of comparison level $b$) from 3 realizations (seeds), starting at anchors $(0,33,67,100)$. Five generator constructions: Baseline (a), Fixed-pD (b), Fixed-F (c), Fixed-FpD (d), Fixed-kpD (e). $Q = 100$, $D \in \{200,10000\}$.

## 4.2 RMSE of Similarity: Experiments versus Theory

This section reports additional RMSE tables corresponding to the experiments summarized in section 7.2, provides additional parameter settings and the interpretation of the trends.

### 4.2.1 Exponential Similarity Family

Table S1: Exponential similarity family. $\mathrm{RMSE}_{\mathrm{emp}}(a)$ and $\mathrm{RMSE}_{\mathrm{th}}(a)$ for anchors $a \in \{0,33,67,100\}$ and $r \in \{0.01,0.02,0.05,0.10,0.20\}$. 'Mean' averages the anchor-wise RMSE values over the listed anchors. Parameters: $D = 10000$, $Q = 100$, $10000$ realizations. Theory rows are shown directly below the corresponding experimental rows.

| Variant | Generator | $r$ | $a = 0$ | $a = 33$ | $a = 67$ | $a = 100$ | Mean |
|---|---|---|---|---|---|---|---|
| 1 | Baseline | 0.01 | 0.009534 | 0.009757 | 0.009727 | 0.009519 | 0.009634 |
| 1 | Baseline | 0.02 | 0.009236 | 0.009438 | 0.009506 | 0.009132 | 0.009328 |
| 1 | Baseline | 0.05 | 0.008861 | 0.009081 | 0.009096 | 0.008866 | 0.008976 |
| 1 | Baseline | 0.10 | 0.008774 | 0.008838 | 0.008806 | 0.008760 | 0.008795 |
| 1 | Baseline | 0.20 | 0.008709 | 0.008727 | 0.008716 | 0.008675 | 0.008707 |
| 1 theory | Baseline | 0.01 | 0.009507 | 0.009748 | 0.009748 | 0.009507 | 0.009628 |
| 1 theory | Baseline | 0.02 | 0.009166 | 0.009463 | 0.009463 | 0.009166 | 0.009314 |
| 1 theory | Baseline | 0.05 | 0.008868 | 0.009050 | 0.009050 | 0.008868 | 0.008959 |
| 1 theory | Baseline | 0.10 | 0.008763 | 0.008850 | 0.008850 | 0.008763 | 0.008807 |
| 1 theory | Baseline | 0.20 | 0.008709 | 0.008744 | 0.008744 | 0.008709 | 0.008727 |
| | | | | | | | |
| 2 | Fixed-pD | 0.01 | 0.006169 | 0.008439 | 0.009085 | 0.009076 | 0.008192 |
| 2 | Fixed-pD | 0.02 | 0.006613 | 0.009058 | 0.009305 | 0.008985 | 0.008490 |
| 2 | Fixed-pD | 0.05 | 0.006897 | 0.008964 | 0.008992 | 0.008799 | 0.008413 |
| 2 | Fixed-pD | 0.10 | 0.006980 | 0.008779 | 0.008847 | 0.008734 | 0.008335 |
| 2 | Fixed-pD | 0.20 | 0.007017 | 0.008663 | 0.008693 | 0.008675 | 0.008262 |
| 2 theory | Fixed-pD | 0.01 | 0.006141 | 0.008440 | 0.009036 | 0.008997 | 0.008153 |
| 2 theory | Fixed-pD | 0.02 | 0.006610 | 0.009029 | 0.009246 | 0.008981 | 0.008466 |
| 2 theory | Fixed-pD | 0.05 | 0.006884 | 0.008968 | 0.008978 | 0.008795 | 0.008406 |
| 2 theory | Fixed-pD | 0.10 | 0.006973 | 0.008811 | 0.008811 | 0.008724 | 0.008330 |
| 2 theory | Fixed-pD | 0.20 | 0.007016 | 0.008722 | 0.008722 | 0.008687 | 0.008287 |
| | | | | | | | |
| 3 | Fixed-F | 0.01 | 0.009292 | 0.009429 | 0.009429 | 0.009288 | 0.009360 |
| 3 | Fixed-F | 0.02 | 0.009046 | 0.009245 | 0.009183 | 0.009034 | 0.009127 |
| 3 | Fixed-F | 0.05 | 0.008794 | 0.008939 | 0.008960 | 0.008834 | 0.008882 |
| 3 | Fixed-F | 0.10 | 0.008737 | 0.008739 | 0.008783 | 0.008734 | 0.008748 |
| 3 | Fixed-F | 0.20 | 0.008698 | 0.008698 | 0.008706 | 0.008693 | 0.008699 |
| 3 theory | Fixed-F | 0.01 | 0.009202 | 0.009372 | 0.009372 | 0.009202 | 0.009287 |
| 3 theory | Fixed-F | 0.02 | 0.008993 | 0.009169 | 0.009169 | 0.008993 | 0.009081 |
| 3 theory | Fixed-F | 0.05 | 0.008795 | 0.008906 | 0.008906 | 0.008795 | 0.008850 |
| 3 theory | Fixed-F | 0.10 | 0.008724 | 0.008772 | 0.008772 | 0.008724 | 0.008748 |
| 3 theory | Fixed-F | 0.20 | 0.008687 | 0.008699 | 0.008699 | 0.008687 | 0.008693 |
| | | | | | | | |
| 4 | Fixed-FpD | 0.01 | 0.005671 | 0.007896 | 0.008605 | 0.008640 | 0.007703 |
| 4 | Fixed-FpD | 0.02 | 0.006368 | 0.008629 | 0.008839 | 0.008782 | 0.008155 |
| 4 | Fixed-FpD | 0.05 | 0.006779 | 0.008770 | 0.008844 | 0.008677 | 0.008268 |
| 4 | Fixed-FpD | 0.10 | 0.006934 | 0.008749 | 0.008737 | 0.008658 | 0.008270 |
| 4 | Fixed-FpD | 0.20 | 0.006987 | 0.008707 | 0.008662 | 0.008649 | 0.008251 |
| 4 theory | Fixed-FpD | 0.01 | 0.005657 | 0.008003 | 0.008630 | 0.008674 | 0.007741 |
| 4 theory | Fixed-FpD | 0.02 | 0.006367 | 0.008721 | 0.008946 | 0.008804 | 0.008209 |
| 4 theory | Fixed-FpD | 0.05 | 0.006789 | 0.008822 | 0.008832 | 0.008720 | 0.008291 |
| 4 theory | Fixed-FpD | 0.10 | 0.006924 | 0.008733 | 0.008733 | 0.008684 | 0.008268 |
| 4 theory | Fixed-FpD | 0.20 | 0.006989 | 0.008677 | 0.008677 | 0.008665 | 0.008252 |
| | | | | | | | |
| 5 | Fixed-kpD | 0.01 | 0.003614 | 0.002724 | 0.002733 | 0.003597 | 0.003167 |
| 5 | Fixed-kpD | 0.02 | 0.004323 | 0.003675 | 0.003691 | 0.004299 | 0.003997 |
| 5 | Fixed-kpD | 0.05 | 0.004744 | 0.004490 | 0.004486 | 0.004734 | 0.004613 |
| 5 | Fixed-kpD | 0.10 | 0.004857 | 0.004740 | 0.004738 | 0.004860 | 0.004799 |
| 5 | Fixed-kpD | 0.20 | 0.004920 | 0.004865 | 0.004873 | 0.004923 | 0.004895 |
| 5 theory | Fixed-kpD | 0.01 | 0.003626 | 0.002748 | 0.002748 | 0.003626 | 0.003187 |
| 5 theory | Fixed-kpD | 0.02 | 0.004324 | 0.003698 | 0.003698 | 0.004324 | 0.004011 |
| 5 theory | Fixed-kpD | 0.05 | 0.004733 | 0.004478 | 0.004478 | 0.004733 | 0.004605 |
| 5 theory | Fixed-kpD | 0.10 | 0.004861 | 0.004743 | 0.004743 | 0.004861 | 0.004802 |
| 5 theory | Fixed-kpD | 0.20 | 0.004922 | 0.004869 | 0.004869 | 0.004922 | 0.004895 |

Table S2: Exponential similarity family. $\mathrm{RMSE}_{\mathrm{emp}}(a)$ and $\mathrm{RMSE}_{\mathrm{th}}(a)$ for anchors $a \in \{0,7,13,20\}$ and $r \in \{0.01,0.02,0.05,0.10,0.20\}$. 'Mean' averages RMSE over the anchors. Parameters: $D = 200$, $Q = 20$, $10000$ realizations. Theory rows are shown directly below the corresponding experimental rows.

| Variant | Generator | $r$ | $a = 0$ | $a = 7$ | $a = 13$ | $a = 20$ | Mean |
|---|---|---|---|---|---|---|---|
| 1 | Baseline | 0.01 | 0.069990 | 0.070443 | 0.070471 | 0.070029 | 0.070233 |
| 1 | Baseline | 0.02 | 0.069192 | 0.070471 | 0.070222 | 0.069411 | 0.069824 |
| 1 | Baseline | 0.05 | 0.067159 | 0.069265 | 0.068889 | 0.067177 | 0.068123 |
| 1 | Baseline | 0.10 | 0.064580 | 0.066283 | 0.066297 | 0.064391 | 0.065388 |
| 1 | Baseline | 0.20 | 0.062802 | 0.063900 | 0.063765 | 0.062674 | 0.063285 |
| 1 theory | Baseline | 0.01 | 0.070344 | 0.070574 | 0.070574 | 0.070344 | 0.070459 |
| 1 theory | Baseline | 0.02 | 0.069576 | 0.070243 | 0.070243 | 0.069576 | 0.069909 |
| 1 theory | Baseline | 0.05 | 0.067087 | 0.068808 | 0.068808 | 0.067087 | 0.067947 |
| 1 theory | Baseline | 0.10 | 0.064610 | 0.066552 | 0.066552 | 0.064610 | 0.065581 |
| 1 theory | Baseline | 0.20 | 0.062886 | 0.063988 | 0.063988 | 0.062886 | 0.063437 |
| | | | | | | | |
| 2 | Fixed-pD | 0.01 | 0.027994 | 0.037622 | 0.042760 | 0.046808 | 0.038796 |
| 2 | Fixed-pD | 0.02 | 0.035522 | 0.048167 | 0.053695 | 0.056630 | 0.048504 |
| 2 | Fixed-pD | 0.05 | 0.043497 | 0.060161 | 0.063309 | 0.062809 | 0.057444 |
| 2 | Fixed-pD | 0.10 | 0.046591 | 0.063618 | 0.064789 | 0.062933 | 0.059483 |
| 2 | Fixed-pD | 0.20 | 0.048205 | 0.063254 | 0.063188 | 0.062249 | 0.059224 |
| 2 theory | Fixed-pD | 0.01 | 0.027939 | 0.037313 | 0.042415 | 0.046449 | 0.038529 |
| 2 theory | Fixed-pD | 0.02 | 0.035425 | 0.048147 | 0.053584 | 0.056716 | 0.048468 |
| 2 theory | Fixed-pD | 0.05 | 0.043370 | 0.060265 | 0.063834 | 0.063591 | 0.057765 |
| 2 theory | Fixed-pD | 0.10 | 0.046576 | 0.063916 | 0.065021 | 0.063262 | 0.059694 |
| 2 theory | Fixed-pD | 0.20 | 0.048104 | 0.063184 | 0.063254 | 0.062142 | 0.059171 |
| 3 | Fixed-F | 0.01 | 0.067897 | 0.068502 | 0.068681 | 0.067513 | 0.068148 |
| 3 | Fixed-F | 0.02 | 0.066774 | 0.068452 | 0.068461 | 0.067436 | 0.067781 |
| 3 | Fixed-F | 0.05 | 0.064483 | 0.066176 | 0.066365 | 0.065058 | 0.065520 |
| 3 | Fixed-F | 0.10 | 0.063150 | 0.064193 | 0.064421 | 0.063198 | 0.063741 |
| 3 | Fixed-F | 0.20 | 0.062106 | 0.062484 | 0.062649 | 0.062134 | 0.062343 |
| 3 theory | Fixed-F | 0.01 | 0.068162 | 0.069050 | 0.069050 | 0.068162 | 0.068606 |
| 3 theory | Fixed-F | 0.02 | 0.066808 | 0.067900 | 0.067900 | 0.066808 | 0.067354 |
| 3 theory | Fixed-F | 0.05 | 0.064891 | 0.065926 | 0.065926 | 0.064891 | 0.065409 |
| 3 theory | Fixed-F | 0.10 | 0.063325 | 0.064251 | 0.064251 | 0.063325 | 0.063788 |
| 3 theory | Fixed-F | 0.20 | 0.062146 | 0.062528 | 0.062528 | 0.062146 | 0.062337 |
| 4 | Fixed-FpD | 0.01 | 0.021855 | 0.034246 | 0.039686 | 0.043175 | 0.034740 |
| 4 | Fixed-FpD | 0.02 | 0.029659 | 0.044771 | 0.050428 | 0.052926 | 0.044446 |
| 4 | Fixed-FpD | 0.05 | 0.039868 | 0.056955 | 0.060582 | 0.061224 | 0.054657 |
| 4 | Fixed-FpD | 0.10 | 0.044793 | 0.061338 | 0.062682 | 0.062097 | 0.057728 |
| 4 | Fixed-FpD | 0.20 | 0.047131 | 0.061349 | 0.061958 | 0.061586 | 0.058006 |
| 4 theory | Fixed-FpD | 0.01 | 0.021906 | 0.034416 | 0.039906 | 0.043148 | 0.034844 |
| 4 theory | Fixed-FpD | 0.02 | 0.029654 | 0.044737 | 0.050548 | 0.053348 | 0.044572 |
| 4 theory | Fixed-FpD | 0.05 | 0.039907 | 0.057010 | 0.060754 | 0.061293 | 0.054741 |
| 4 theory | Fixed-FpD | 0.10 | 0.044786 | 0.061540 | 0.062674 | 0.061957 | 0.057739 |
| 4 theory | Fixed-FpD | 0.20 | 0.047137 | 0.061710 | 0.061781 | 0.061396 | 0.058006 |
| 5 | Fixed-kpD | 0.01 | 0.009452 | 0.005695 | 0.005666 | 0.009426 | 0.007560 |
| 5 | Fixed-kpD | 0.02 | 0.015877 | 0.010271 | 0.010282 | 0.015737 | 0.013042 |
| 5 | Fixed-kpD | 0.05 | 0.025580 | 0.019329 | 0.019330 | 0.025739 | 0.022495 |
| 5 | Fixed-kpD | 0.10 | 0.030465 | 0.026075 | 0.026240 | 0.030640 | 0.028355 |
| 5 | Fixed-kpD | 0.20 | 0.032633 | 0.030875 | 0.030685 | 0.032577 | 0.031692 |
| 5 theory | Fixed-kpD | 0.01 | 0.009408 | 0.005695 | 0.005695 | 0.009408 | 0.007551 |
| 5 theory | Fixed-kpD | 0.02 | 0.015851 | 0.010254 | 0.010254 | 0.015851 | 0.013052 |
| 5 theory | Fixed-kpD | 0.05 | 0.025572 | 0.019262 | 0.019262 | 0.025572 | 0.022417 |
| 5 theory | Fixed-kpD | 0.10 | 0.030413 | 0.026196 | 0.026196 | 0.030413 | 0.028305 |
| 5 theory | Fixed-kpD | 0.20 | 0.032708 | 0.030718 | 0.030718 | 0.032708 | 0.031713 |

Table S3: Exponential similarity family. $\mathrm{RMSE_{emp}}(a)$ and $\mathrm{RMSE_{th}}(a)$ for anchors $a \in \{0,7,13,20\}$ and $r \in \{0.01,0.02,0.05,0.10,0.20\}$. 'Mean' averages RMSE over the anchors. Parameters: $D = 10000$, $Q = 20$, $10000$ realizations. Theory rows are shown directly below the corresponding experimental rows.

| Variant | Generator | $r$ | $a = 0$ | $a = 7$ | $a = 13$ | $a = 20$ | Mean |
|---|---|---|---|---|---|---|---|
| 1 | Baseline | 0.01 | 0.009998 | 0.010024 | 0.010011 | 0.009966 | 0.010000 |
| 1 | Baseline | 0.02 | 0.009877 | 0.009943 | 0.009906 | 0.009813 | 0.009885 |
| 1 | Baseline | 0.05 | 0.009497 | 0.009730 | 0.009750 | 0.009485 | 0.009616 |
| 1 | Baseline | 0.10 | 0.009148 | 0.009402 | 0.009400 | 0.009131 | 0.009270 |
| 1 | Baseline | 0.20 | 0.008907 | 0.009100 | 0.009054 | 0.008902 | 0.008991 |
| 1 theory | Baseline | 0.01 | 0.009948 | 0.009981 | 0.009981 | 0.009948 | 0.009964 |
| 1 theory | Baseline | 0.02 | 0.009840 | 0.009934 | 0.009934 | 0.009840 | 0.009887 |
| 1 theory | Baseline | 0.05 | 0.009488 | 0.009731 | 0.009731 | 0.009488 | 0.009609 |
| 1 theory | Baseline | 0.10 | 0.009137 | 0.009412 | 0.009412 | 0.009137 | 0.009275 |
| 1 theory | Baseline | 0.20 | 0.008893 | 0.009049 | 0.009049 | 0.008893 | 0.008971 |
| | | | | | | | |
| 2 | Fixed-pD | 0.01 | 0.003993 | 0.005334 | 0.006073 | 0.006626 | 0.005507 |
| 2 | Fixed-pD | 0.02 | 0.005027 | 0.006832 | 0.007622 | 0.008076 | 0.006889 |
| 2 | Fixed-pD | 0.05 | 0.006181 | 0.008545 | 0.009080 | 0.009006 | 0.008203 |
| 2 | Fixed-pD | 0.10 | 0.006575 | 0.009017 | 0.009182 | 0.008926 | 0.008425 |
| 2 | Fixed-pD | 0.20 | 0.006807 | 0.008980 | 0.008996 | 0.008817 | 0.008400 |
| 2 theory | Fixed-pD | 0.01 | 0.003951 | 0.005277 | 0.005998 | 0.006569 | 0.005449 |
| 2 theory | Fixed-pD | 0.02 | 0.005010 | 0.006809 | 0.007578 | 0.008021 | 0.006854 |
| 2 theory | Fixed-pD | 0.05 | 0.006133 | 0.008523 | 0.009027 | 0.008993 | 0.008169 |
| 2 theory | Fixed-pD | 0.10 | 0.006587 | 0.009039 | 0.009195 | 0.008947 | 0.008442 |
| 2 theory | Fixed-pD | 0.20 | 0.006803 | 0.008936 | 0.008946 | 0.008788 | 0.008368 |
| | | | | | | | |
| 3 | Fixed-F | 0.01 | 0.009719 | 0.009842 | 0.009872 | 0.009715 | 0.009787 |
| 3 | Fixed-F | 0.02 | 0.009505 | 0.009640 | 0.009605 | 0.009543 | 0.009573 |
| 3 | Fixed-F | 0.05 | 0.009136 | 0.009311 | 0.009283 | 0.009172 | 0.009226 |
| 3 | Fixed-F | 0.10 | 0.008943 | 0.009039 | 0.009107 | 0.008931 | 0.009005 |
| 3 | Fixed-F | 0.20 | 0.008788 | 0.008843 | 0.008885 | 0.008857 | 0.008843 |
| 3 theory | Fixed-F | 0.01 | 0.009639 | 0.009765 | 0.009765 | 0.009639 | 0.009702 |
| 3 theory | Fixed-F | 0.02 | 0.009447 | 0.009602 | 0.009602 | 0.009447 | 0.009525 |
| 3 theory | Fixed-F | 0.05 | 0.009176 | 0.009322 | 0.009322 | 0.009176 | 0.009249 |
| 3 theory | Fixed-F | 0.10 | 0.008955 | 0.009085 | 0.009085 | 0.008955 | 0.009020 |
| 3 theory | Fixed-F | 0.20 | 0.008788 | 0.008842 | 0.008842 | 0.008788 | 0.008815 |
| | | | | | | | |
| 4 | Fixed-FpD | 0.01 | 0.003085 | 0.004834 | 0.005607 | 0.006038 | 0.004891 |
| 4 | Fixed-FpD | 0.02 | 0.004179 | 0.006270 | 0.007050 | 0.007473 | 0.006243 |
| 4 | Fixed-FpD | 0.05 | 0.005630 | 0.008035 | 0.008585 | 0.008652 | 0.007726 |
| 4 | Fixed-FpD | 0.10 | 0.006341 | 0.008764 | 0.008834 | 0.008749 | 0.008172 |
| 4 | Fixed-FpD | 0.20 | 0.006648 | 0.008676 | 0.008741 | 0.008658 | 0.008181 |
| 4 theory | Fixed-FpD | 0.01 | 0.003092 | 0.004857 | 0.005633 | 0.006091 | 0.004918 |
| 4 theory | Fixed-FpD | 0.02 | 0.004187 | 0.006315 | 0.007138 | 0.007535 | 0.006294 |
| 4 theory | Fixed-FpD | 0.05 | 0.005639 | 0.008053 | 0.008585 | 0.008663 | 0.007735 |
| 4 theory | Fixed-FpD | 0.10 | 0.006331 | 0.008698 | 0.008861 | 0.008760 | 0.008163 |
| 4 theory | Fixed-FpD | 0.20 | 0.006665 | 0.008725 | 0.008736 | 0.008682 | 0.008202 |
| | | | | | | | |
| 5 | Fixed-kpD | 0.01 | 0.001325 | 0.000800 | 0.000800 | 0.001328 | 0.001063 |
| 5 | Fixed-kpD | 0.02 | 0.002228 | 0.001438 | 0.001449 | 0.002224 | 0.001835 |
| 5 | Fixed-kpD | 0.05 | 0.003614 | 0.002709 | 0.002697 | 0.003626 | 0.003162 |
| 5 | Fixed-kpD | 0.10 | 0.004265 | 0.003712 | 0.003688 | 0.004280 | 0.003986 |
| 5 | Fixed-kpD | 0.20 | 0.004625 | 0.004344 | 0.004348 | 0.004625 | 0.004486 |
| 5 theory | Fixed-kpD | 0.01 | 0.001325 | 0.000802 | 0.000802 | 0.001325 | 0.001063 |
| 5 theory | Fixed-kpD | 0.02 | 0.002232 | 0.001444 | 0.001444 | 0.002232 | 0.001838 |
| 5 theory | Fixed-kpD | 0.05 | 0.003604 | 0.002713 | 0.002713 | 0.003604 | 0.003159 |
| 5 theory | Fixed-kpD | 0.10 | 0.004289 | 0.003692 | 0.003692 | 0.004289 | 0.003990 |
| 5 theory | Fixed-kpD | 0.20 | 0.004613 | 0.004331 | 0.004331 | 0.004613 | 0.004472 |

#### 4.2.2 Discussion

Across all exponential derandomization variants and flip-rate settings $r$, the theoretical predictions closely track the experimental RMSE values in Table 3, both in overall magnitude and in the systematic anchor-to-anchor differences. RMSE decreases in the order Baseline $\approx$ Fixed-F $>$ Fixed-pD $>$ Fixed-FpD $>$ Fixed-kpD. The main reason for the pattern Baseline $\approx$ Fixed-F $>$ Fixed-pD is constant-weight initialization $|\mathbf{x}_0| = D/2$ of Fixed-pD. In Baseline and Fixed-F, random initialization produces a realization-dependent offset in the similarity curve, which increases variance across realizations. This appears to be the dominant error contribution, keeping Fixed-F close to Baseline.

Anchor-dependent variations arise because anchor-wise RMSE averages over comparison levels $b$, so different anchors produce different distributions of separations $\Delta$. Interior anchors contain a larger proportion of small-separation comparisons, whereas endpoint anchors contain a larger proportion of large-separation comparisons. Therefore, the direction and magnitude of the anchor effect depend on how each generator's per-separation squared-error contribution varies with $\Delta$.

Baseline and Fixed-F use $\mathrm{Bernoulli}(1/2)$ initialization, so the separation dependence in the finite-D variance contains an explicit Bernoulli-start term $\mathrm{Var}(A_a{+}A_b) = 2D{+}2D{\cdot}\rho_{\mathrm{step}}^{\Delta}$. This term is largest at $\Delta = 0$ and decays as $\Delta$ increases, while the remaining contribution $\mathrm{Var}(C_{\mathrm{ab}})$ approaches a nearly $\Delta$-independent long-separation regime. Consequently, over the relevant range of $\Delta$, the per-separation error contribution is slightly larger at small $\Delta$ than at large $\Delta$, so interior anchors (which contain a larger proportion of small-$\Delta$ comparisons) exhibit slightly larger anchor-wise $\mathrm{RMSE}_{\mathrm{emp}}(a)$ than endpoint anchors.

Fixed-pD and Fixed-FpD use constant Hamming-weight initialization. Here the dominant anchor-to-anchor difference is driven by the special role of level 0. $\mathrm{Var}(A_a{+}A_b)$ takes the form $D(2 - \gamma_{\mathrm{step}}^{a} - \gamma_{\mathrm{step}}^{b}{+}2\rho_{\mathrm{step}}^{\Delta}(1 - \gamma_{\mathrm{step}}^{z}))$ with $\mathrm{z} = \min(a{,}b)$. With $a = 0$, we have $z = 0$ for every $b$, so $(1 - \gamma_{\mathrm{step}}^{z}) = 0$ and this distance-dependent contribution is suppressed uniformly across $b$, reducing overlap variability and lowering $\mathrm{RMSE}_{\mathrm{emp}}(0)$. For $a > 0$, this suppression does not apply in the same way, so the remaining anchors cluster at a higher, typical RMSE level; any additional effect of the separation distribution for endpoints versus interior anchors is secondary to this $a = 0$ effect.

Fixed-kpD preserves constant $|\mathbf{x}_q| = D/2$ for all levels $q$, so $S(a{,}a) = 1$ exactly and the $\Delta = 0$ error contribution is strongly suppressed. As $\Delta$ increases, successive equal-swap updates make the overlap increasingly variable and it approaches its long-separation regime, so the per-separation error contribution is larger at larger $\Delta$. Endpoint anchors (which contain a larger proportion of large-$\Delta$ comparisons) therefore have larger anchor-wise $\mathrm{RMSE_{emp}}(a)$ than interior anchors, opposite to Baseline and Fixed-F.

**Dependence on $r$.** As $r$ increases, $\rho_{\mathrm{step}}$ decreases in magnitude, so the mean/variance terms involving powers of $\rho_{\mathrm{step}}$ (and $\gamma_{\mathrm{step}}$) reach their long-separation regime at smaller $\Delta$. As a result, the pair-wise error depends less on separation $\Delta$ over the evaluated range, and the differences in separation distributions of endpoint and interior anchors have less impact; the spread across endpoint and interior anchors narrows.

### 4.2.3 Linear Similarity Family

Table S4: Linear similarity family. $\mathrm{RMSE_{emp}}(a)$ and $\mathrm{RMSE_{th}}(a)$ for anchors $a \in \{0,33,67,100\}$. Parameters: $D = 10000$, $Q = 100$, $10000$ realizations. Bias is reported for Baseline and Fixed-$pD$. Theory rows are shown directly below the corresponding experimental rows for each generator.

RMSE

| Variant | Generator | $a = 0$ | $a = 33$ | $a = 67$ | $a = 100$ | Mean |
|---|---|---|---|---|---|---|
| 1 | Baseline (Pool-Bernoulli) | 0.009980 | 0.009991 | 0.009964 | 0.009925 | 0.009965 |
| 1 | Baseline (Perm-Scan) | 0.009987 | 0.009980 | 0.009993 | 0.010007 | 0.009992 |
| 1 theory | Baseline (Perm-Scan) | 0.009937 | 0.009970 | 0.009993 | 0.009956 | 0.009964 |
| 2 | Fixed-pD (Pool-Bernoulli) | 0.008067 | 0.009322 | 0.009308 | 0.008035 | 0.008683 |
| 2 | Fixed-pD (Perm-Scan) | 0.008114 | 0.009450 | 0.009438 | 0.008157 | 0.008790 |
| 2 theory | Fixed-pD (Perm-Scan) | 0.008077 | 0.009381 | 0.009405 | 0.008125 | 0.008747 |
| 3 | Fixed-F (Pool-Uniform) | 0.007113 | 0.008513 | 0.008507 | 0.007107 | 0.007810 |
| 3 | Fixed-F (Perm-Scan) | 0.007025 | 0.008446 | 0.008523 | 0.007063 | 0.007764 |
| 3 theory | Fixed-F (Perm-Scan) | 0.007071 | 0.008479 | 0.008479 | 0.007071 | 0.007775 |
| 4 | Fixed-FpD (Pool-Uniform) | 0.004069 | 0.007824 | 0.007778 | 0.004069 | 0.005935 |
| 4 | Fixed-FpD (Perm-Scan) | 0.004074 | 0.007766 | 0.007817 | 0.004074 | 0.005933 |
| 4 theory | Fixed-FpD (Perm-Scan) | 0.004062 | 0.007779 | 0.007779 | 0.004062 | 0.005920 |
| 5 | Fixed-kpD (Block-Swap) | 0.000000 | 0.000000 | 0.000000 | 0.000000 | 0.000000 |
| 5 | Fixed-kpD (Float) | 0.000000 | 0.000000 | 0.000000 | 0.000000 | 0.000000 |
| 5 theory | Fixed-kpD (Float) | 0.000000 | 0.000000 | 0.000000 | 0.000000 | 0.000000 |

Bias

| Variant | Generator | $a = 0$ | $a = 33$ | $a = 67$ | $a = 100$ | Mean |
|---|---|---|---|---|---|---|
| 1 | Baseline (Pool-Bernoulli) | 0.000396 | 0.000411 | 0.000410 | 0.003784 | 0.001250 |
| 1 | Baseline (Perm-Scan) | 0.000407 | 0.000412 | 0.000417 | 0.003994 | 0.001308 |
| 1 theory | Baseline (Perm-Scan) | 0.000403 | 0.000403 | 0.000403 | 0.003942 | 0.001288 |
| 2 | Fixed-pD (Pool-Bernoulli) | 0.000393 | 0.000400 | 0.000398 | 0.003880 | 0.001268 |
| 2 | Fixed-pD (Perm-Scan) | 0.000412 | 0.000419 | 0.000416 | 0.003952 | 0.001300 |
| 2 theory | Fixed-pD (Perm-Scan) | 0.000403 | 0.000403 | 0.000403 | 0.003942 | 0.001288 |

Table S5: Linear similarity family. $\mathrm{RMSE_{emp}}(a)$ and $\mathrm{RMSE_{th}}(a)$ for anchors $a \in \{0,7,13,20\}$. Parameters: $D = 200$, $Q = 20$, $10000$ realizations. Bias is reported for Baseline and Fixed-$pD$. Theory rows are shown directly below the corresponding experimental rows for each generator.

RMSE

| Variant | Generator | $a = 0$ | $a = 7$ | $a = 13$ | $a = 20$ | Mean |
|---|---|---|---|---|---|---|
| 1 | Baseline (Pool-Bernoulli) | 0.068257 | 0.069488 | 0.070334 | 0.068268 | 0.069087 |
| 1 | Baseline (Perm-Scan) | 0.068546 | 0.069521 | 0.070535 | 0.069300 | 0.069476 |
| 1 theory | Baseline (Perm-Scan) | 0.068332 | 0.069673 | 0.070474 | 0.069197 | 0.069419 |
| 2 | Fixed-pD (Pool-Bernoulli) | 0.053810 | 0.064403 | 0.065629 | 0.054900 | 0.059686 |
| 2 | Fixed-pD (Perm-Scan) | 0.054832 | 0.065661 | 0.066467 | 0.057382 | 0.061086 |
| 2 theory | Fixed-pD (Perm-Scan) | 0.054348 | 0.065366 | 0.066158 | 0.056626 | 0.060625 |
| 3 | Fixed-F (Pool-Uniform) | 0.049897 | 0.059967 | 0.059958 | 0.049912 | 0.054934 |
| 3 | Fixed-F (Perm-Scan) | 0.050201 | 0.060263 | 0.059898 | 0.049782 | 0.055036 |
| 3 theory | Fixed-F (Perm-Scan) | 0.050000 | 0.059861 | 0.059861 | 0.050000 | 0.054930 |
| 4 | Fixed-FpD (Pool-Uniform) | 0.028381 | 0.055694 | 0.055334 | 0.028381 | 0.041947 |
| 4 | Fixed-FpD (Perm-Scan) | 0.028074 | 0.054487 | 0.054903 | 0.028074 | 0.041384 |
| 4 theory | Fixed-FpD (Perm-Scan) | 0.028207 | 0.055024 | 0.055024 | 0.028207 | 0.041616 |
| 5 | Fixed-kpD (Block-Swap) | 0.000000 | 0.000000 | 0.000000 | 0.000000 | 0.000000 |
| 5 | Fixed-kpD (Float) | 0.000000 | 0.000000 | 0.000000 | 0.000000 | 0.000000 |
| 5 theory | Fixed-kpD (Float) | 0.000000 | 0.000000 | 0.000000 | 0.000000 | 0.000000 |

Bias

| Variant | Generator | $a = 0$ | $a = 7$ | $a = 13$ | $a = 20$ | Mean |
|---|---|---|---|---|---|---|
| 1 | Baseline (Pool-Bernoulli) | 0.005961 | 0.006220 | 0.006282 | 0.025398 | 0.010965 |
| 1 | Baseline (Perm-Scan) | 0.006569 | 0.006411 | 0.006444 | 0.026695 | 0.011530 |
| 1 theory | Baseline (Perm-Scan) | 0.006329 | 0.006329 | 0.006329 | 0.026345 | 0.011333 |
| 2 | Fixed-pD (Pool-Bernoulli) | 0.005841 | 0.005883 | 0.005665 | 0.024420 | 0.010452 |
| 2 | Fixed-pD (Perm-Scan) | 0.006588 | 0.006495 | 0.006460 | 0.026964 | 0.011627 |
| 2 theory | Fixed-pD (Perm-Scan) | 0.006329 | 0.006329 | 0.006329 | 0.026345 | 0.011333 |

Table S6: Linear similarity family. $\mathrm{RMSE_{emp}}(a)$ and $\mathrm{RMSE_{th}}(a)$ for anchors $a \in \{0,7,13,20\}$. Parameters: $D = 10000$, $Q = 20$, $10000$ realizations. Bias is reported for Baseline and Fixed-

$pD$. Theory rows are shown directly below the corresponding experimental rows for each generator.

RMSE

| Variant | Generator | $a = 0$ | $a = 7$ | $a = 13$ | $a = 20$ | Mean |
|---|---|---|---|---|---|---|
| 1 | Baseline (Pool-Bernoulli) | 0.009668 | 0.009825 | 0.009844 | 0.009526 | 0.009716 |
| 1 | Baseline (Perm-Scan) | 0.009788 | 0.009894 | 0.009972 | 0.009883 | 0.009884 |
| 1 theory | Baseline (Perm-Scan) | 0.009760 | 0.009896 | 0.009964 | 0.009778 | 0.009849 |
| 2 | Fixed-pD (Pool-Bernoulli) | 0.007701 | 0.009185 | 0.009283 | 0.007492 | 0.008415 |
| 2 | Fixed-pD (Perm-Scan) | 0.007789 | 0.009342 | 0.009403 | 0.007885 | 0.008605 |
| 2 theory | Fixed-pD (Perm-Scan) | 0.007815 | 0.009301 | 0.009374 | 0.007864 | 0.008589 |
| 3 | Fixed-F (Pool-Uniform) | 0.007042 | 0.008374 | 0.008420 | 0.007118 | 0.007739 |
| 3 | Fixed-F (Perm-Scan) | 0.007025 | 0.008436 | 0.008493 | 0.007063 | 0.007754 |
| 3 theory | Fixed-F (Perm-Scan) | 0.007071 | 0.008466 | 0.008466 | 0.007071 | 0.007768 |
| 4 | Fixed-FpD (Pool-Uniform) | 0.003960 | 0.007685 | 0.007728 | 0.003960 | 0.005834 |
| 4 | Fixed-FpD (Perm-Scan) | 0.003991 | 0.007772 | 0.007799 | 0.003991 | 0.005888 |
| 4 theory | Fixed-FpD (Perm-Scan) | 0.003979 | 0.007762 | 0.007762 | 0.003979 | 0.005871 |
| 5 | Fixed-kpD (Block-Swap) | 0.000000 | 0.000000 | 0.000000 | 0.000000 | 0.000000 |
| 5 | Fixed-kpD (Float) | 0.000000 | 0.000000 | 0.000000 | 0.000000 | 0.000000 |
| 5 theory | Fixed-kpD (Float) | 0.000000 | 0.000000 | 0.000000 | 0.000000 | 0.000000 |

Bias

| Variant | Generator | $a = 0$ | $a = 7$ | $a = 13$ | $a = 20$ | Mean |
|---|---|---|---|---|---|---|
| 1 | Baseline (Pool-Bernoulli) | 0.000795 | 0.000803 | 0.000796 | 0.003547 | 0.001485 |
| 1 | Baseline (Perm-Scan) | 0.000842 | 0.000868 | 0.000877 | 0.003826 | 0.001603 |
| 1 theory | Baseline (Perm-Scan) | 0.000849 | 0.000849 | 0.000849 | 0.003795 | 0.001585 |
| 2 | Fixed-pD (Pool-Bernoulli) | 0.000796 | 0.000814 | 0.000805 | 0.003597 | 0.001503 |
| 2 | Fixed-pD (Perm-Scan) | 0.000848 | 0.000857 | 0.000862 | 0.003791 | 0.001589 |
| 2 theory | Fixed-pD (Perm-Scan) | 0.000849 | 0.000849 | 0.000849 | 0.003795 | 0.001585 |

### 4.2.4 Discussion

Across the generators listed for each derandomization variant in Table 4, the theoretical predictions match the experiments well in both overall RMSE and anchor-dependent trends. RMSE decreases in the order Baseline $>$ Fixed-pD $>$ Fixed-F $>$ Fixed-FpD $>$ Fixed-kpD. The fixed-count variants (Fixed-F, Fixed-FpD, Fixed-kpD) are essentially unbiased, whereas the random-count variants (Baseline, Fixed-pD) exhibit nonzero bias due to finite-capacity effects near level $Q$. Fixed-kpD (Block-Swap / Float) is deterministic by construction and matches the linear target exactly, so both bias and variance vanish (RMSE is zero up to numerical precision).

To understand Fixed-pD $>$ Fixed-F, note that the key difference is whether the number $K$ of newly flipped components between levels is deterministic. Fixed-F has $K = \Delta\text{F}$, removing the additional $\text{Var}(K)/D^2$ contribution present when $K$ is random, so bias is near zero and variance is smaller. Fixed-pD retains random per-level proposed count together with the "cannot exceed $D$" constraint near level $Q$, which induces extra bias near the end of the axis and extra variance through randomness in $K$.

Concerning anchor dependence, for Fixed-F and Fixed-FpD, the finite-D variance decreases linearly with separation $\Delta$ through the factor $D - \Delta F$: as $\Delta$ grows, fewer components remain unchanged and able to contribute overlap, so $\text{Var}(S(a,b))$ drops accordingly. In Fixed-F, since interior anchors contain a larger proportion of small-separation comparisons under the $b$-averaging protocol, their anchor-wise average therefore contains a larger proportion of the higher-variance small-$\Delta$ regime and therefore exhibits larger $\text{RMSE}_{\text{emp}}(a)$ than endpoint anchors. In Fixed-FpD, constant Hamming-weight initialization further suppresses overlap variability when one of the two levels is an endpoint $a = 0$ or $a = Q$ (both have $D/2$ ones by construction). Consequently $\text{RMSE}_{\text{emp}}(a)$ is noticeably smaller at endpoint anchors, and the endpoint/interior contrast is stronger than in Fixed-F.

In Baseline and Fixed-pD variants, the per-level flip count is random and constrained by the remaining pool of not-yet-flipped components. This induces bias when one of the levels is close to $Q$: the expected number of new distinct flips between two nearby levels falls below the ideal linear count $\Delta \cdot (D/Q)$, so the mean similarity deviates from $T_{\text{lin}}$. This effect is most visible for anchors near $Q$, where many $(a,b)$ pairs involve an "almost fully flipped" level. At the same time, the remaining-pool constraint reduces run-to-run randomness near pool exhaustion because fewer admissible choices remain, so the variance contribution decreases near $Q$ even while bias increases. Relative to Baseline, the constant Hamming-weight initialization of Fixed-pD reduces the overall variance while leaving the qualitative terminal-bias pattern unchanged.

Across dimensionality $D$, RMSE follows the expected $1/\sqrt{D}$ scaling.